%% file: paper.tex
\documentclass[a4paper,11pt]{article}
\usepackage{placeins}
\usepackage{graphicx}
\usepackage{xcolor}
\usepackage{float}
\usepackage{afterpage}
\usepackage{amssymb,amsmath,bm}
\usepackage{multirow,booktabs,ulem}
\usepackage{cite}
\usepackage{jheppub}
\usepackage{hyperref}
\usepackage{slashed}
\usepackage{enumitem}
\usepackage{tabularx,booktabs}
\usepackage{multicol}
\usepackage{array,multirow}
\usepackage{booktabs}
\usepackage{colortbl}
\usepackage{float}
\usepackage{pifont}

\usepackage{caption}
\usepackage{subcaption}

\newcommand{\mtc}{\mathcal}
\newcommand{\ep}{\epsilon}

\newcommand{\expval}{\mathbb{E}}

\renewcommand{\to}{\rightarrow}

\newcommand{\be}{\begin{equation}}
\newcommand{\ee}{\end{equation}}
\DeclareMathOperator*{\argmin}{argmin} 

\title{Evaluating the faithfulness of PDF uncertainties in the presence of inconsistent data}

\preprint{TIF-UNIMI-2025-7}
\author[a]{Andrea Barontini,}
\author[c]{Mark N. Costantini,}
\author[b]{Giovanni De Crescenzo,}
\author[a]{Stefano Forte,}
\author[c]{ and Maria Ubiali}
\affiliation[a]{TIF Lab, Dipartimento di Fisica, Università degli Studi di Milano and INFN Sezione di
Milano, Via Celoria 16, 20133, Milano, Italy}
\affiliation[b]{Institut für Theoretische Physik, Universität Heidelberg, Germany}
\affiliation[c]{DAMTP, University of Cambridge, Wilberforce Road, Cambridge CB3 0WA, UK }

\emailAdd{andrea.barontini@unimi.it}
\emailAdd{g.de.crescenzo@thphys.uni-heidelberg.de}
\emailAdd{mnc33@cam.ac.uk}
\emailAdd{M.Ubiali@damtp.cam.ac.uk}

\date{\today}

\abstract{
  We critically assess the
  robustness of uncertainties on parton distribution functions (PDFs)
  determined using neural networks from global sets of experimental
  data collected from  multiple experiments. We view the
  determination of PDFs as an inverse problem, and we study the
  way the neural network model tackles it when inconsistencies
  between input datasets are present. We use a closure test approach,
  in which the regression model is applied to  artificial  data
  produced from a known underlying truth, to which the output of the
  model can be compared and its accuracy can be assessed in a
  statistically reliable way. We explore various
  phenomenologically relevant scenarios in
  which inconsistencies arise due to incorrect estimation of
  correlated systematic uncertainties. We show that the neural network
  generally corrects for the inconsistency except in cases of extreme
  uncertainty underestimation. When the inconsistency is not
  corrected, we propose and validate a procedure to detect inconsistencies.}


\begin{document}
\maketitle

\input{sec-intro.tex}
\input{sec-definition.tex}

\input{sec-results.tex}

\input{sec-prescription.tex}
\input{sec-summary.tex}

\section*{Acknowledgments}
We are extremely grateful to the members of the NNPDF collaboration for their suggestions regarding
this work; in particular, we are indebted to Luigi Del Debbio and Juan M. Cruz-Martinez for
useful discussions and comments. We thank Lucian Harland-Lang for
discussions on the subject of the paper, Felix Hekhorn and Ramon Winterhalder for a
critical reading of the manuscript.
S.~F. is partly funded by the European Union NextGeneration EU program,
NRP Mission 4 Component 2 Investment 1.1 – MUR PRIN 2022 – CUP
G53D23001100006 through the Italian Ministry of University and
Research.
G.~D.~C. is supported by the KISS consortium (05D2022) funded by the German Federal Ministry of Education and Research BMBF in the ErUM-Data action plan.
M.~N.~C. and M.~U. are supported by the European Research Council under the European Union’s
Horizon 2020 research and innovation Programme (grant agreement n.950246), and partially by the STFC
consolidated grant ST/T000694/1 and ST/X000664/1.

\appendix
\input{app-comparison-est.tex}
\input{app-bootstrap_def.tex}

\input{app-pdfobscorr.tex}

\bibliographystyle{JHEP}
\bibliography{references}

\end{document}

%% file: sec-intro.tex
\section{Introduction}
\label{sec:intro}

The accurate estimate of uncertainties on
Parton Distribution Functions (PDFs) is necessary for precision phenomenology, see~\cite{Amoroso:2022eow,Ubiali:2024pyg} for recent reviews.
Because PDFs are extracted by comparing to experimental data theory
predictions obtained using them, their determination falls into
the general category of inverse problems, {\it i.e.}\ problems in which a model
is inferred starting from a finite set of observations, which are both noisy and sometimes
incompatible with one another. Machine Learning is currently
widely used in this context
see Refs.~\cite{Guest:2018yhq,Albertsson:2018maf,Plehn:2022ftl} for comprehensive reviews.

The NNPDF collaboration uses machine learning techniques for the determination of
PDFs and their uncertainties, by combining a parametrization of PDFs through a feed--forward neural network with a 
Monte Carlo importance sampling procedure to propagate the data
uncertainty in the space of PDFs. The faithfulness of a regression model's uncertainties 
can be assessed by means of a closure test. In the
context of PDFs, this has been first done by NNPDF in
Ref.~\cite{Ball:2014uwa}, following a suggestion in
Ref.~\cite{demortier} and previous studies in
Ref.~\cite{Watt:2012tq}. A detailed theoretical discussion of the statistical underpinnings of the
closure testing methodology is given in
Ref.~\cite{DelDebbio:2021whr}. Similar studies have been recently
also performed in the context of PDF fitting methodologies based on a
fixed polynomial parametrization and Hessian
approach for uncertainty propagation in parameter space~\cite{Harland-Lang:2024kvt}. 
The idea of closure tests is to apply the regression model to
artificial data generated from a known underlying truth. In this case the
underlying law of Nature ({\it i.e.}\ the PDFs)
is known. The model is however run without exploiting this knowledge,
as would be the case in a realistic situation.
By performing the procedure repeatedly it is
possible to test whether the distribution of results about the
underlying truth faithfully reproduces their nominal uncertainties
(including their correlation), thereby validating the accuracy of the
model uncertainty estimate.

Closure tests so far have been performed on a set of consistent artificial data, meaning that the theory predictions and
the uncertainties that are used in the generation of the data are the
same ones that the model then uses for regression. However, in a realistic
case both theory prediction and uncertainty could be inconsistent. The
former case corresponds to a situation in which the theory used in
regression is the Standard Model (SM), while the data actually contain
physics beyond the SM.  This situation has been studied in   
Refs.~\cite{Hammou:2023heg,Costantini:2024xae,Hammou:2024xuj}, in which closure tests
were used to investigate whether in such a situation possible new
physics would be detected. The latter case corresponds to a situation
in which the actual statistical distribution of data differs from the covariance
matrix that is supposed to describe them, for instance because some
sources of systematic uncertainty have not been correctly estimated.

In this work we explore what happens when 
inconsistencies of experimental origin are present in the training dataset.
The use of 
closure tests in a context in which an inconsistency is injected in
the data in a controlled
way allows testing the behavior of the model in a situation that may
well happen in
real-world situations.
As  by-products of our investigation, we consolidate and improve
the closure test methodology, and we validate
a procedure aimed at singling out inconsistent datasets, improving
over a previously proposed procedure~\cite{NNPDF:2021njg}.

The paper is organised as follows. In Sect.~\ref{sec:definition} we
summarise our methodology: first, we briefly review the
NNPDF methodology for PDF determination, then we summarise the
principles of closure testing it in the approach of Refs.~\cite{Ball:2014uwa,DelDebbio:2021whr}, and finally we discuss how
inconsistencies can be introduced in the closure testing methodology.
In Sect.~\ref{sec:results} we present
results in different representative inconsistency scenarios that differ both
in the degree of inconsistency introduced, in the relative
size of the inconsistent dataset compared to the global dataset, and
in the presence or absence of other consistent datasets similar in
nature to the inconsistent one, specifically assessing the extent to which
the model manages to correct for the inconsistency. In Sect.~\ref{sec:recipe} we construct
and validate a diagnostic tool that
can be used to detect situations in which the inconsistency is not
corrected, improving upon a methodology previously proposed and adopted
by the NNPDF collaboration in Ref.~\cite{NNPDF:2021njg}. 

%% file: sec-definition.tex
\section{Closure tests and inconsistent data}
\label{sec:definition}

We discuss here the idea of closure testing the impact of
inconsistent data. We start by presenting in
Sect.~\ref{subsec:nn} a brief review of the way PDFs and their
uncertainties are obtained within the
NNPDF approach, in
Sect.~\ref{subsec:est} we then briefly summarise the way closure tests can
be used to test it, specifically by defining suitable statistical
estimators, and finally in Sect.~\ref{sec:approach} we present the way
the effect of data inconsistencies can be studied through closure tests.

\subsection{The NNPDF approach to PDF determination}
\label{subsec:nn}

The NNPDF approach uses the Monte Carlo (MC) replica method for the
propagation of experimental and model uncertainties onto the PDF
functional space. The MC replica method consists in determining a set of fit outcomes to approximate
the posterior probability distribution of the PDF model given a set of
experimental input data. The input data are in turn represented as a
MC sample of $N_{\rm rep}$ pseudodata replicas whose distribution (typically a multivariate normal)
reproduces the covariance matrix of the experimental data. The fit
outcomes are determined by comparing to the data replicas predictions
that depend on PDFs via a forward
map ${\cal G}$.
The MC replica method shares similarities with the parametric
bootstrap~\cite{Hall1992TheBA} approach in statistical literature, see
Ref.~\cite{Costantini:2024wby} for an analytical expression
for the posterior distribution of the model derived from this method.

Following the notation of Ref.~\cite{DelDebbio:2021whr}, if we
assume that the observational noise of the experimental
data can be approximated as a vector drawn
from a multivariate normal distribution with a given covariance matrix
$C$, 
the central experimental values, $y_0$, are given by
\begin{align}
    y_0 &= f + \eta,
    \label{eq:l1_data}
\end{align}
where $f \in \mathbb{R}^{N_{\rm data}}$ is the vector of true,
thus unknown, observable values and $\eta$ is the
observational noise drawn from a Gaussian  $\mtc{N}(0,C)$ centred on zero with
covariance matrix $C$. This covariance matrix contains both
statistical uncertainties (mostly but not always uncorrelated) and
systematic uncertainties (mostly correlated). Note that some
systematic uncertainties are
actually of theoretical origin, in that they affect the
relation of the quantity that is actually measured to the experimental
(pseudo)-observable  for which theory predictions are
obtained. Examples are  electroweak corrections, sometimes already
subtracted by experimental collaborations in order to obtain their
published data, or nuclear corrections that relate nuclear to proton
structure functions.
The  vector $\eta$ accordingly induces correlations among data points
that reflect those contained in the $C$ matrix.

In the NNPDF approach, pseudodata replicas of the data are 
generated by augmenting $y_0$ with noise $\epsilon^{(k)}$
\begin{align}
   \mu^{(k)} &= y_0 + \epsilon^{(k)} = f + \eta +  \epsilon^{(k)}, \quad k=1,...,N_{\rm rep},
    \label{eq:l2_pseudodata}
\end{align}
where $k$ is the replica index and $N_{\rm rep}$ is the number of pseudodata replicas.
Each instance of the noise $\epsilon^{(k)}$ is drawn independently from the same
multi-Gaussian distribution $\mtc{N}(0, C)$, which the
observational noise is  drawn from. This  implies that 
the covariance of any two data points over the replica sample in the limit of large  
$N_{\rm rep}$ tends to the corresponding $C$ matrix element.

An optimal model replica is then determined for each data replica by
minimizing a loss function computed on a training subset of data and 
stopping the training conditionally on the loss computed on the
remaining (validation) data subset. Namely, by determining for each
replica $k$ the values $u_{*,k}$  of
the parameters $u$ of a model $w(u)$ of the PDFs $w$ such that
\begin{align}
  u_{*,k} &= \argmin_{u_k} \left[\chi_{\rm val}^{2 (k)}| \argmin_{u_k}\chi_{\rm tr}^{2 (k)}\right], \quad k=1,...,N_{\rm rep}, 
    \label{eq:minimisation_problem}
\end{align}
where $\chi_{\rm tr}^{2 (k)}$ and $\chi_{\rm val}^{2 (k)}$ are the
training and validation loss, and the notation means that the minimum
validation loss is picked conditional to the minimisation of the
training loss.

The loss function is chosen as a maximum likelihood estimator, namely
\begin{align}
  \chi^{2 (k)}(C) & = \frac{1}{N_{\rm data}} \left({\cal G}(w(u_k)) - \mu^{(k)}\right)^T
                 C^{-1} \left({\cal G}(w(u_k)) - \mu^{(k)}\right),
    \label{eq:chi2_definition}
\end{align}
where ${\cal G}(w)$ is  the forward map,
that maps the PDF  $w$ to the measured observable to which the
pseudodata replica  $\mu^{(k)}$ values Eq.~(\ref{eq:l2_pseudodata})
refer to. The forward map is constructed by evolving the PDF to the
appropriate scale and folding them with partonic cross-sections to get
observables. The training and validation losses
$\chi_{\rm tr}^{2 (k)}$ and $\chi_{\rm val}^{2 (k)}$ are
respectively defined by only including in
Eq.~(\ref{eq:chi2_definition}) the data in the respective subset.
The final outcome of the procedure is thus a set of PDF model replicas
$w(u_{*,k})$ from which the predicted PDFs can be determined as
\begin{align}\label{eq:result}
  w^{\rm pred}=\expval_{\ep}\left[w(u_{*,k})\right],
\end{align}
where here and henceforth $\expval_{\ep}$ denotes the average over the
replica sample. Similarly, one can determine the prediction for any
set of PDF-dependent physical observables as
\begin{align}\label{eq:pdfresult}
 {\cal G}^{\rm pred} =\expval_{\ep}\left[ {\cal G}(w(u_{*,k}))\right].
\end{align}
Uncertainties on the predictions and their correlations can be further
determined from higher moments of the distribution $w(u_{*,k})$ of PDF
replicas. In particular, the  predicted  PDF uncertainties are given by the diagonal  
elements of the covariance matrix
\begin{align}
  \label{eq:pdf_covmat_replicas}
\left(C_{\rm PDF}\right)_{ij} = \frac{N_{\rm reps}}{N_{\rm reps}-1} \expval_{\ep} \left[\left({\cal
          G}(u_{*,k})_i -  \expval_{\ep} {\cal G}(u_{*,k})_i\right)\left({\cal
          G}(u_{*,k})_j -  \expval_{\ep} {\cal G}(u_{*,k})_j\right)\right]
\end{align}
whose off-diagonal elements provide the PDF-induced correlations.

The model adopted by NNPDF is a feed-forward neural network, whose
weights and thresholds are the parameters $u$. The architecture,
initialisation and activation function  of the
neural network are determined through a hyperoptimisation
procedure. The loss minimisation is performed using a gradient descent
algorithm. The algorithm and its parameters are also part of the
hyperoptimisation.

In order to compute quantities of physical interest, it is
necessary to invert 
the covariance matrix $C_{\rm PDF}$, which 
has dimension $N_{\rm data}\times N_{\rm data}$, and is
computed from the underlying PDFs. The full
dataset contains several thousand datapoints. On the other hand, it is
known~\cite{Carrazza:2015aoa,PDF4LHCWorkingGroup:2022cjn} that for
current PDF sets an accurate representation of the full PDF covariance matrix can be
obtained by means of a Hessian representation with a much smaller
number of eigenvectors, between $50$ and $100$~\cite{Carrazza:2015hva,Carrazza:2015aoa,Carrazza:2016htc}. 
This means that the full $C_{\rm PDF}$ matrix is necessarily poorly conditioned.
Namely, the condition number, defined as
\begin{align}
    \kappa(C_{\rm PDF}) = \frac{|\lambda_{\rm max}(C_{\rm PDF})|}{|\lambda_{\rm min}(C_{\rm PDF})|}, 
\end{align}
where $\lambda_{\rm max}$ and $\lambda_{\rm min}$ are respectively
the largest and smallest eigenvalue of the covariance matrix $C_{\rm PDF})$,
is very large. Furthermore, even the
restriction of $C_{\rm PDF}$ to individual datasets, with a
number of datapoints of order of a hundred, or even ten, is
generally poorly conditioned, because individual datasets are
typically correlated only to a small number of PDFs or
PDF combinations in a limited kinematic region (see
e.g.~\cite{Forte:2010dt}).

\begin{figure}
  \centering
  \includegraphics[width=0.8\linewidth]{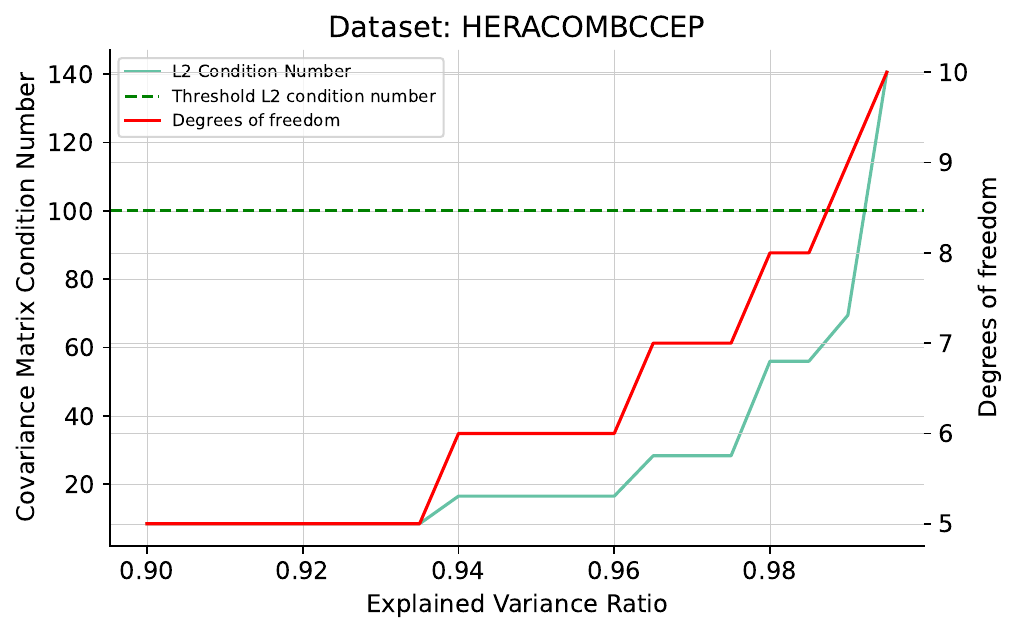}
  \caption{Condition number for the HERA I+II inclusive CC dataset
    (see Table~\ref{tab:ratio_bv_dis} ) as a function of the
    explained variance ratio, namely the ratio of  the traces of the regularised to full
covariance matrix.}
  \label{fig:l2_cond_heracombccep}
\end{figure}
This problem can be addressed by only retaining the contribution
to the covariance matrix of the eigenvectors with largest eigenvalues,
i.e., to regularize it through a Principal Component Analysis (PCA). 
The PCA requires a criterion in order to determine how many
eigenvectors should be kept, by balancing accuracy against
stability. We will  require the trace of
the regularised covariance matrix to differ at most by 1\% from the
trace of the starting matrix. This roughly corresponds to condition number
$ \kappa(\rho_{\rm PDF})\lesssim100$, as illustrated in
Fig.~\ref{fig:l2_cond_heracombccep}, where we plot the condition
number vs. the ratio of  the traces of the regularised to full
covariance matrix, computed for the data included in a specific
dataset (see
Table~\ref{tab:ratio_bv_dis} below). When dealing with the covariance
matrix of the full dataset, we will apply this criterion to the
correlation matrix rather 
than to the covariance matrix, in order to deal with eigenvectors that
are dimensionless and normalised and can thus be compared even when
referring to heterogeneous observables.

\subsection{The closure test and its statistical estimators}
\label{subsec:est}

The closure test is based on assuming knowledge of the true PDFs
$w^0$, so that the true experimental values $f$ of Eq.~\eqref{eq:l1_data} are given by
$f = {\cal G}(w^0)$. This knowledge is not shared by the NNPDF
algorithm, so the accuracy of the result $w^{\rm pred}$ Eq.~(\ref{eq:result})
can  be tested by comparing it to the truth $w^0$.

Specifically, closure test experimental data are generated
according to Eq.~\eqref{eq:l1_data} with $f = G(w^0)$
\begin{align}
    y_0 &= {\cal G}(w^0) + \eta,
    \label{eq:l1_ctdata}
\end{align}
where the observational noise $\eta$ is pseudo-randomly generated from
the assumed distribution. We refer to the  closure test data
for short (following Ref.~\cite{Ball:2014uwa}) as Level-1 ($L_1$) data.
The power of the closure test consists in the fact that it is possible
to generate several $L_1$ data, and run the full methodology on each of
them, thereby obtaining several instances of predicted model and model
uncertainties, all based on different ``runs of the universe''
(i.e. $L_1$ data) based on a fixed underlying truth.

We thus produce $N_{\rm fit}$ sets of $L_1$ data. 
For each of those we perform a full PDF determination, thereby ending 
up with $N_{\rm fit}$ PDF sets, each made of $N_{\rm rep}$ PDF replicas. Henceforth we will
consistently use the index $(l)$ to label the $N_{\rm fit}$ instances
of $L_1$ data and corresponding PDF sets, and the index $k$ to label the $N_{\rm rep}$ replicas
in each of these $N_{\rm fit}$ PDF set. We can then construct estimators that
assess whether the PDF uncertainties in Eq.~\eqref{eq:pdf_covmat_replicas} are faithful.
The main estimator that we consider is the normalised bias. This
measures the actual mean square deviation of the prediction from the
truth in units of its predicted standard deviation. The normalised bias for each of the
$N_{\rm fit}$ PDF sets is defined as
\begin{align}
    \label{eq:bias_definition}
        B^{(l)} (C_{\rm PDF}) & = \frac{1}{N_{\rm data}}\sum_{i,j=1}^{N_{\rm data}} \left(\expval_{\ep}{\cal G}(u_{*,k}^{(l)})_i - f_i\right) (\overline{C}_{\rm PDF})_{ij}^{-1} \left(\expval_{\ep}{\cal G}(u_{*,k}^{(l)})_j - f_j\right) ,
\end{align}
where, as in Eq.~(\ref{eq:l1_data}), $f_i$ are the components of the vector of
true values of the $N_{\rm data}$ observables; $\overline{C}_{\rm  PDF}$ is the average 
PDF covariance matrix
\begin{align}
  \label{eq:avcovmat}
\left(\overline{C}_{\rm PDF}\right)_{ij} = \expval_{\eta} \left(C^{(l)}_{\rm PDF}\right)_{ij},
\end{align}
where $\expval_{\eta}$ here and henceforth denotes the average over the $N_{\rm fit}$ instances of 
$L_1$ data.

The reason why in Eq.~(\ref{eq:bias_definition}), the average
covariance matrix is used instead of the
covariance matrix $\left(C^{(l)}_{\rm PDF}\right)_{ij}$ for the $l$-th
$L_1$ data instance is that
the covariance matrix for the full dataset
computed for individual instances of $L_1$ data is numerically unstable,
so it is impossible to achieve our target value of the explained
variance ratio. The problem goes away when considering the
average covariance matrix in Eq.~(\ref{eq:avcovmat}). We have 
checked that for individual datasets, for which our desired target of accuracy can be
achieved, the PCA covariance matrix computed for individual instances
of $L_1$ data varies little, with percent-level
differences between the covariance
matrix $\left(C^{(l)}_{\rm PDF}\right)_{ij}$ computed for each $L_1$
instance and the average $\left(\overline{C}_{\rm
  PDF}\right)_{ij}$. We conclude that hence replacing individual
covariance matrices with their average does not affect our results in
a significant way, while solving the numerical instability issue.

The meaning of the normalised bias is manifest by rewriting it
in the basis of
eigenvectors of the PDF covariance matrix $C^{(l)}_{\rm PDF}$:
\begin{align}
    \label{eq:bias_diag}
        B^{(l)} (\overline{C}_{\rm PDF}) & = \frac{1}{N_{\rm data}}\sum_{i=1}^{N_{\rm data}}
        \frac{({\Delta^{(l)}_i})^2}{({\sigma_{\rm PDF}}_i)^2},
\end{align}
where ${\sigma_{\rm PDF}}_i$ is the $i$-th eigenvalue of $\overline{C}_{\rm PDF}$ and $\Delta^{(l)}_i$ is the projection of the vector
$\big(\expval_{\ep}{\cal G}(u_{*,k}^{(l)})_j - f_j\big)$ along the $i$-th
normalised eigenvector $v^{(i)}_j$ of  $\overline{C}_{\rm  PDF}$:
\begin{equation}\label{eq:deltadef}
 \Delta^{(l)}_i=\sum_{j=1}^{N_{\rm data}}\left(\expval_{\ep}{\cal
    G}(u_{*,k}^{(l)})_j - f_j\right)v^{(i)}_j.
\end{equation}
If uncertainties are Gaussian and faithful, for each eigenvector $i$
the distribution of $\Delta^{(l)}_i$ over the $N_{\rm fits}$ is a
  univariate Gaussian. The simplest test that this is the case is to
  compute the variance of this Gaussian, namely
the root-mean square normalised bias
\begin{align}
  \label{eq:bias_variance_ratio_definition}
    R_{b} = \sqrt{\expval_{\eta} B^{(l)}(\overline{C}_{\rm PDF})},
\end{align}
where the average runs over $L_1$ instances $l=1,\ldots,N_{\rm fit}$.
For faithful (Gaussian) uncertainties  $R_{b}$ is the
variance of an univariate Gaussian so $R_{b}=1$. We will henceforth
refer to $R_b$ as the normalised bias. 

The indicator $R_{b}$ is very close to the bias-variance ratio
$R_{bv}$ that
was used as a closure test indicator in
Refs.~\cite{Ball:2014uwa,DelDebbio:2021whr,NNPDF:2021uiq}, the 
difference being that in the bias-variance ratio the experimental $C$
covariance matrix appears, instead of the PDF covariance matrix
$C_{\rm PDF}$ Eq.~(\ref{eq:pdf_covmat_replicas}). Whereas
both estimators must equal one for a consistent methodology, the new
estimator  $R_{b}$ has certain advantages, that are discussed in
Appendix~\ref{app:comp}, essentially related to the fact that the
bias-variance ratio, unlike the new estimator $R_b$, gives equal weight to experimentally uncorrelated
data even when they include very different numbers of data points, so
when computing its value
small datasets and their fluctuations weigh disproportionately.

A more detailed way of testing that $\Delta^{(l)}_i$ are Gaussianly
  distributed is to consider the $n$-th quantile of their
  distributions over $L_1$ data, namely
\begin{align}
\label{eq:xi1sigma_estimator}
\xi_{n\sigma} = \frac{1}{N_{\rm fits}} \frac{1}{N_{\rm data}} \sum_{i=1}^{N_{\rm data}}  \sum_{l=1}^{N_{\rm fits}} I_{[-n,n]}
        \bigg( \frac{(\Delta^{(l)}_i)^2}{({\sigma_{\rm PDF}}_i)^2}\bigg),
\end{align}
where  $I_{A}(x)$ denotes the indicator function of the interval $A$, which
is equal to one if its argument lies within the interval $A$, and zero
otherwise. For a Gaussian distribution, $\xi_1=0.68$ (to two
significant digits).
A yet more detailed test consists of comparing directly
$\Delta^{(l)}_i$ to a normal Gaussian distribution with mean 0 and variance 1.

\subsection{Modeling inconsistencies}
\label{sec:approach}

In a realistic situation it may happen that some sources of
experimental systematics are overlooked or underestimated for a given
dataset. In this 
case, the measured experimental values for this dataset may deviate
from the true value 
by an amount that is not reflected by the experimental covariance
matrix. This will then generate tensions between this dataset and the
rest of the data. We dub such a dataset ``inconsistent''. It is
interesting to ask how the NNPDF methodology behaves in such case, and
whether the inconsistency can be detected.

To study this situation in a closure test, we model the
inconsistency as follows. We separate off the uncorrelated and
correlated parts of the experimental covariance matrix $C$
\begin{align}
    \left(C\right)_{ij}  = \delta_{ij} \sigma_i^{(\rm uncorr)} \sigma_j^{(\rm uncorr)} + \sum_{k=1}^{N_{\rm corr}} \sigma_{i,k}^{(\rm corr)} \sigma_{j,k}^{(\rm corr)},  
    \label{eq:experimental covmat}
\end{align}
where $\sigma_i^{\rm (uncorr)}$  and  $\sigma_i^{\rm (corr)}$  denote respectively
the uncorrelated and correlated systematics. 
We then define a rescaled covariance matrix
\begin{align}
    \left(C^\lambda\right)_{ij}  = \delta_{ij} \sigma_i^{(\rm uncorr)}
    \sigma_j^{(\rm uncorr)} + \sum_{k=1}^{N_{\rm corr}} \lambda_{i,k}
    \sigma_{i,k}^{(\rm corr)} \lambda_{j,k} \sigma_{j,k}^{(\rm corr)}  
    \label{eq:clambda}
\end{align}
in which correlated uncertainties have been rescaled.

\begin{figure}[htb]
  \centering
\includegraphics[width=0.48\textwidth]{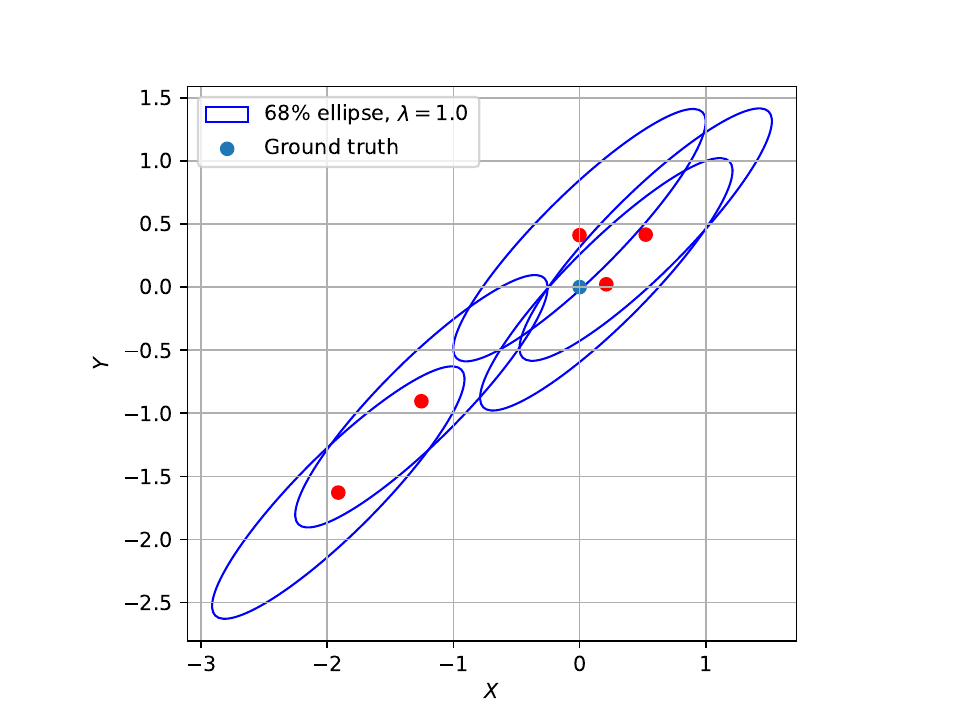}
\includegraphics[width=0.48\textwidth]{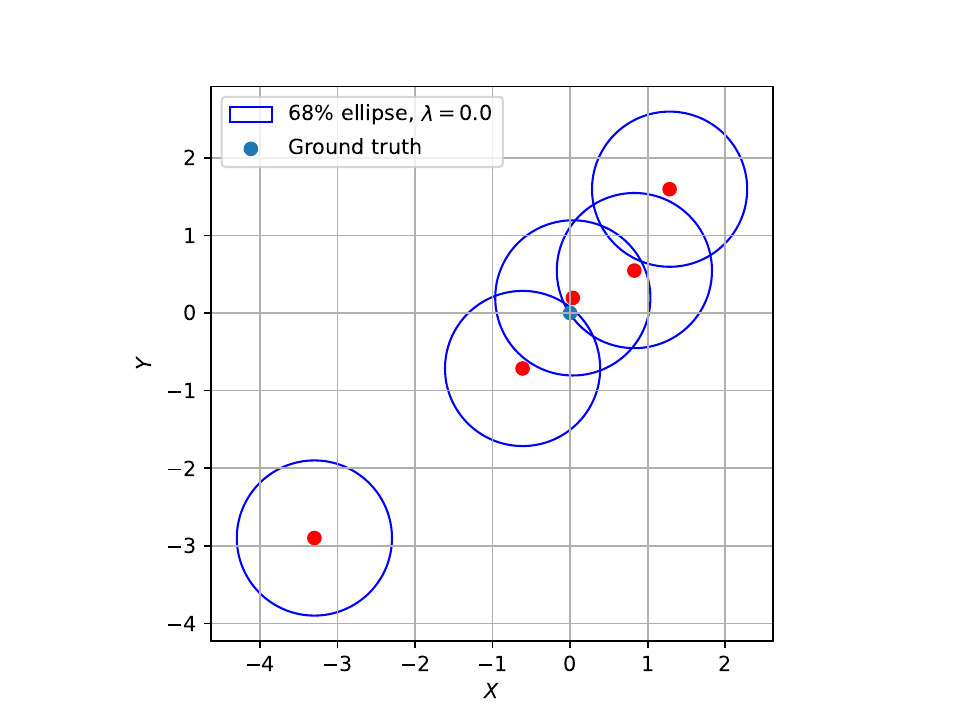}
\caption{Data points for a pair of observables generated with a
  nontrivial correlated covariance matrix, shown on the left with
  their correct error ellipse, and on the right with an error circle
  in which all correlated uncertainties have been set to zero. }
\label{fig:data_generation}
\end{figure}

This then allows the modeling of a situation in which the correlated
systematics have been incorrectly estimated either for some sources of
uncertainty, or for some data, or both. To this purpose, we generate $L_1$
closure test
data  according to Eq.~(\ref{eq:l1_ctdata}) using shifts  $\eta $
based on  the covariance matrix $C$ Eq.~(\ref{eq:experimental covmat}).
This corresponds to the ``true'' data uncertainties
Eq.~(\ref{eq:l1_data}) that the closure test simulates. However, we then
determine PDFs using the rescaled matrix $C^\lambda$ everywhere in the
NNPDF methodology of Sect.~\ref{subsec:nn}: we generate pseudodata
replicas using Eq.~(\ref{eq:l2_pseudodata}) with  $\epsilon^{(k)}$
drawn from a Gaussian distribution 
$\mtc{N}(0, C^\lambda)$, and we determine the optimal model by
minimising the loss function $\chi^2(C^\lambda)$
Eq.~(\ref{eq:chi2_definition}). This corresponds to the incorrectly
estimated uncertainties. 
A simple visualisation of this situation for a pair of data for which
all $\lambda_{i,k}=0$ is provided in Fig.~\ref{fig:data_generation},
where on the left we show data generated using a nontrivial covariance
matrix, with the corresponding true error ellipse, while on the right
each datapoint is shown with the uncorrelated error circle obtained by
setting all correlated uncertainties to zero.

It is important to understand that each datapoint in Fig.~\ref{fig:data_generation}
corresponds to a different run of the universe. This means that 
in a real PDF determination only one occurrence is actually realised: 
it is only within the closure test that one has a distribution of events. 
This means that our model of inconsistency reproduces  both
the situation in which an uncertainty has been incorrectly estimated,
but also the situation in which datapoints are systematically offset from
their true values, because each particular instance of $L_1$ data
corresponds to a specific value of the nuisance parameter
associated to each correlated systematics, i.e. to a systematic shift.

%% file: sec-results.tex
\section{Results}
\label{sec:results}

We will now present the results of this study. First, we summarise the
settings that we have adopted and introduce the three inconsistency
scenarios that we have considered, then we present results for each
scenario.

\subsection{Settings and scenarios}
\label{sec:settings}

We have performed a number of closure tests with various inconsistency
scenarios, starting each time from a baseline consistent closure
test. In all cases, for ease of comparison with previous work we adopt
the
same underlying truth adopted for the closure 
test of Ref.~\cite{DelDebbio:2021whr}, namely a specific random replica
from the  {\tt NNPDF4.0} set. Individual NNPDF replicas have more
structure than the average over replicas, so this choice is somewhat
more general than that of any current best-fit PDF.
We perform $N_{\rm fit}=25$ closure tests, each based on a different
randomly generated set of $L_1$ data, and each containing 
$N_{\rm rep}=100$ replicas. Previous
studies~\cite{NNPDF:2021njg} suggest that these sample sizes are
sufficient to ensure reliable results; as in these previous references,
uncertainties associated to the finite size of the samples will be estimated 
through a bootstrap procedure (see App.~\ref{app:bootstrap_def} for details).

We specifically consider three inconsistency scenarios, in each of which we
divide the dataset into two subsets, one used for PDF
determination (in-sample data) and the other used for testing the
accuracy of predictions (out-of-sample data). The two samples are
in each case constructed in such a way that the in-sample and
out-of-sample datasets are equally representative of the full
dataset in terms of kinematics. We furthermore introduce inconsistencies in each case in
a different subset of the in-sample data. The scenarios differ in the
choice of full dataset and inconsistent dataset. In all cases, the
inconsistency is introduced following Eq.~(\ref{eq:clambda}): the
baseline consistent case corresponds to choosing $\lambda=1$, and an
increasingly strong inconsistency is introduced by choosing
increasingly small values of $\lambda$, until with $\lambda=0$ we end
up with a situation similar to that depicted in
Fig.~\ref{fig:data_generation} (right), in which all correlated
uncertainties are set to zero in the covariance matrix used in the
PDF determination.
Results shown for the statistical indicators follow in all cases the
definitions given in Sect.~\ref{subsec:est}.

The three scenarios illustrate three possible
situations. The first is a determination of PDFs based only on DIS
data, in which all the in-sample neutral-current (NC) HERA data are made inconsistent.
This illustrates a case of bulk inconsistency, in which the majority
of the data that determine the PDFs are inconsistent. The second is a
determination based on a global dataset, in which we introduce the
inconsistency in a single double-differential Drell-Yan dataset. This
is representative of a situation in which the inconsistency is
localised in a single, though highly precise and relevant,
dataset. The third is again based on a global PDF determination, in
which we make single-inclusive jet data inconsistent. This is
illustrative of the inconsistency being present in a dataset that
almost entirely determines one specific PDF (in this case the gluon at medium
$x$).

\input{subsec-dis.tex}

\input{subsec-dy.tex}

\input{subsec-jets.tex}

%% file: subsec-dis.tex
\subsection{Bulk inconsistency: Deep Inelastic Scattering}
\label{subsec:dis}

\begin{figure}[htb]
  \centering
    \includegraphics[width=0.9\textwidth]{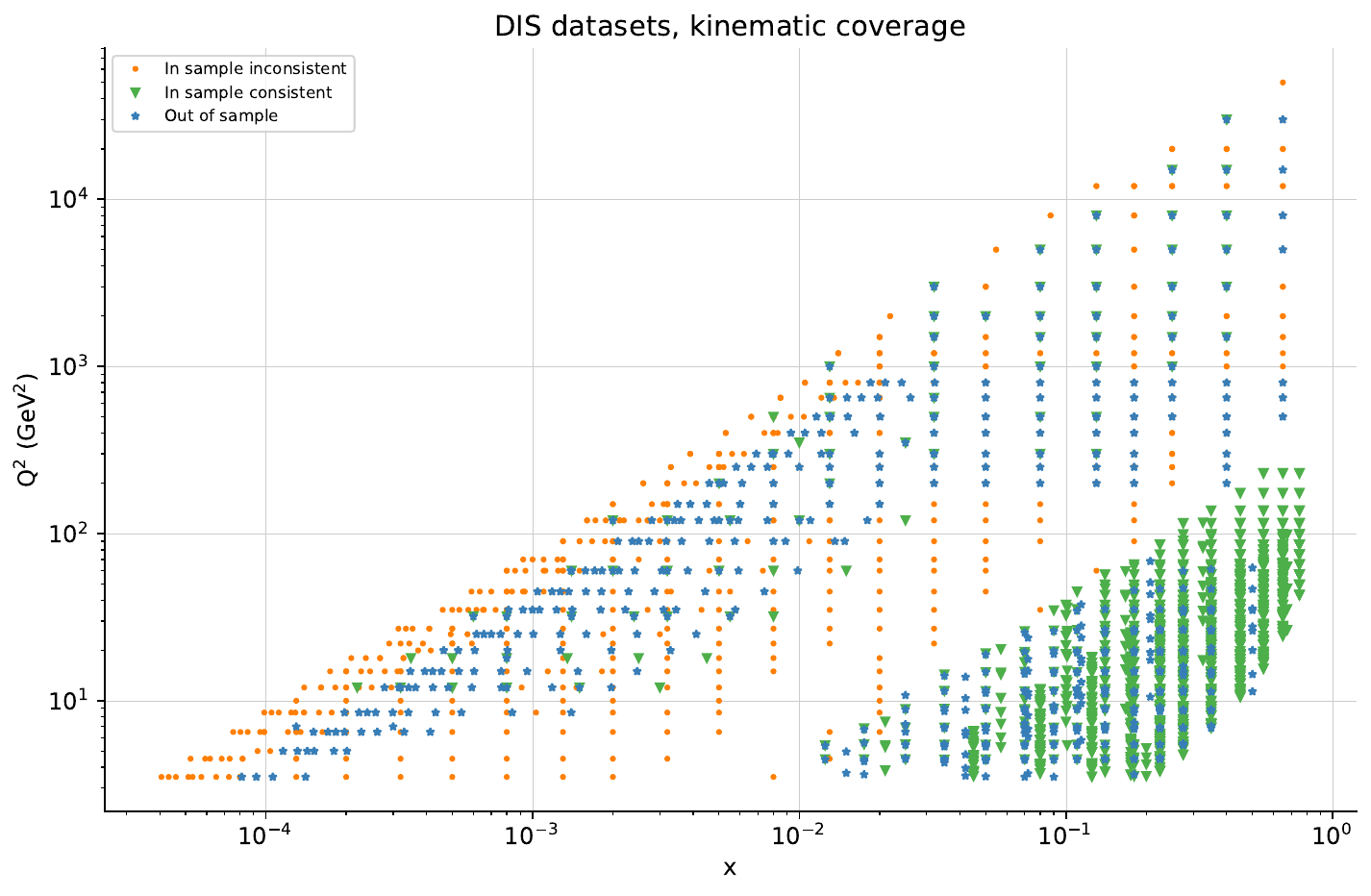}
    \caption{Kinematic coverage of the data of Table~\ref{tab:ratio_bv_dis} in the $(x,Q^2)$ plane. In-sample
      consistent data are denoted by green inverted triangles,
      in-sample inconsistent by orange dots, and out-of-sample by
      blue stars. }
    \label{fig:DISpartition}
\end{figure}

The first scenario, of a bulk inconsistency, is a PDF determination
using DIS data only.
The datasets correspond to all the DIS data used 
for the {\tt NNPDF4.0}~\cite{NNPDF:2021njg} PDF determination, with the same
kinematic cuts. They are listed in
Table~\ref{tab:ratio_bv_dis}, where the split between in-sample and
out-of-sample is also indicated.
The kinematic coverage and data split are displayed in Fig.~\ref{fig:DISpartition}, 
which shows that the in- and out-of-sample datasets have similar coverage.
\begin{table}[]
  \centering
  \small
    \begin{tabular}{l|c|cccc}
      \hline
    \multirow{ 2}{*}{Datasets} &  \multirow{ 2}{*}{$N_{\text{data}}$}  & \multicolumn{4}{c}{$R_{b}$} \\
     &   & $\lambda=1.0$ & $\lambda=0.7$ & $\lambda=0.4$ & $\lambda=0.0$ \\
      \hline
      \hline
      NMC $F_2^d/F_2^p$ \cite{Arneodo:1996kd} &  121 & 0.8 & 0.7 & 0.8 & 1.0\\
      SLAC $F_2^p$  \cite{Whitlow:1991uw}   & 33 & 0.7 & 0.7 & 0.8 & 1.1\\
      SLAC $F_2^d$  \cite{Whitlow:1991uw}   & 34 & 0.8 & 0.8 & 0.8 & 0.9\\
      BCDMS $F_2^p$ \cite{Benvenuti:1989rh} & 333 & 0.8 & 0.8 & 0.8 & {\bf 1.1}\\
      BCDMS $F_2^d$ \cite{Benvenuti:1989rh} & 248 & 0.9 & 0.9 & 0.9 & {\bf 1.1}\\
      CHORUS $\sigma_{CC}^{\nu}$ \cite{Onengut:2005kv}  & 416 & 0.8 & 0.9 & 0.8 & 0.9\\
      CHORUS $\sigma_{CC}^{\bar{\nu}}$   \cite{Onengut:2005kv}  & 416 & 0.9 & 1.0 & 1.0 & {\bf 1.2}\\
      NuTeV $\sigma_{CC}^{\nu}$ (dimuon) \cite{Goncharov:2001qe,MasonPhD}  & 39 & 0.8 & 0.9 & 0.9 & {\bf 1.2}\\
      HERA I+II $\sigma_{\rm CC}^{e^+p}$ \cite{Abramowicz:2015mha} & 39 & 0.9 & 1.0 & 1.0 & {\bf 1.3}\\
      HERA I+II $\sigma_{\rm NC}^{\rm charm}$ \cite{H1:2018flt} & 37 & 1.0  & {\bf 1.1} & {\bf 1.1} & {\bf 1.2}\\
      \hline
            $(^{*})$ HERA I+II $\sigma_{\rm NC}^{e^- p}$  $E_p=320$ GeV \cite{Abramowicz:2015mha} & 159 & 0.9 & 1.0 & {\bf 1.2} & {\bf 2.2}\\
      $(^{*})$ HERA I+II $\sigma_{\rm NC}^{e^+ p}$  $E_p=575$ GeV \cite{Abramowicz:2015mha} & 254 & 0.8 & 0.9 & {\bf 1.3} & {\bf 2.4}\\
      $(^{*})$ HERA I+II $\sigma_{\rm NC}^{e^+ p}$  $E_p=820$ GeV \cite{Abramowicz:2015mha} & 70  & 0.8 & 0.9 & {\bf 1.2} & {\bf 2.3}\\
      $(^{*})$ HERA I+II $\sigma_{\rm NC}^{e^+ p}$  $E_p=920$ GeV \cite{Abramowicz:2015mha} & 377 & 0.9 & 0.9 & {\bf 1.3} & {\bf 2.4}\\
      \hline
      Total (in-sample) & 2576 & 0.9 & 1.0 & {\bf 1.1} & {\bf 1.7} \\
      \hline
      NMC $\sigma^{{\rm NC},p}$ \cite{Arneodo:1996qe}& 204 & 0.9 & 0.9 & {\bf 1.1} & {\bf 1.6}\\
      NuTeV $\sigma_{CC}^{\bar\nu}$ (dimuon) \cite{Goncharov:2001qe,MasonPhD} &37  & 0.8 & 0.9 & 0.9 & {\bf 1.1}\\
      HERA I+II $\sigma_{\rm NC}^{e^+ p}$  $E_p=460$ GeV \cite{Abramowicz:2015mha} & 204 & 0.9 & 1.0 & {\bf 1.4} & {\bf 2.7}\\
      HERA I+II $\sigma_{\rm CC}^{e^-p}$ \cite{Abramowicz:2015mha}              & 42 & 0.9 & 1.0 & {\bf 1.2}  & {\bf 1.4}\\
      HERA I+II $\sigma_{\rm NC}^{\rm bottom}$ \cite{H1:2018flt} & 26 & 0.9 & 1.0 & {\bf 1.2} & {\bf 1.9}\\
      \hline
      Total (out-sample) & 513 & 0.9 & 0.9 & {\bf 1.1} & {\bf 1.9} \\
      \hline
      Total & 3089 & 0.9 & 1.0 & {\bf 1.1} & {\bf 1.6} \\
      \hline
    \end{tabular}
\caption{Datasets included in the DIS-only inconsistent closure
test. The in-sample datasets are in the upper part and the
out-of-sample datasets in the
lower part of the table. The inconsistent datasets are denoted by an asterisk and
collected at the bottom of the upper part of the table.
For each dataset we indicate the number of data points after kinematic cuts, and give the value of the normalised bias $R_b$,  
Eq.~(\ref{eq:bias_variance_ratio_definition}), obtained in a closure test
as a function of $\lambda$. All instances in which $R_{b}>1$ are highlighted in boldface.}
\label{tab:ratio_bv_dis}
\end{table}

The inconsistency is introduced in all of the HERA neutral-current
inclusive structure function data, starting with consistent
$\lambda=1$ and then with increasingly small
$\lambda=0.7,\, 0.6,\, 0.4,\, 0.2,\, 0$. The inconsistent datasets are
flagged in Table~\ref{tab:ratio_bv_dis} and 
Fig.~\ref{fig:DISpartition}. The total number of inconsistent
datapoints is $N_{\rm inc}=860$ out of $N_{\rm data}=2576$ in-sample datapoints.

\begin{figure}[htb!]
  \centering
    \includegraphics[width=0.8\textwidth]{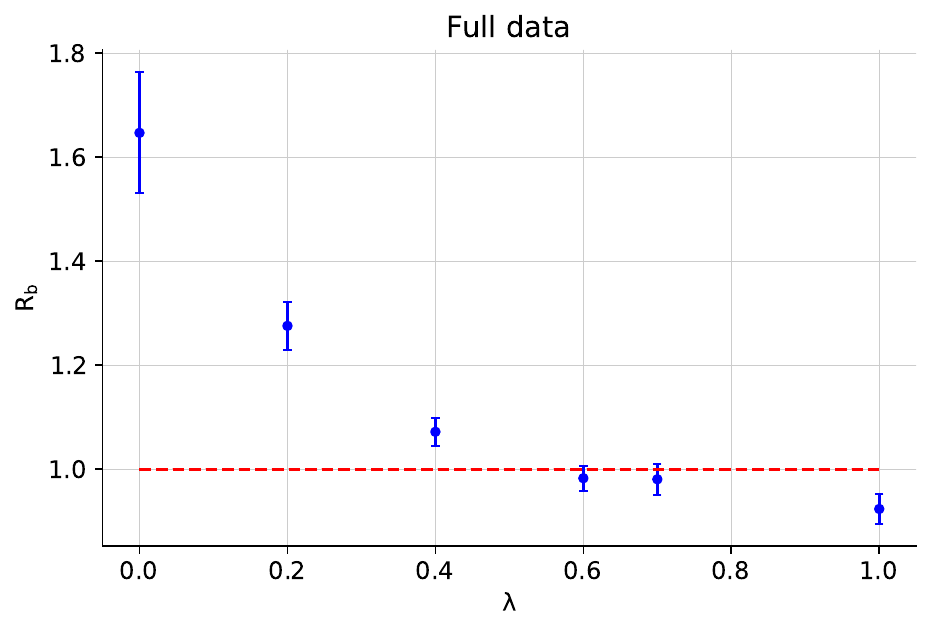}
    \caption{The normalised bias $R_b$
Eq.~(\ref{eq:bias_variance_ratio_definition}), as given in
Table~\ref{tab:ratio_bv_dis} for the full
dataset (both in-sample and out-of-sample),
as a function of  $\lambda$. The uncertainty shown reflects the finite
    size of the replica sample, estimated by bootstrap (see App.~\ref{app:bootstrap_def} for details).}
    \label{fig:rbv_scan_all_dis_data}
\end{figure}
\begin{figure}
  \centering
  \includegraphics[width=0.31\textwidth]{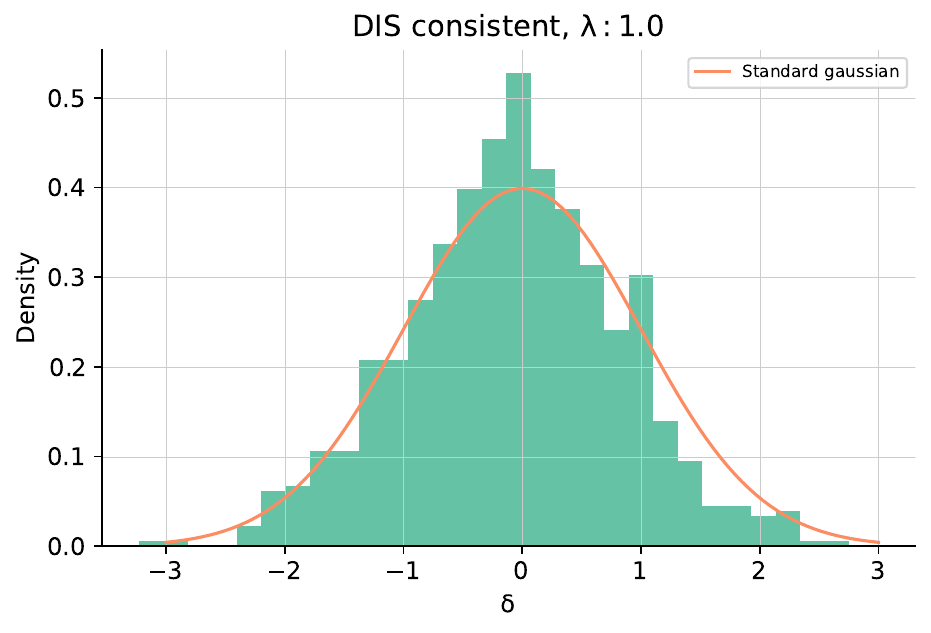}
  \includegraphics[width=0.31\textwidth]{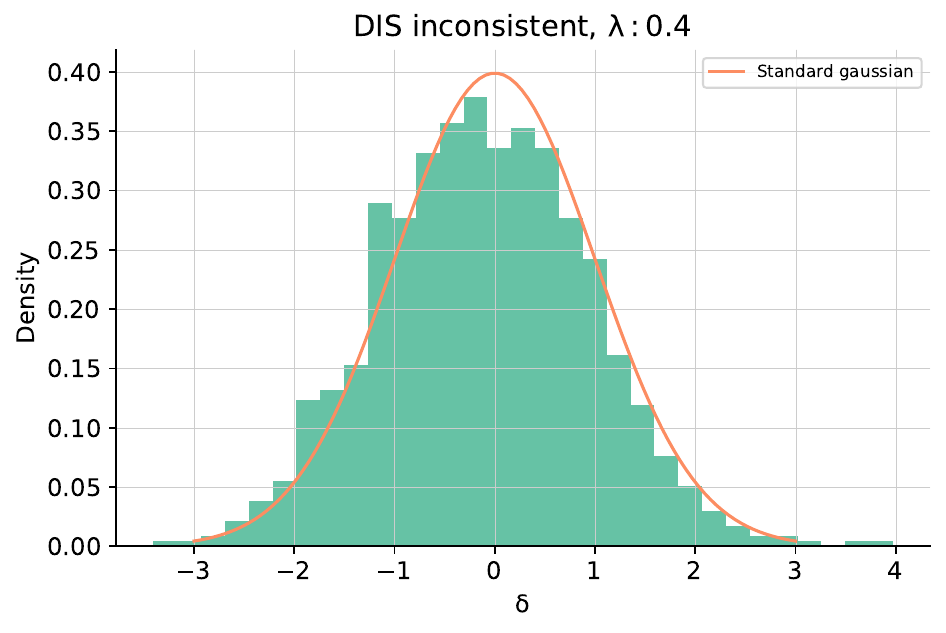}
  \includegraphics[width=0.31\textwidth]{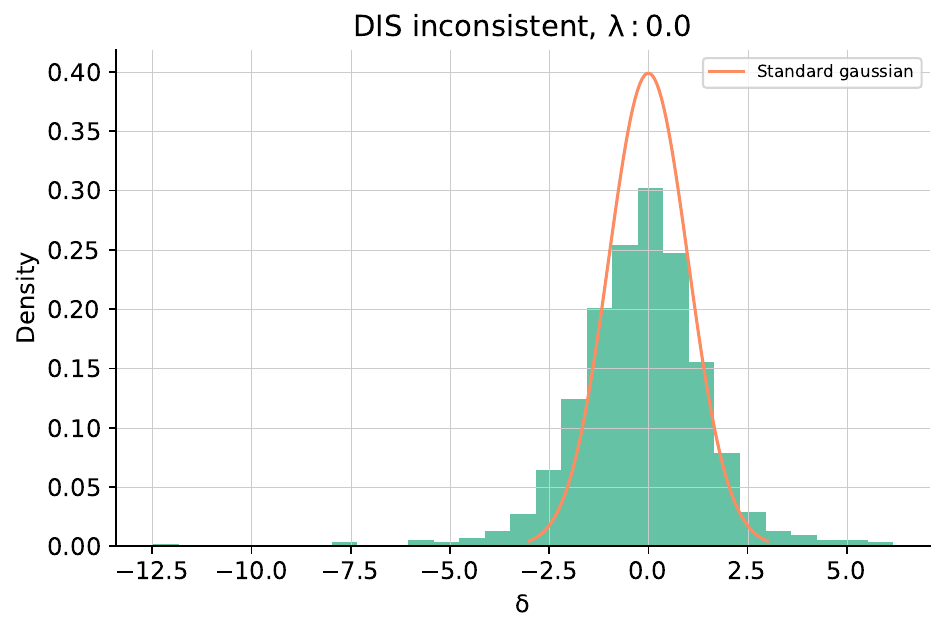}
\caption{The normalised distribution of relative differences
$\delta_i^{(l)}$ in the observables space,~Eq. \eqref{eq:deltadef}, for
$\lambda=1$ (left), $\lambda= 0.4$ (centre) and $\lambda= 0$ (right). 
In all cases, a univariate zero-mean Gaussian distribution is plotted for reference.}
\label{fig:delta_histograms_dis}
\end{figure}

The values of the normalised bias $R_b$, 
Eq.~(\ref{eq:bias_variance_ratio_definition}), found in various
cases are collected in Table~\ref{tab:ratio_bv_dis}
both for all individual datasets and the total dataset. All values for
which $R_b>1$ are highlighted in boldface. The values
for the full dataset are also displayed 
in Fig.~\ref{fig:rbv_scan_all_dis_data} as a function of
$\lambda$. The uncertainty on each value is determined by bootstrap
(see App.~\ref{app:bootstrap_def} for details). In
Table~\ref{tab:xi1sigma_dis_full_data} we also collect the
values of the  $\xi_{1\sigma}$
quantile estimator Eq.~(\ref{eq:xi1sigma_estimator}). Finally, the normalised distribution of relative differences $\delta_{i}^{(l)}$
Eq.~(\ref{eq:deltadef}) is displayed in Fig.~\ref{fig:delta_histograms_dis} for 
the fully consistent $\lambda=1$ case, the intermediate  case
$\lambda=0.4$, and  the extreme inconsistency  $\lambda=0$.

It is interesting to observe that, in the consistent $\lambda=1$ case,
$R_b\lesssim 1$, and accordingly  $\xi_{1\sigma}\gtrsim0.68$, indicating that uncertainties are
somewhat overestimated. Note that the bias-variance ratio $R_{b}$ was
instead found to equal one within uncertainty in
Ref.~\cite{NNPDF:2021njg}, however, as mentioned in
Sect.~\ref{subsec:est} (see also App.~\ref{app:comp}), the
normalised bias is a somewhat more accurate estimator as it is less
subject to fluctuations between different datasets.

The most remarkable feature of the behavior of the normalised bias and
corresponding quantile estimator for the total dataset  is
that, contrary to what one might expect, they  display very nonlinear
behavior when viewed as a function of $\lambda$: for $\lambda \ge 0.4$ they are almost flat, indicating that,
despite the inconsistency, PDF uncertainties remain faithful, but
then when $\lambda$ approaches zero $R_b$ sharply rises and
$\xi_{1\sigma}$ rapidly drops, indicating substantial uncertainty underestimate.

\begin{table}[htb]
    \centering
    \begin{tabular}{|c|c|}
    \hline
    $\lambda$ & $\xi_{1\sigma}$ \\
    \hline
      1.0 & 0.72 $\pm$ 0.02 \\
      0.7 & 0.69 $\pm$ 0.02 \\
      0.6 & 0.69 $\pm$ 0.02 \\
      0.4 & 0.64 $\pm$ 0.02 \\
      0.2 & 0.60 $\pm$ 0.02 \\
      0.0 & 0.52 $\pm$ 0.03 \\
    \hline
    \end{tabular}
\caption{The $\xi_{1\sigma}$ quantile estimator, Eq.~(\ref{eq:xi1sigma_estimator}), as a function of
$\lambda$. The uncertainty shown reflects the finite size of the replica sample, 
 (see App.~\ref{app:bootstrap_def} for details).}
\label{tab:xi1sigma_dis_full_data}
\end{table}

\begin{figure}
  \centering
  \includegraphics[width=0.45\textwidth]{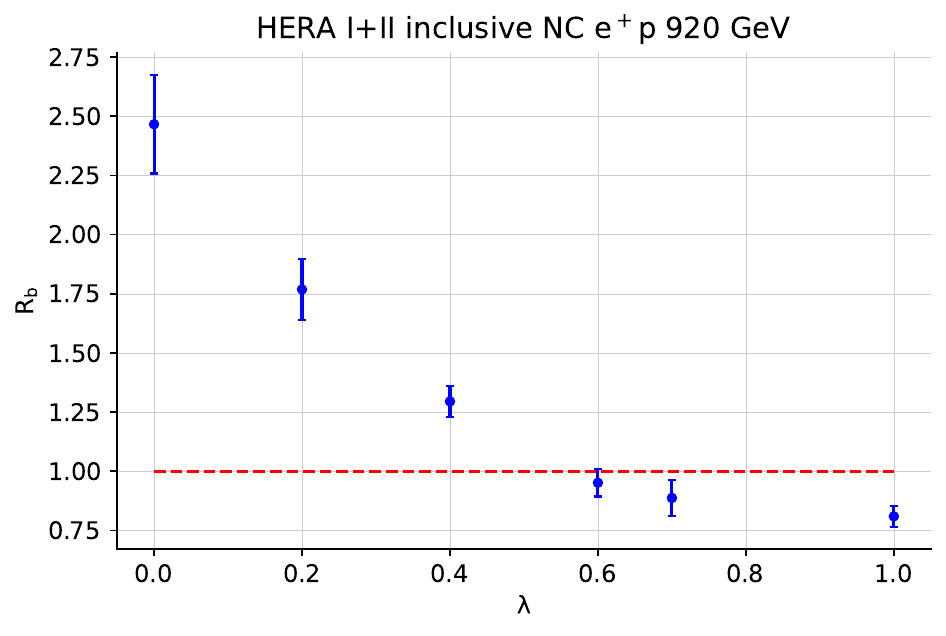}
  \includegraphics[width=0.45\textwidth]{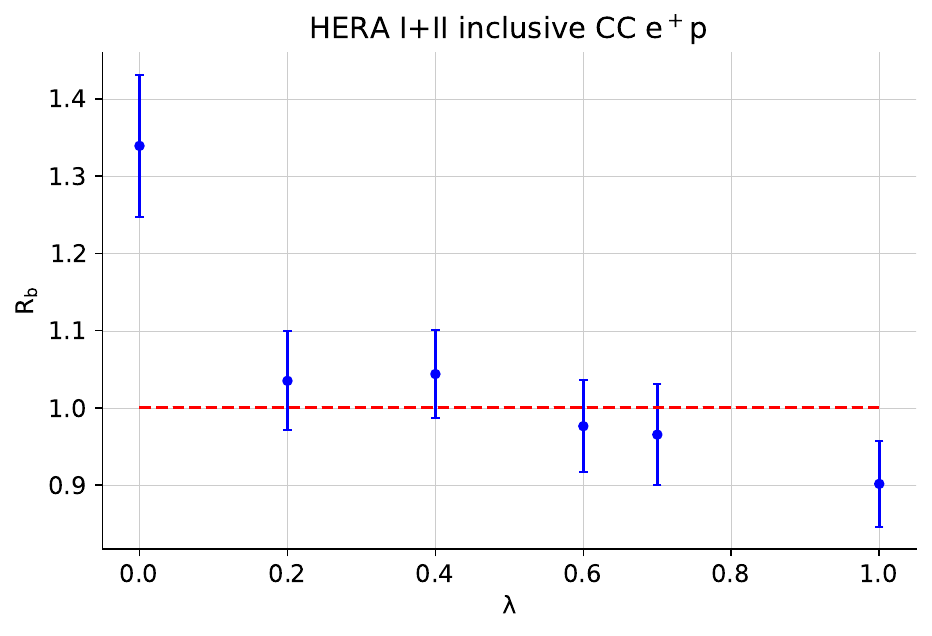}
  \includegraphics[width=0.45\textwidth]{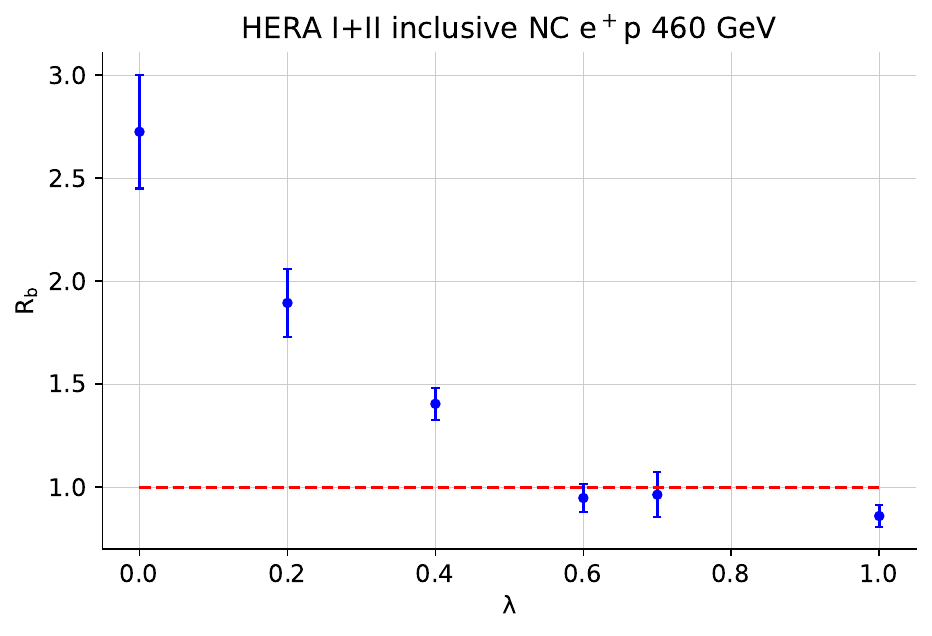}
  \includegraphics[width=0.45\textwidth]{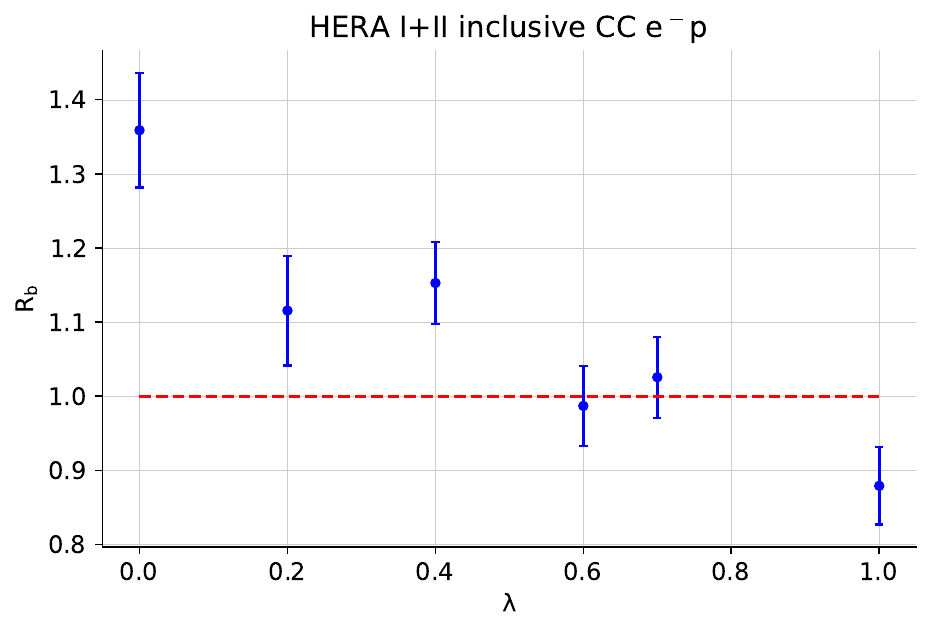}
    \caption{Same as Fig.~\ref{fig:rbv_scan_all_dis_data}, but now for
      some individual datasets: HERA I+II $\sigma_{\rm NC}^{e^+ p}$ 
$E_p=920$~GeV (in-sample, inconsistent, top left); HERA I+II $\sigma_{\rm
        CC}^{e^+ p}$ (in-sample, consistent, top right); HERA I+II $\sigma_{\rm NC}^{e^+ p}$ 
$E_p=420$~GeV and HERA I+II $\sigma_{\rm CC}^{e^- p}$ (both out-of-sample, bottom).}
    \label{fig:ratio_bv_dis}
  \end{figure}

Coming now to individual datasets, we can see from
Table~\ref{tab:ratio_bv_dis} that for $\lambda>0.4$ essentially all
datasets remain consistent, while for $\lambda=0.4$ all in-sample data
remain consistent while marginal inconsistency starts appearing both in
the inconsistent datasets, as well as in out-of-sample datasets that
probe similar kinematics. When $\lambda=0$ almost all in-sample and
all out-of-sample datasets display significant inconsistencies.
Quite in general, the fact that in-sample and out-of-sample
datasets behave in a similar way shows that the model is effective at generalising.
To better illustrate the behavior of individual datasets, in Fig.~\ref{fig:ratio_bv_dis} we
also show graphically  some of the results of
Table~\ref{tab:ratio_bv_dis}
(with additional values of $\lambda$,
to ease visualization).
 \begin{figure}
  \centering
  \includegraphics[width=0.45\textwidth]{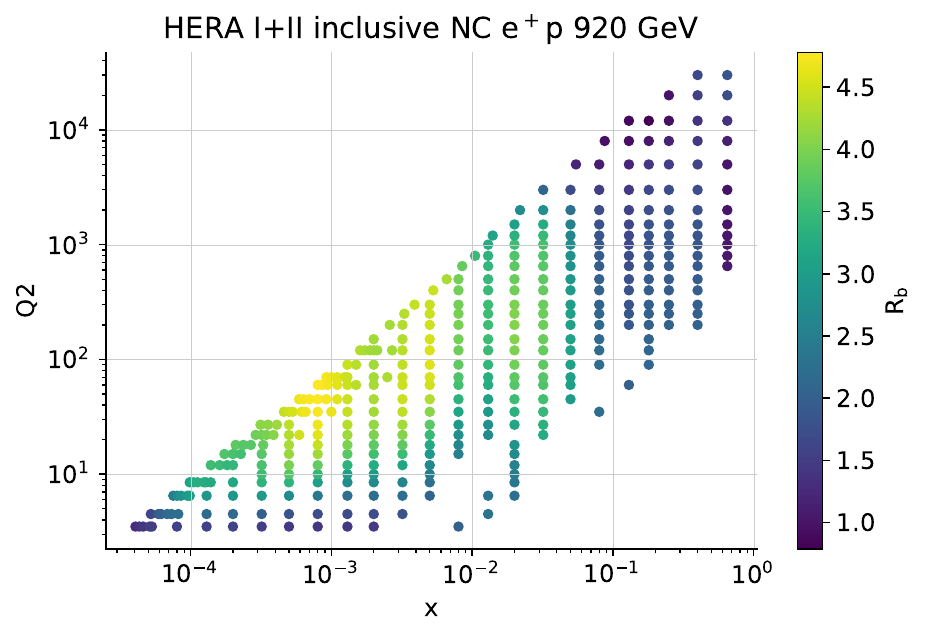}
  \includegraphics[width=0.45\textwidth]{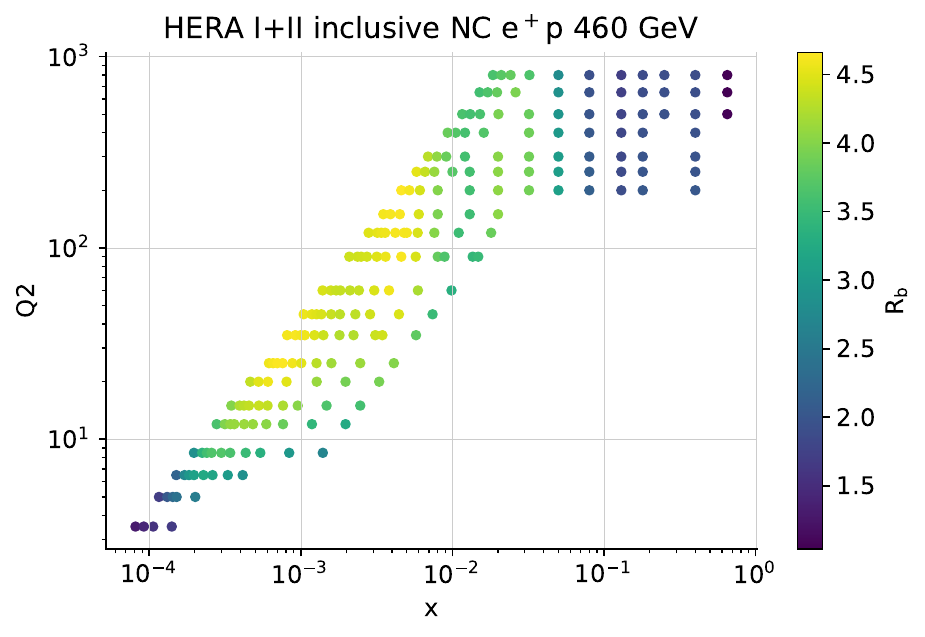}
    \caption{Scatter plot of the normalised bias $R_{b}$ for individual datapoints
      included in the two HERA I+II $\sigma_{\rm NC}^{e^+ p}$ datasets shown in the left column of
      Fig.~\ref{fig:ratio_bv_dis}  in the 
      maximally inconsistent scenario ($\lambda=0$): $E_p=920$~GeV
      (left), in-sample, inconsistent and $E_p=460$~GeV (right), out-of-sample.}
    \label{fig:ratio_bv_dis_singledp}
  \end{figure}
Finally, we look at individual datapoints. This misses the information
on  PDF-induced correlations, but it allows for a more fine-grained
understanding of the behavior of the PDF model.
  In Fig.~\ref{fig:ratio_bv_dis_singledp} we show the normalised bias
  $R_{b}$ for each datapoint in the maximally inconsistent $\lambda=0$
  scenario, for the two HERA I+II dataset 
  $\sigma_{\rm NC}^{e^+ p}$ datasets shown in the left column
  of Fig.~\ref{fig:ratio_bv_dis}, which are both affected by the
  inconsistency:   $E_p=920$~GeV  (top left in
  Fig.~\ref{fig:ratio_bv_dis}, left in
  Fig.~\ref{fig:ratio_bv_dis_singledp}), in-sample, directly affected,
  and $E_p=460$ GeV (bottom left in
  Fig.~\ref{fig:ratio_bv_dis}, left in
  Fig.~\ref{fig:ratio_bv_dis_singledp}), out-of-sample and thus
  indirectly affected.
It is clear that the pattern of inconsistency is quite similar for
both datasets. This can be understood by studying the correlation
between data and individual PDFs (see App.~\ref{app:corr}): the
PDFs that are most strongly correlated to the inconsistent dataset
will be affected, and in turn lead to inconsistent predictions for
data that are also strongly correlated to them. In this case indeed
both datasets are strongly correlated to the gluon (see
App.~\ref{app:corr}). It should be noted however that the
similarity is also partly due to the fact that in both cases
datapoints with $10^{-3}\lesssim x\lesssim 10^{-2}$ have the smallest
statistical uncertainties and thus the effect of the inconsistency is
more clearly visible.

 \begin{figure}
  \centering
  \includegraphics[width=0.45\textwidth]{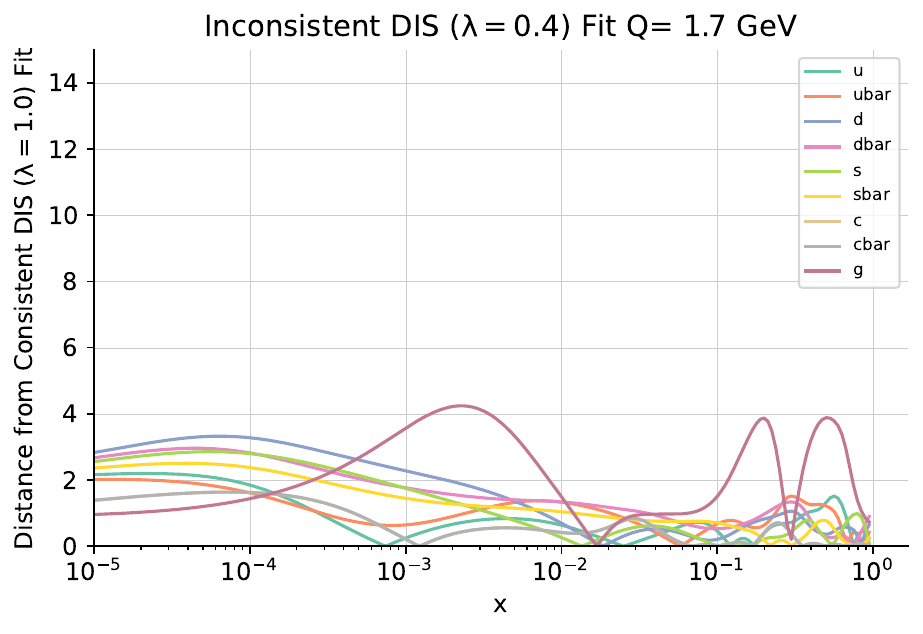}
  \includegraphics[width=0.45\textwidth]{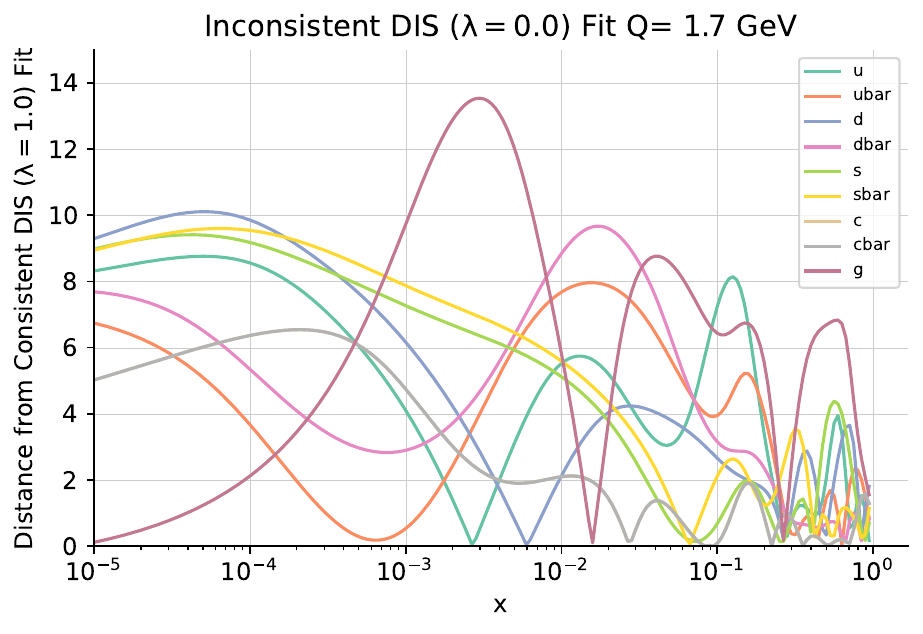}
  \caption{Distance between the PDFs obtained in an inconsistent and
    consistent fit to the same underlying data. Results are shown for
    $\lambda=0.4$ (left) and $\lambda=0$ (right). 
    \label{fig:pdf_dist_dis}}
  \end{figure}
In conclusion, it is interesting to ask how the PDFs behave upon
introducing an inconsistency, both in the case (such as for $\lambda=0.4$) in which
the model largely corrects for the inconsistency, and in the case (such as for $\lambda=0$) in which it
does not. In Fig.~\ref{fig:pdf_dist_dis} we plot the distance between the central
values of all PDFs determined from a consistent and inconsistent fit
to the same underlying
$L_1$ data, which hence correspond to one particular random fluctuation
of the data about the true underlying value.  In the
inconsistent case, the fluctuations are  not reflected correctly by
the size of the uncertainty used for fitting, which is smaller than what it ought to
be.
The distance
(defined in App.~B of Ref.~\cite{Ball:2014uwa}) is
the mean-square difference of central values in units of the
standard deviation of the mean, so that $d=1$ corresponds to
statistical equivalence; with $N_{\rm rep}=100$ replicas a one-sigma
shift corresponds to $d=10$. Results are shown both for $\lambda=0.4$ and
$\lambda=0$. It is clear that for  $\lambda=0.4$ the PDFs found in a
consistent and inconsistent fit are almost statistically equivalent,
with at most a localized quarter sigma deviation in the gluon. On the
contrary, for $\lambda=0$ there is clear inconsistency.

 \begin{figure}
  \centering
  \includegraphics[width=0.45\textwidth]{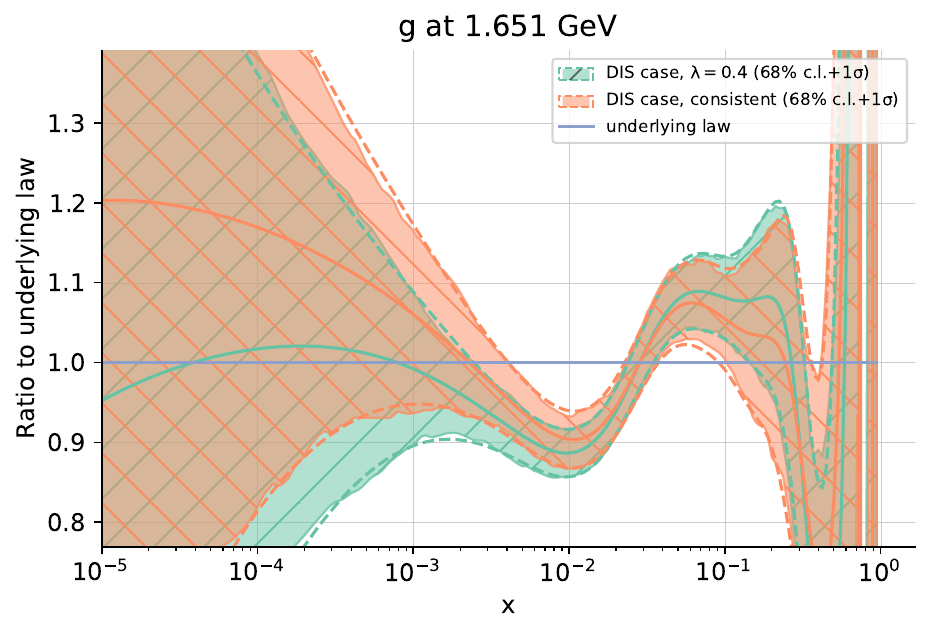}
  \includegraphics[width=0.45\textwidth]{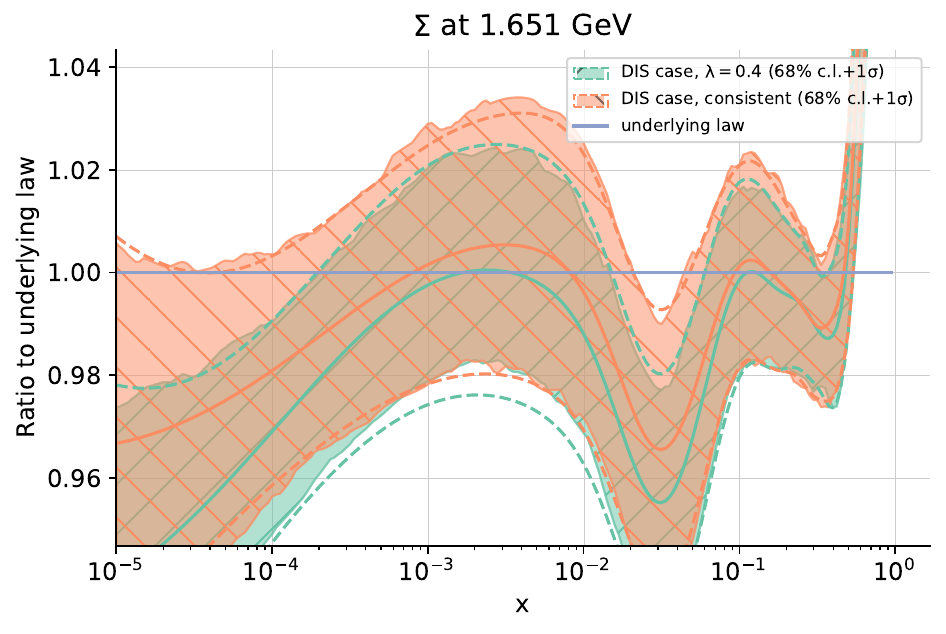}
  \includegraphics[width=0.45\textwidth]{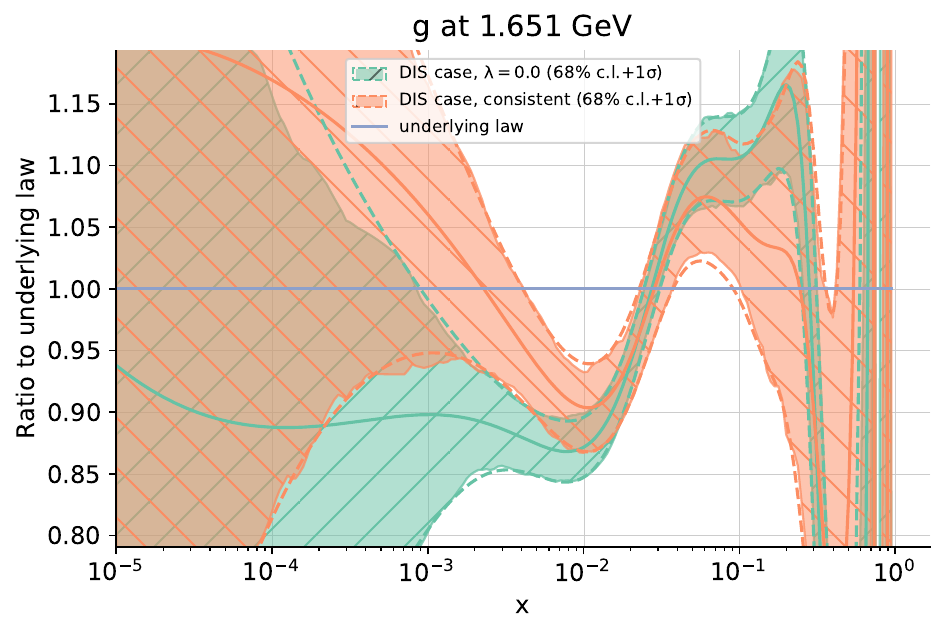}
  \includegraphics[width=0.45\textwidth]{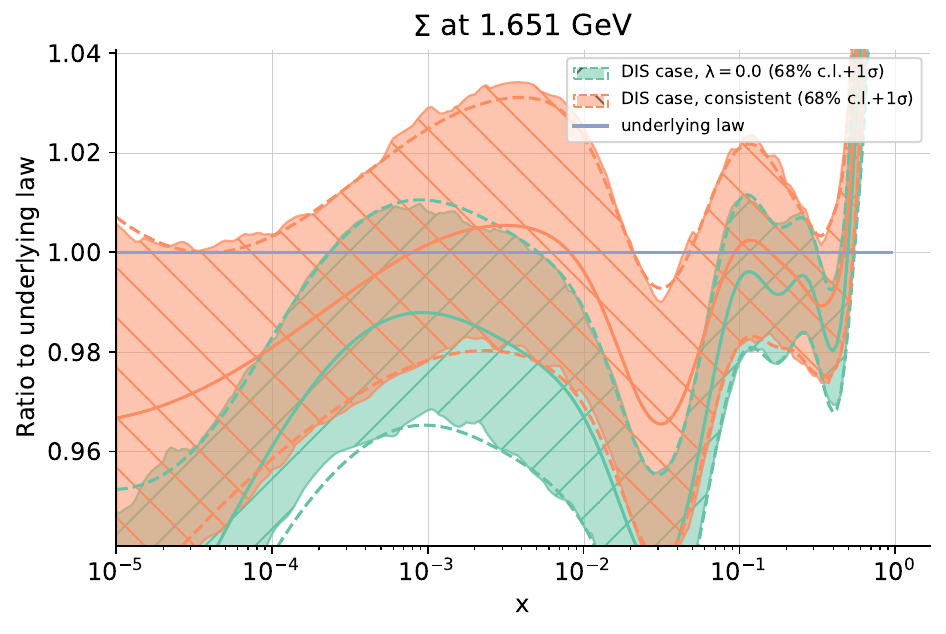}
  \caption{Comparison between the inconsistent and consistent closure
    test results, shown as a ratio to the true underlying law. The
    gluon (left) and singlet (right) are shown, for $\lambda=0.4$ (top row) and
    $\lambda=0$ (bottom row).
    \label{fig:pdf_comp_dis}}
  \end{figure}
This then begs the question of whether the inconsistency is due to
shrinking of uncertainties with fixed bias, or increased bias with
shrinking uncertainty, or both, and how when $\lambda=0.4$ the model
corrects for this. In order to address this question, in
Fig.~\ref{fig:pdf_comp_dis} we directly compare the gluon and
singlet PDF, shown as a ratio to the true
underlying law. It is clear that with $\lambda=0.4$ the
model leads to results that are essentially unchanged in comparison to
the consistent case, except perhaps a marginal reduction in
uncertainty, which explains the increase of $R_b$ by about 10\% in
comparison to the fully consistent case, and a change in the central
value of the gluon which is compatible with a statistical fluctuation.
When  $\lambda=0$  instead, there is a certain shift in central value,
but more importantly a significant reduction in uncertainty. This
means that, when $\lambda=0.4$, the model does correct for the
underestimated uncertainty in the inconsistent data, by not reducing
the PDF uncertainty despite the reduced data uncertainty. When $\lambda=0$,
it does not and the PDF uncertainty shrinks, resulting in underestimated uncertainties on the PDFs.
The behavior of other PDFs is similar, though the largest effects are
seen in the gluon and quark singlet combinations.

%% file: subsec-dy.tex
\subsection{Single dataset inconsistency: Drell-Yan}
\label{subsec:dy}

\begin{figure}[htb]
  \centering
    \includegraphics[width=0.9\textwidth]{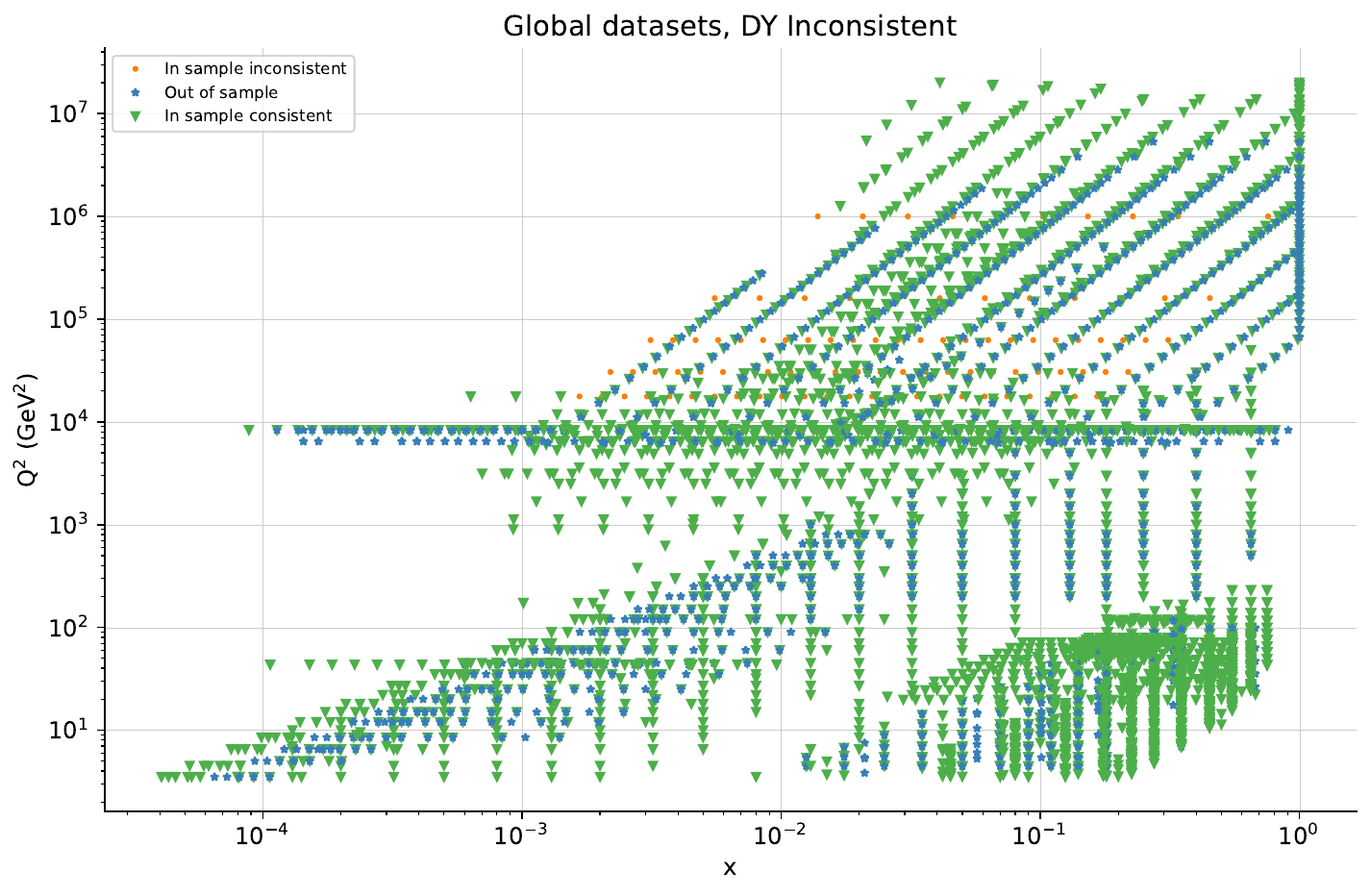}
    \caption{Same as Fig.~\ref{fig:DISpartition} for the datasets
      included in the closure test discussed in Sect.~\ref{subsec:dy}.}
    \label{fig:DYpartition}
\end{figure}

We now consider the full {\tt NNPDF4.0} dataset~\cite{NNPDF:2021njg}. This is an extensive global
collection of data, which includes, on top of all of the DIS datasets
listed in Table~\ref{tab:ratio_bv_dis}, also a wide set of hadronic
data, which are collected for ease of reference in
Tables~\ref{tab:full_dataset_dy}-\ref{tab:full_dataset_jets}. Because
we will use the same dataset for the inconsistency scenarios discussed
in this section and also in Sect.~\ref{subsec:jets} below, which are
based on different in-sample and out-of-sample partitions, we
indicate in Tables~\ref{tab:full_dataset_dy}-\ref{tab:full_dataset_jets} whether
each dataset is in-sample or out-of-sample, and whether
it is affected by an inconsistency, in either of the two scenarios
discussed in this section and the next. In all cases the DIS data of Tab.~\ref{tab:ratio_bv_dis}
are split between in-sample and out-of-sample as indicated in that
table, and always treated as consistent.
The kinematic coverage  and in- and out-of-sample split for the closure
test discussed in this Section are displayed in Fig.~\ref{fig:DYpartition}. 

\input{tab-dyjets.tex}
      
We consider the case in which the inconsistency is
introduced in a single dataset, namely the double differential 
high-mass Drell-Yan cross section at LHC measured by ATLAS at $\sqrt{s}=8 \, \text{TeV}$~\cite{Aad:2016zzw}. 
As for the DIS case, we introduce different degrees of inconsistency,
parametrized by the value of $\lambda$. The total number of in-sample
datapoints is $N_{\text{data}} = 3772$ , while
the dataset which is made inconsistent only includes $N_{\text{inc}}
= 48$ datapoints. However, because all correlated uncertainties affecting
this data are affected by the inconsistency, and  some of them (such
as the luminosity uncertainty) are actually
correlated to other datasets, the total number of data that are
affected by the inconsistency, at least to some extent,
is  $N_{\text{tot, inc}} = 607$. 

\begin{figure}[htb]
  \centering
  \includegraphics[width=0.8\textwidth]{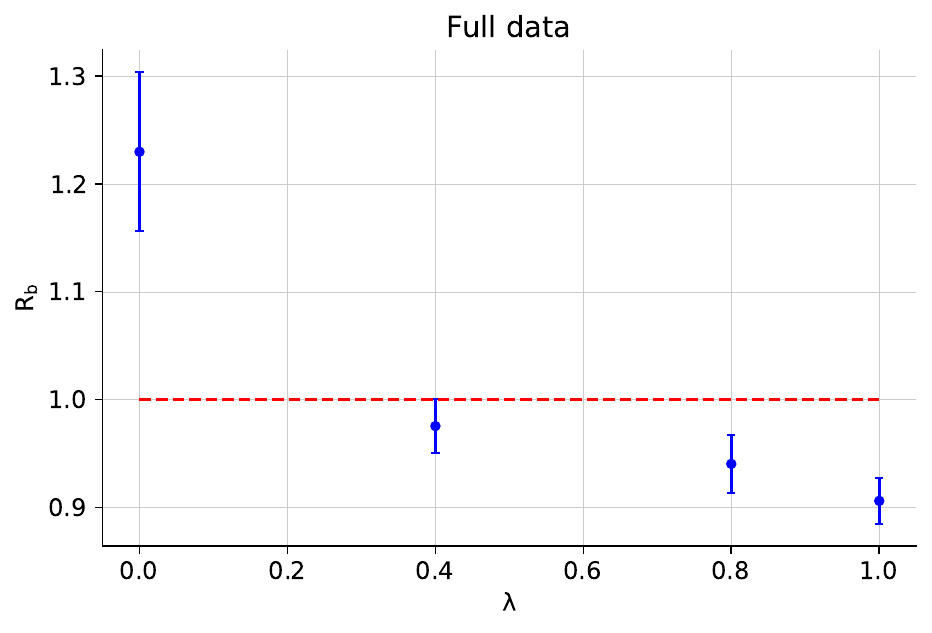}
    \caption{Same as Fig.~\ref{fig:rbv_scan_all_dis_data} for the
    closure test of Sect.~\ref{subsec:dy}.}
    \label{fig:glo_trend_dy}
\end{figure}

\begin{figure}[hb]
  \centering
  \includegraphics[width=0.31\textwidth]{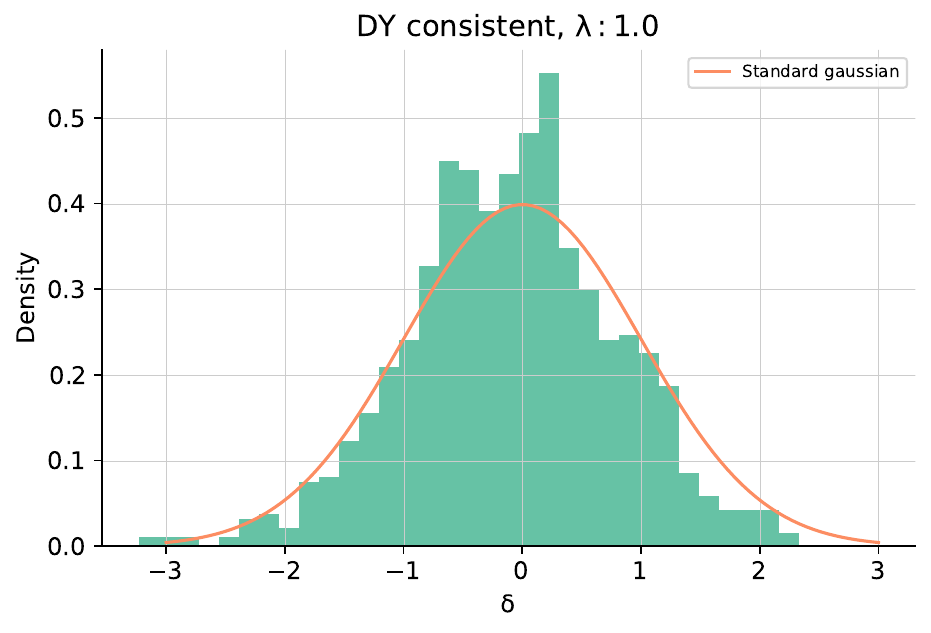}
  \includegraphics[width=0.31\textwidth]{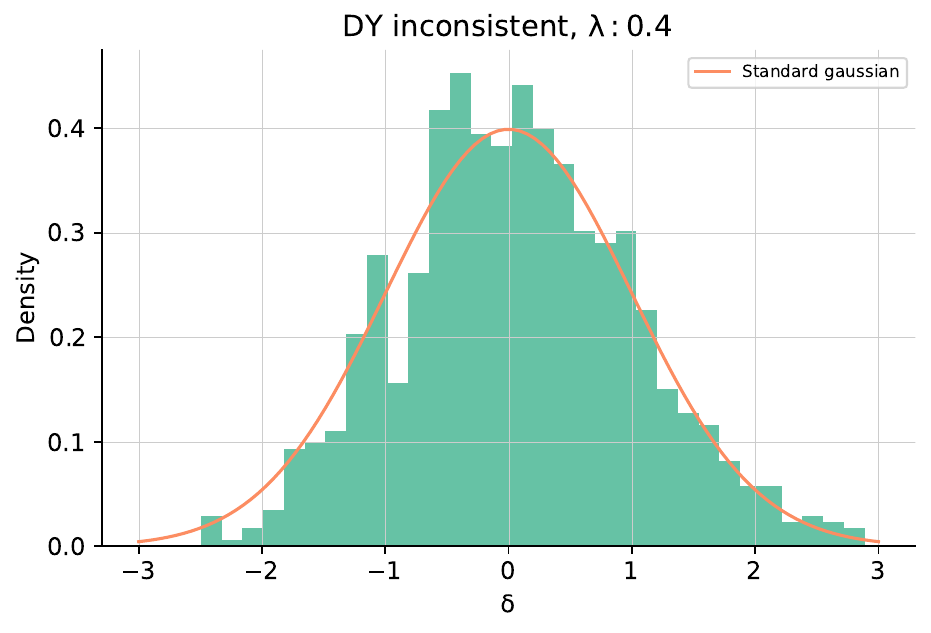}
  \includegraphics[width=0.31\textwidth]{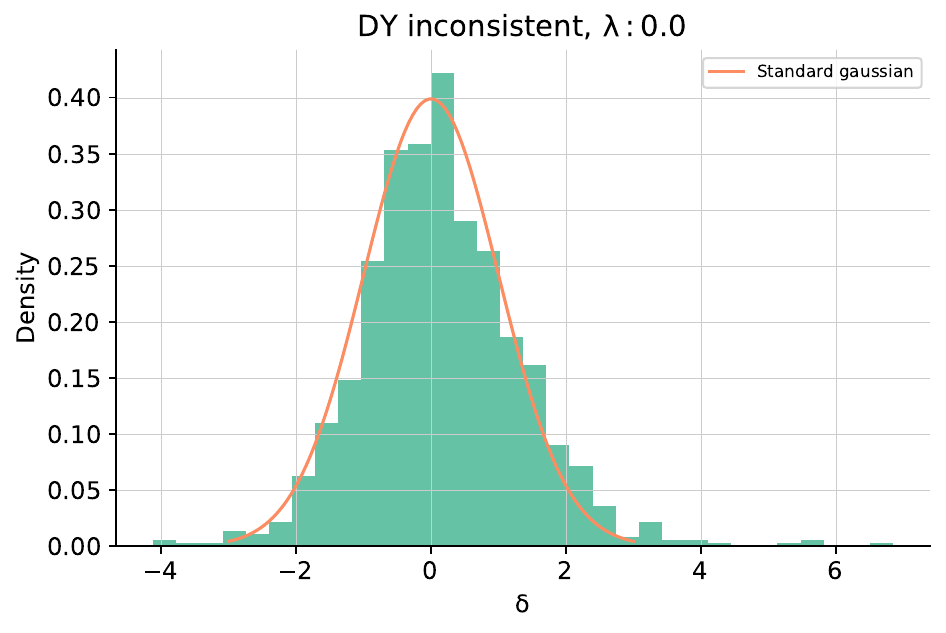}
    \caption{Same as Fig.~\ref{fig:delta_histograms_dis} for the
      closure test of Sect.~\ref{subsec:dy}.}
    \label{fig:delta_dy}
\end{figure}

The values of the normalised bias $R_b$
are collected in Table~\ref{tab:ratio_bv_dy}, with  values for
which $R_b>1$ highlighted in boldface. We show values for
all of the out-of-sample data, for the inconsistent in-sample ATLAS data, and
only for the  consistent in-sample data for which $R_b>1$ for at least
one value of $\lambda$. 
The values of $R_b$ for the full dataset are plotted as a function of  $\lambda$ in Fig.~\ref{fig:glo_trend_dy}, the
values of the  $\xi_{1\sigma}$ are in Table~\ref{tab:xi_dy} and the normalised distribution of relative differences $\delta_{i}^{(l)}$
is in Fig.~\ref{fig:delta_dy} for 
the fully consistent $\lambda=1$ case, the intermediate  case
$\lambda=0.4$, and the extreme inconsistency $\lambda=0$.
\begin{table}[htb]
  \centering
  \begin{tabular}{l|c|cccc}
    \hline
  \multirow{ 2}{*}{Datasets} &  \multirow{ 2}{*}{$N_{\text{data}}$}  & \multicolumn{4}{c}{$R_{b}$} \\
   &   & $\lambda=1.0$ & $\lambda=0.8$ & $\lambda=0.4$ & $\lambda=0.0$ \\
    \hline
    DY E866 $\sigma^p_{\rm DY}$  & 29 & 0.9 & 0.9 & 0.9 & {\bf 1.6} \\
    ATLAS $W,Z$ 7 TeV ($\mathcal{L}=35$~pb$^{-1}$)  & 30 & 0.7& 0.7& 0.9& {\bf 1.6}\\
    ATLAS low-mass DY 2D 8 TeV  & 60 & 0.8 & 0.9 & 0.9 & {\bf 1.8}\\
    ATLAS $Z$ $p_T$ 8 TeV $(p_T , m_{ll})$  & 44 & 0.9 & 1.0 & {\bf 1.5} & {\bf 2.8}\\
    CMS $Z$ $p_T$ 8 TeV  & 28 & 0.8 & 0.9 & 1.0 & {\bf 1.7}\\
    (*) ATLAS high mass DY 8 TeV \cite{ATLAS_2016} & 48 & 0.9 & 1.0 & {\bf 1.3} & {\bf 2.5} \\
    \hline
    Total (in-sample) & 3772 & 0.8 & 0.9 & 0.9 & {\bf 1.5} \\
    \hline
    HERA I+II $\sigma_{\rm CC}^{e^+p}$ \cite{Abramowicz:2015mha} & 39 & 0.9 & 1.0 & 1.0 & {\bf 1.1} \\
    HERA I+II $\sigma_{\rm NC}^{e^\pm p}$  $E_p=575$ GeV \cite{Abramowicz:2015mha} & 254 & 0.7 & 0.7 & 0.8 & 0.8 \\
    NMC $F_2^d/F_2^p$ \cite{Arneodo:1996kd} &  121 & 0.9 & 0.9 & 0.9 & 0.8  \\
    NuTeV $\sigma_{CC}^{\nu}$ (dimuon) \cite{Goncharov:2001qe,MasonPhD}  & 39 & 1.0 & {\bf 1.1} & {\bf 1.1} & {\bf 1.1} \\
    LHCb $W,Z \to \mu$ 7 TeV \cite{LHCb:2015okr} & 29 & 0.9 & 0.9& 1.0 & {\bf 1.4} \\
    LHCb $Z\to \mu\mu$ 13 TeV \cite{LHCb:2016fbk} & 15 & 0.9 & 0.9 & 1.0 & {\bf 1.1} \\
    ATLAS $W^{+}$+jet 8 TeV \cite{ATLAS:2017irc} & 15 & 0.7 & 0.7 & 1.0 & {\bf 1.5} \\
    CMS $W$ muon asymmetry 7 TeV \cite{CMS:2013pzl} & 11 & 0.7 & 0.7 & 0.7 & 0.8 \\
    ATLAS $\sigma^{tot}_{tt}$ 8 TeV \cite{ATLAS:2014nxi} & 1 & 0.92 & 0.8 & 0.9 & 0.9\\
    ATLAS high mass DY 7 TeV \cite{ATLAS:2013xny} & 5 & 0.3 & 0.4 & 0.7 & {\bf 1.6}\\
    ATLAS single $t$ 8 TeV ($1/\sigma d\sigma/dy_{t}$) \cite{ATLAS:2017rso} & 3 & 0.9 & 0.9 & 0.8& 0.9\\
    CMS $\sigma_{tt}^{\rm tot}$ 5 TeV \cite{CMS:2017zpm} & 1 &  0.8 & 0.8 & 0.9 & 0.7 \\
    CMS $t\bar{t}$ 2D $2l$ 8 TeV ($1/\sigma d\sigma/dy_{t}dm_{t\bar{t}}$) \cite{CMS:2017iqf} & 15 & 0.7 & 0.7 & 0.8 & 0.8 \\
    CMS single-inclusive jets 8 TeV \cite{CMS:2016lna} & 185 & 0.7 & 0.7 & 0.7 & 0.9 \\
    \hline
    Total (out-sample) & 734 & 0.9 & 0.9 & 1.0 & {\bf 1.3}\\
    \hline
    Total & 4506 & 0.9 & 0.9 & 1.0 & {\bf 1.2} \\
    \hline
  \end{tabular}
  \caption{Normalised bias $R_b$, Eq.~\eqref{eq:bias_variance_ratio_definition}, as a function of
$\lambda$, for the closure test discussed in Sect.~\ref{subsec:dy}. As in Table~\ref{tab:ratio_bv_dis}, the
in-sample datasets are in the upper part and the out-of-sample in the
lower part of the table, with the inconsistent datasets denoted by an
asterisk and  collected at the bottom of the upper part of the table. Only the
in-sample datasets for which $R_{b}>1$ for at least one value of $\lambda$ are included.}
  \label{tab:ratio_bv_dy}
\end{table}

\begin{table}[htb]
  \centering  
  \begin{tabular}{|c|c|}
    \hline
    $\lambda$ & $\xi_{1\sigma}$ \\
    \hline
    $1.00$ & $0.74 \pm 0.02$\\
    $0.80$ &  $0.71 \pm 0.02$\\
    $0.40$ &  $0.69 \pm 0.02$\\
    $0.00$ &  $0.63 \pm 0.03$\\
    \hline
  \end{tabular}
  \caption{Same as Table~\ref{tab:xi1sigma_dis_full_data} for the
    closure test of Sect.~\ref{subsec:dy}.}
  \label{tab:xi_dy}
\end{table}
The same qualitative behaviour that was seen in the closure test of
Sect.~\ref{subsec:dis} is observed: namely, for $\lambda\gtrsim 0.4$
the model largely corrects for the inconsistency and all the
estimators are qualitatively similar  in the consistent and inconsistent case,
while when $\lambda=0$ the inconsistency shows up
clearly. Quite in general, even in the most inconsistent case,
the effect of the inconsistency is milder than in the case discussed
in the previous Sect.~\ref{subsec:dis}, as is especially clear
comparing the distribution of relative differences
$\delta_i^{(l)}$~Eq. \eqref{eq:deltadef} respectively shown in
Fig.\ref{fig:delta_histograms_dis} and in Fig.\ref{fig:delta_dy}. 

Looking at individual datasets, the normalised bias $R_b$
remains below one and thus faithful for 
almost all datasets even with $\lambda=0.4$, the only exceptions being
the dataset which is made inconsistent, and  one particular Drell-Yan
transverse momentum
distribution,
which starts showing some degree of  inconsistency. Interestingly, all
out-of-sample datasets remain faithful, with two marginal exceptions of
$R_b\sim 1.1$, meaning
that the inconsistency has not affected the reliability of
predictions. Specifically, the out-of-sample ATLAS $\sqrt{s}=7$~TeV DY dataset,
which is closely related to the inconsistent dataset, and indeed shows
the higher degree of inconsistency in the maximally inconsistent
$\lambda=0$ case, remains completely consistent when $\lambda=0.4$.

\begin{figure}[htb]
  \centering
  \includegraphics[width=0.45\textwidth]{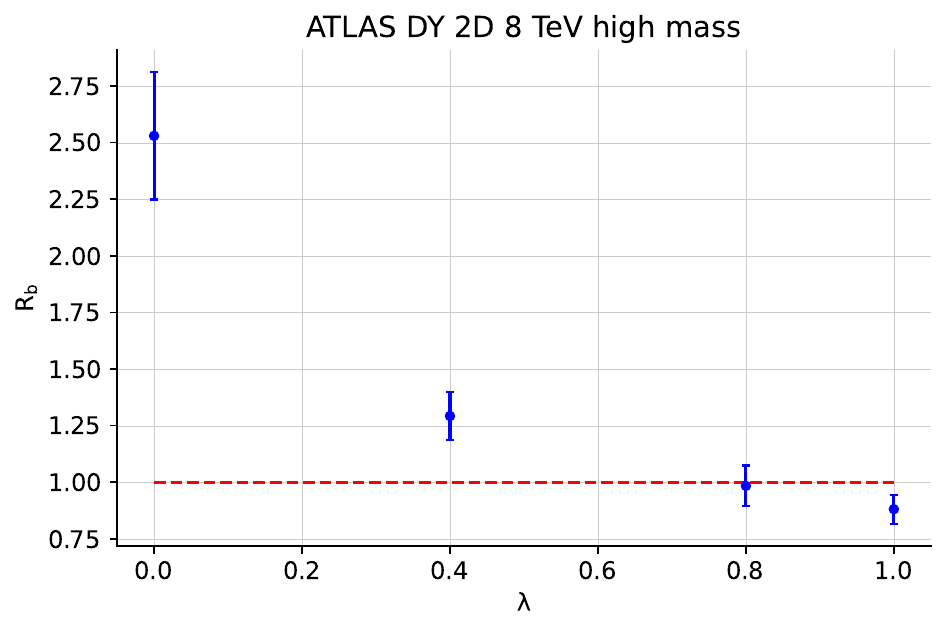}
  \includegraphics[width=0.45\textwidth]{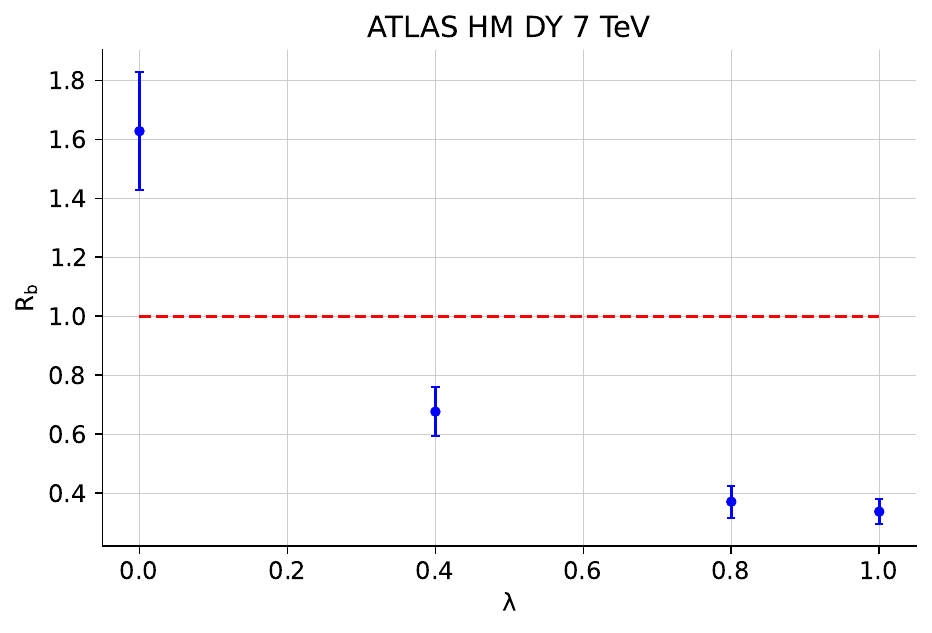}
    \caption{Same as Fig.~\ref{fig:glo_trend_dy} but now for
      the inconsistent ATLAS
      $\sqrt{s}=8$~TeV DY dataset (left) and the ATLAS
      $\sqrt{s}=7$~TeV DY dataset which shows the largest
      out-of-sample inconsistency when $\lambda=0$ (right).}
    \label{fig:ratio_trend_in_out}
  \end{figure}

\begin{figure}[htb]
  \centering
  \includegraphics[width=0.45\textwidth]{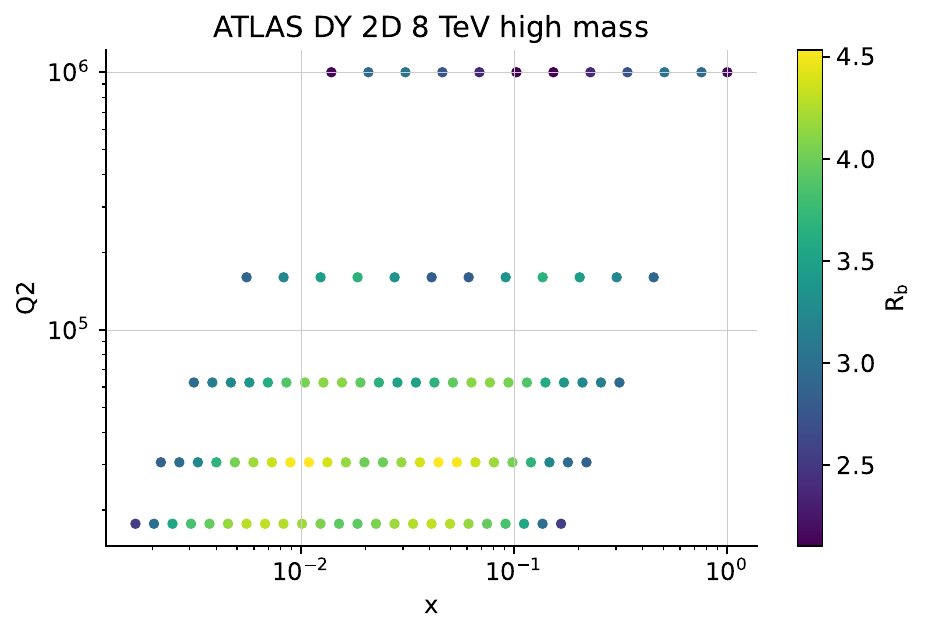}
    \includegraphics[width=0.45\textwidth]{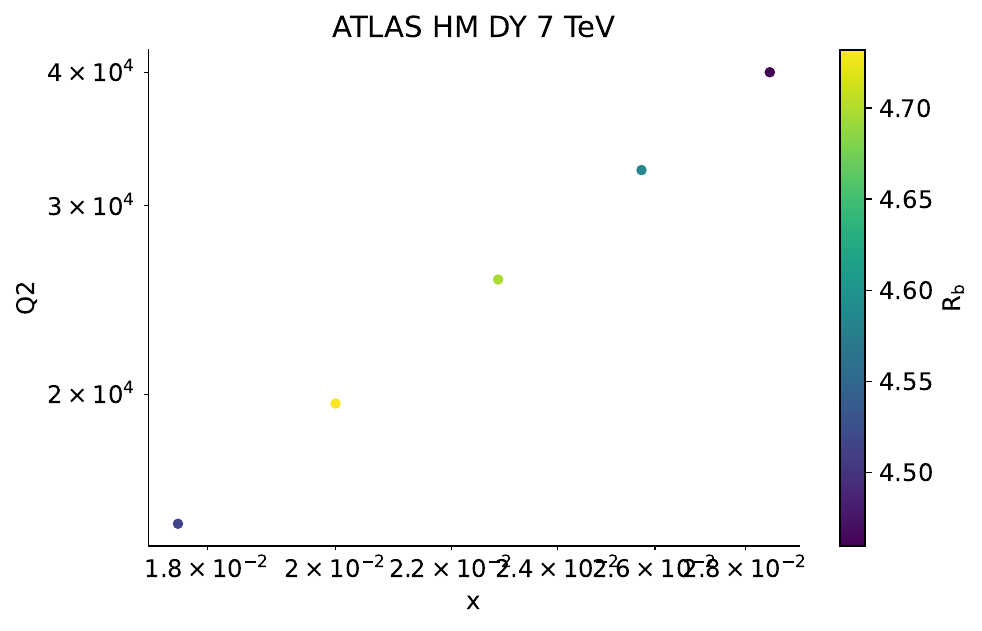}
    \caption{Same as Fig.~\ref{fig:ratio_bv_dis_singledp} but for the
      two datasets shown in Fig.~\ref{fig:ratio_trend_in_out} (left
      and right, respectively).}
    \label{fig:ratio_bv_dy_oos}
  \end{figure}

We finally look specifically at the inconsistent dataset and the
out-of-sample dataset which displays the largest inconsistency
in the maximally inconsistent $\lambda=0$ case, namely the
aforementioned  ATLAS $\sqrt{s}=7$~TeV DY dataset.  For
these datasets in Fig.~\ref{fig:ratio_trend_in_out} we show the normalised bias as a
function of $\lambda$, and in
Fig.~\ref{fig:ratio_bv_dy_oos} the normalised bias for each individual
datapoint. The behaviour of the normalised bias as a function of
$\lambda$ for these datasets is in fact quite similar to that of the
full dataset shown in Fig.~\ref{fig:glo_trend_dy}. This means that the
model generalises well, with no significant difference between
in-sample, out-of-sample data that are most correlated to the
in-sample ones, and the rest of the data. 

The behaviour of the normalised bias for individual datapoints
shows that both in sample and out of sample  the largest inconsistency
is found in the low-mass bins and 
the region $10^{-2}\lesssim x \lesssim 10^{-1}$. In this region, the
data are most strongly correlated to the antiquark distributions (see
App.~\ref{app:corr}), which are thus driving the inconsistent behaviour.


 \begin{figure}
  \centering
  \includegraphics[width=0.45\textwidth]{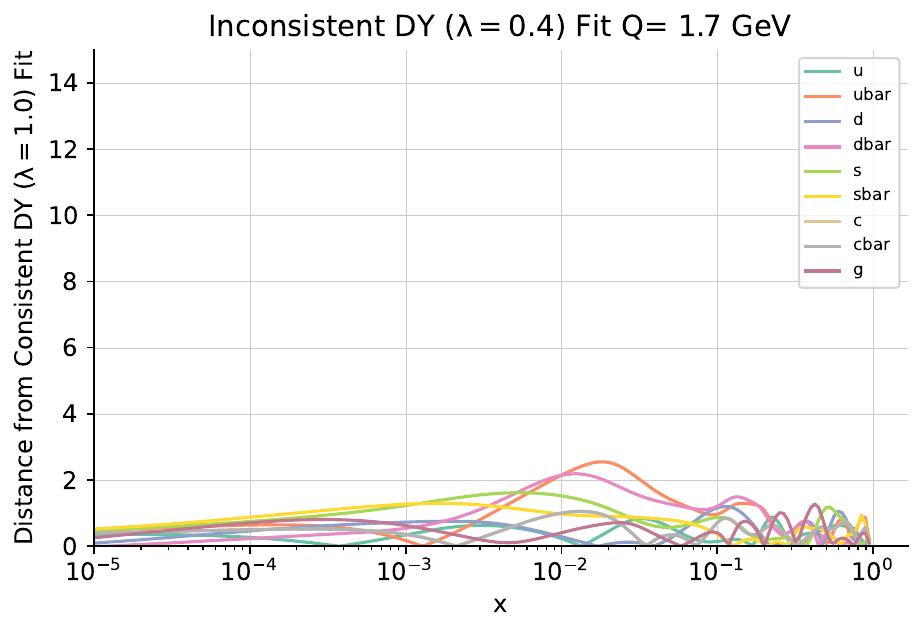}
  \includegraphics[width=0.45\textwidth]{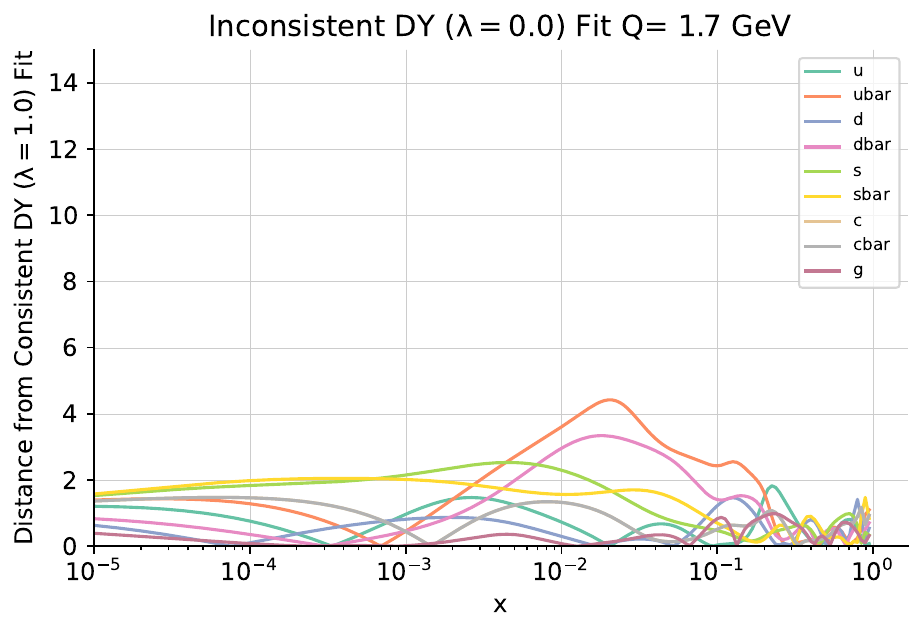}
  \caption{Same as Fig.~\ref{fig:pdf_dist_dis} for the
      closure test of  Sect.~\ref{subsec:dy}. 
    \label{fig:pdf_dist_dy}}
  \end{figure}
 In order to check that this is indeed the case, in
 Fig.~\ref{fig:pdf_dist_dy} we  plot the distance between the central
values of all PDFs determined from a consistent and inconsistent fit
to the same underlying data, both for $\lambda=0.4$ and
$\lambda=0$. It is clear that when $\lambda=0$ indeed it is the up and
down antiquark distributions that display a statistically significant
distance from the consistent ones, with all other PDFs being
essentially unaffected. On the other hand when $\lambda=0.4$ it is
clear that there is no statistically significant difference between
PDFs found in the consistent and inconsistent cases, so indeed the
model is correcting for the inconsistency.

 \begin{figure}
  \centering
  \includegraphics[width=0.45\textwidth]{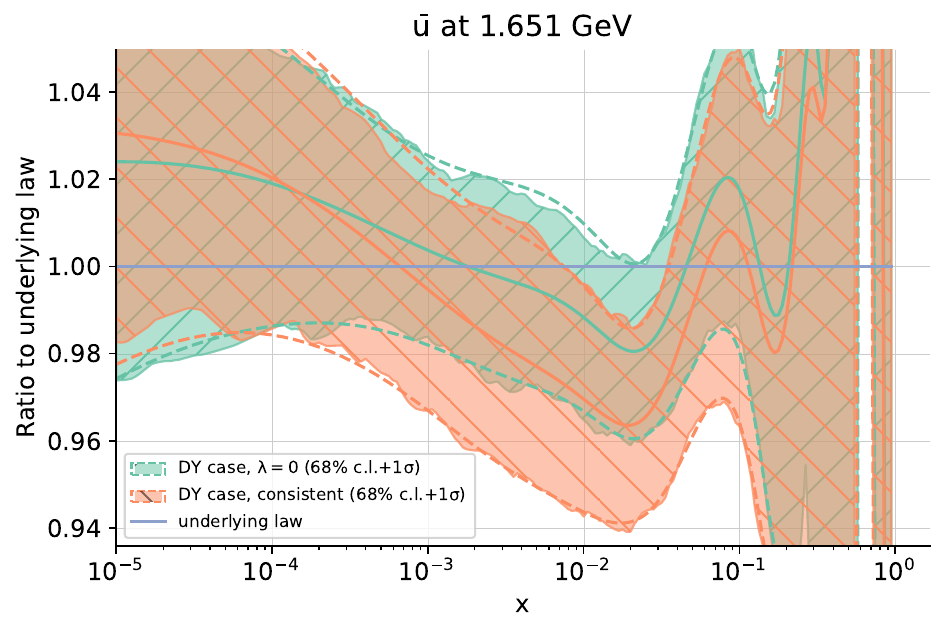}
  \includegraphics[width=0.45\textwidth]{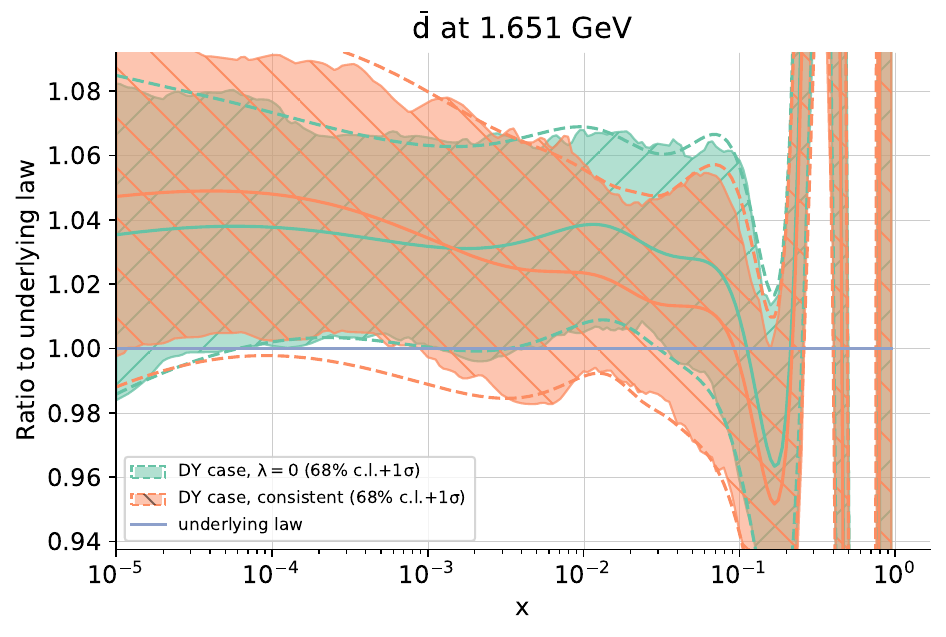}
  \caption{Comparison between the inconsistent and consistent closure
    test results, shown as a ratio to the true underlying law. The
    antiup (left) and antidown (right) are shown, in the maximally
    inconsistent
    $\lambda=0$ case.
    \label{fig:pdf_comp_dy}}
  \end{figure}
Furthermore, inspection of the PDFs shows that, 
while when $\lambda=0.4$ the PDFs are essentially unchanged between
the consistent and inconsistent case, when $\lambda=0$ (see
Fig.~\ref{fig:pdf_comp_dy}) the PDF uncertainties are unchanged, but
especially the antiup and antidown PDFs undergo a significant
shift. The difference in comparison to the case of bulk inconsistency
(recall Fig.~\ref{fig:pdf_comp_dis}) where the central value moved
little but the uncertainty was shrinking can be understood as the
consequence of the fact that in that case there was a large amount of
different data all controlling the same PDFs (specifically the gluon)
and with, in the inconsistent case, underestimated uncertainty,
thereby leading to uncertainty reduction. In this case instead there
is only one dataset that has a high impact on specific PDFs, with
data that are randomly offset by an amount which is larger that their
nominal uncertainty. This leads to a corresponding significant offset
of the PDFs but without significant uncertainty reduction given the
scarcity of the inconsistent data.

%% file: tab-dyjets.tex
\begin{table}
  \centering
  \small
    \begin{tabular}{l|c|c|c}
    \hline
      Datasets &  $N_{\text{data}}$ & in/out sample & Inconsistency\\
      \hline
      DY E866 $\sigma^p_{\rm DY}$  & 29 & in &\\
      DY E605 $\sigma^p_{\rm DY}$  & 85 & in &\\
      DY E906 $\sigma^d_{\rm DY}/\sigma^p_{\rm DY}$ (SeaQuest)  & 6 & in &\\
      D0 $Z$ rapidity  & 28 & in &\\
      D0 $W\to \mu\nu$ asymmetry ($\mathcal{L} = 7.3$~fb${}^{-1}$)  & 9 & in &\\
      ATLAS $W,Z$ 7 TeV ($\mathcal{L}=35$~pb$^{-1}$)  & 30 & in &\\
      ATLAS low mass DY 7 TeV  & 6 & in &\\
      ATLAS $W,Z$ 7 TeV ($\mathcal{L}=4.6$~pb$^{-1}$) CF  & 15 & in &\\
      ATLAS low-mass DY 2D 8 TeV  & 60 & in &\\
      ATLAS $\sigma_{W,Z}$ 13 TeV  & 3 & in &\\
      ATLAS $W^{-}$+jet 8 TeV  & 15 & in &\\
      ATLAS $Z$ $p_T$ 8 TeV $(p_T , m_{ll})$  & 44 & in &\\
      ATLAS $Z$ $p_T$ 8 TeV $(p_T , y_Z)$  & 48 & in &\\
      CMS $W$ electron asymmetry 7 TeV  & 11 & in &\\
      CMS DY 2D 7 TeV  & 110 & in &\\
      CMS $W$ rapidity 8 TeV  & 22 & in &\\
      LHCb $Z \to ee$ 7~TeV  & 9 & in &\\
      LHCb $Z\to ee$ 8~TeV ($\mathcal{L}=2$~fb$^{-1}$)  & 17 & in &\\
      LHCb $W,Z \to \mu$ 8 TeV  & 30 & in &\\
      \hline
            ATLAS high-mass DY 2D 8 TeV  & 48 & in & $\checkmark^{\text{DY}}$\\
      \hline
      ATLAS $W,Z$ 7 TeV ($\mathcal{L}=4.6$~pb$^{-1}$) CC  & 46 & out$^{\text{JETS}}\,$ in$^{\text{DY}}$ &\\
      DY E866 $\sigma^d_{\rm DY}/2\sigma^p_{\rm DY}$ (NuSea) \cite{H1:2018flt} & 15 & out$^{\text{JETS}}\,$ in$^{\text{DY}}$ &\\
      CDF $Z$ rapidity  & 28 & out$^{\text{JETS}}\,$ in$^{\text{DY}}$ &\\
      CMS $Z$ $p_T$ 8 TeV  & 28 & in$^{\text{JETS}}\,$ out$^{\text{DY}}$ &\\
            \hline      
      LHCb $Z \to ee$ 13~TeV  & 16 & out &\\
      \hline
    \end{tabular}
        \caption{Hadronic data entering the  global  NNPDF4.0 dataset,
          and used together with the DIS data of
          Table~\ref{tab:ratio_bv_dis} for the closure tests of
          Sects.~\ref{tab:full_dataset_dy}-\ref{tab:full_dataset_jets}. On
          top of the number of datapoints we indicate whether each
          dataset has been included in or out of sample, and whether it
          is affected by an inconsistency, with DY denoting the
          closure test of  Sect.~\ref{subsec:dy} and JETS
          denoting the closure test of Sect.~\ref{subsec:jets}.}
    \label{tab:full_dataset_dy}
  \end{table}
  \begin{table}
    \small
    \centering
    \begin{tabular}{l|c|c|c}
    \hline
      Datasets &  $N_{\text{data}}$ & in/out sample & Inconsistency\\
      \hline
      ATLAS dijets 7 TeV, R=0.6  & 90 & in &\\
      ATLAS direct photon production 13 TeV  & 53 & in &\\
      ATLAS single $t$ $R_{t}$ 7 TeV  & 1 & in &\\
      ATLAS single $t$ $R_{t}$ 13 TeV   & 1 & in &\\
      ATLAS single $t$ 7 TeV ($1/\sigma d\sigma/dy_{t}$)  & 3 & in &\\
      ATLAS single $t$ 7 TeV ($1/\sigma d\sigma/dy_{\bar{t}}$)  & 3 & in &\\
      ATLAS single $t$ 8 TeV ($1/\sigma d\sigma/dy_{\bar{t}}$)  & 3 & in &\\
      ATLAS $\sigma_{tt}^{\rm tot}$ 13 TeV ($\mathcal{L}=139$~fb$^{-1}$)  & 1 & in &\\
      ATLAS $t\bar{t}$ $l + \text{jets}$ 8 TeV ($1/\sigma d\sigma/dy_{t}$)  & 4 & in &\\
      ATLAS $t\bar{t}$ $l + \text{jets}$ 8 TeV ($1/\sigma d\sigma/dy_{t\bar{t}}$)  & 4 & in &\\
      ATLAS $t\bar{t}$ $2l$ 8 TeV ($1/\sigma d\sigma/dy_{t\bar{t}}$)  & 4 & in &\\
      CMS dijets 7 TeV  & 54 & in &\\
      CMS $\sigma_{tt}^{\rm tot}$ 7, 8, 13 TeV  & 3 & in &\\
      CMS $t\bar{t}$ $l + \text{jets}$ 8 TeV ($1/\sigma d\sigma/dy_{t\bar{t}}$)  & 9 & in &\\
      CMS $t\bar{t}$ $2l$ 13 TeV ($d\sigma/dy_{t}$)  & 10 & in &\\
      CMS $t\bar{t}$ $l$+ jets 13 TeV ($d\sigma/dy_{t}$)  & 11 & in &\\
      CMS single $t$ 7 TeV ($\sigma_{t} + \sigma_{\bar{t}}$)  & 1 & in &\\
      CMS single $t$ 8 TeV $R_{t}$  & 1 & in &\\
      CMS single $t$ 13 TeV  $R_{t}$  & 1 & in &\\
      \hline
            ATLAS single-inclusive jets 8 TeV, R=0.6  & 171 & in & $\checkmark^{\text{JETS}}$\\
\hline
      LHCb $W,Z \to \mu$ 7 TeV  & 29 & out &\\
      LHCb $Z\to \mu\mu$ 13 TeV  & 15 & out &\\
      ATLAS $W^{+}$+jet 8 TeV  & 15 & out &\\
      CMS $W$ muon asymmetry 7 TeV  & 11 & out &\\
      ATLAS $\sigma^{tot}_{tt}$ 7 TeV  & 1 & out &\\
      ATLAS $\sigma^{tot}_{tt}$ 8 TeV  & 1 & out &\\
      ATLAS single $t$ 8 TeV ($1/\sigma d\sigma/dy_{t}$)  & 3 & out &\\
      CMS $\sigma_{tt}^{\rm tot}$ 5 TeV  & 1 & out &\\
      CMS $t\bar{t}$ 2D $2l$ 8 TeV ($1/\sigma d\sigma/dy_{t}dm_{t\bar{t}}$)  & 15 & out &\\
      CMS single-inclusive jets 8 TeV  & 185 & out$^{*}$ &\\
        \hline
    \end{tabular}
    \caption{Table~\ref{tab:full_dataset_dy} continued.}
    \label{tab:full_dataset_jets}
\end{table}

%% file: subsec-jets.tex
\subsection{High-impact inconsistency: jets}
\label{subsec:jets}

We finally consider the case in which, as in the previous section, the
inconsistency is introduced in a single dataset for a global PDF
determination, but now choosing a dataset that has the dominant impact
on a specific PDF in a well-defined kinematic region: namely
single-inclusive jets that have the dominant impact on the gluon in
the intermediate $x$ region. 

\begin{figure}[ht]
  \centering
    \includegraphics[width=0.9\textwidth]{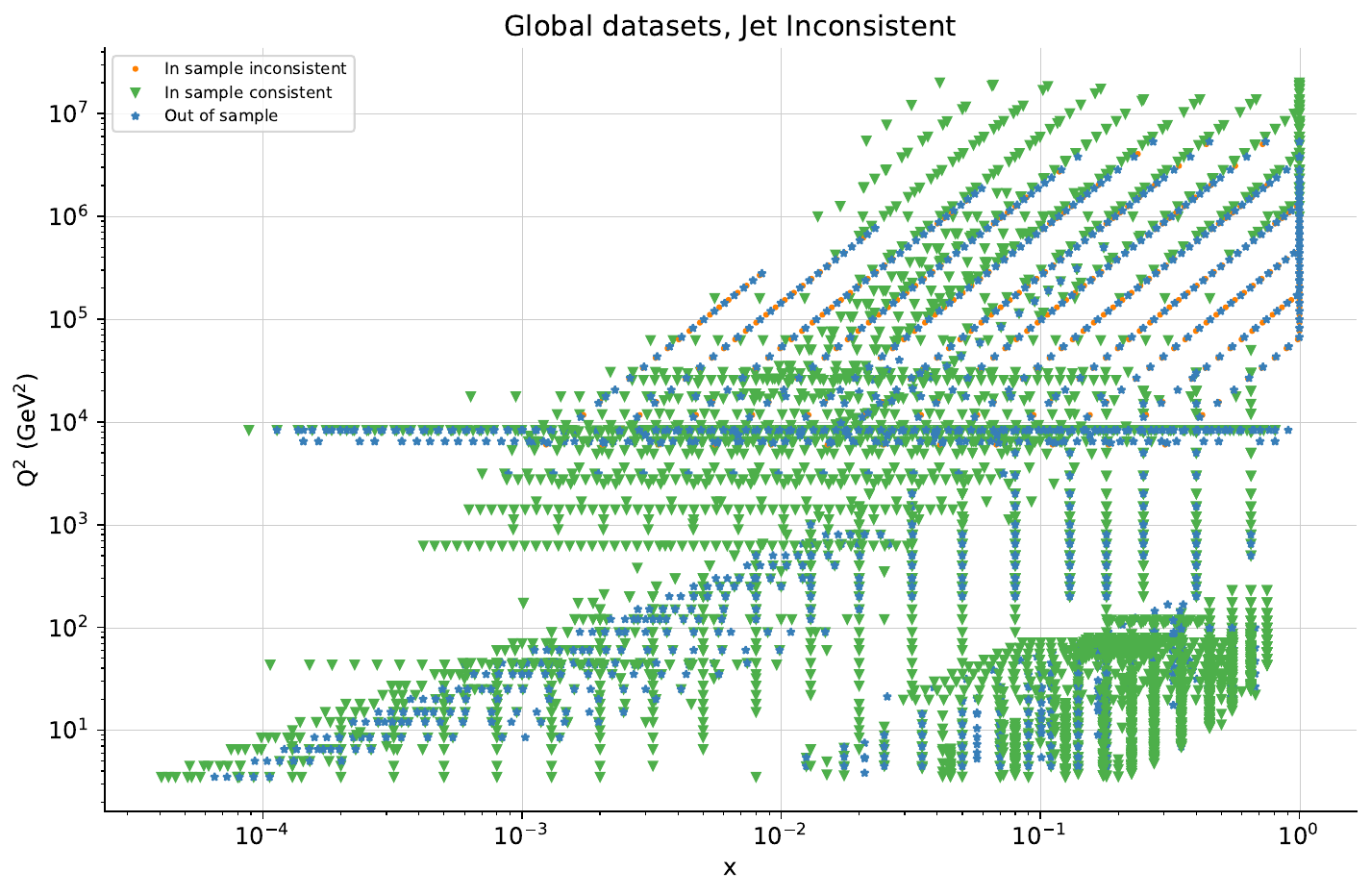}
    \caption{Same as Fig.~\ref{fig:DISpartition} for the datasets
      included in the closure test discussed in Sect.~\ref{subsec:jets}.}
    \label{fig:JETSpartition}
\end{figure}

In particular, we introduce the inconsistency in the ATLAS
$\sqrt{s}=8$~TeV single-inclusive jet data~\cite{Aaboud:2017dvo}. The
corresponding CMS measurement  is kept out-of-sample, so that all the
in-sample data for this process are inconsistent. However, the
in-sample dataset includes other datasets that are also strongly
correlated to the gluon in a similar region, specifically dijet and
top pair production.
The full in-sample and out-of-sample partition  
is shown in
Tables~\ref{tab:full_dataset_dy}-\ref{tab:full_dataset_jets}, and 
displayed in Fig.~\ref{fig:JETSpartition}. We have
$N_{\text{data}}=3793$ in-sample datapoints, with $N_{\text{inc}}=171$  ATLAS
inconsistent data, and keeping into account their correlation to other
datasets, overall $N_{\text{tot, inc}}=607$ datapoints affected by the
inconsistency. 

\begin{table}[ht]
  \centering
  \small
  \begin{tabular}{l|c|cccc}
    \hline
  \multirow{ 2}{*}{Datasets} &  \multirow{ 2}{*}{$N_{\text{data}}$}  & \multicolumn{4}{c}{$R_{b}$} \\
   &   & $\lambda=1.0$ & $\lambda=0.6$ & $\lambda=0.33$ & $\lambda=0.00$\\
    \hline
    CMS $t\bar{t} \; l +$ jets, 8 TeV ($1/\sigma d\sigma/dy_{tt}$) & 10 & 0.8 & 0.9 & 1.0 & {\bf 3.6}\\
    ATLAS direct photon production 13 TeV & 53 & 0.8 & 0.8 & {\bf 1.3}& {\bf 2.0}\\
    ATLAS $t\bar{t} \; l +$ jets, 8 TeV ($1/\sigma d\sigma/dy_{tt}$) & 5 & 0.7& 0.8& 1.0& {\bf 3.3}\\
    NMC $\sigma^{NC,p}$ & 292 & 0.9 & 0.9 & 1.0 & {\bf 1.6} \\
    ATLAS high mass DY 2D 8 TeV	& 48& 0.9 & 0.9 & 0.9 & {\bf 1.2} \\
    (*) ATLAS jets $8$ TeV, $R=0.6$ & 171 & 0.8 & 0.9 & {\bf 1.1} & {\bf 2.3} \\
    \hline
    Total (in-sample) & 3793 & 0.9 & 0.9 & 1.0 & {\bf 2.5}\\
    \hline
    HERA I+II $\sigma_{\rm CC}^{e^+p}$ \cite{Abramowicz:2015mha} & 39 & 1.0  & 0.8 & 1.0 & {\bf 1.2} \\
    HERA I+II $\sigma_{\rm NC}^{e^\pm p}$  $E_p=575$ GeV \cite{Abramowicz:2015mha} & 254 & 0.8 &  0.8 & 0.7 & {\bf 1.6} \\
    NMC $F_2^d/F_2^p$ \cite{Arneodo:1996kd} &  121 & 0.7  & 0.7 & 0.8 & 0.9 \\
    NuTeV $\sigma_{CC}^{\nu}$ (dimuon) \cite{Goncharov:2001qe,MasonPhD}  & 39 & 0.9  & 1.0 & 1.0 & {\bf 1.1}\\
    LHCb $W,Z \to \mu$ 7 TeV \cite{LHCb:2015okr} & 29 & 0.8 & 0.9 & 0.9 & {\bf 1.1}\\
    LHCb $Z\to \mu\mu$ 13 TeV \cite{LHCb:2016fbk} & 15 & 0.8  & 0.8 & 0.8 & {\bf 1.4} \\
    ATLAS $W,Z$ 7 TeV ($\mathcal{L}=4.6$~pb$^{-1}$) CC \cite{ATLAS:2016nqi} & 46 & 0.7  & 0.6 & 0.6 & 0.9 \\
    ATLAS $W^{+}$+jet 8 TeV \cite{ATLAS:2017irc} & 15 & 0.8  & 0.8 & 1.0 & {\bf 3.2}\\
    ATLAS high mass DY 7 TeV \cite{ATLAS:2013xny} & 5 & 1.0  & 0.9 & 1.0 & {\bf 1.2}\\
    CMS $W$ muon asymmetry 7 TeV \cite{CMS:2013pzl} & 11 & 0.7  & 0.7 & 0.7 & 0.7 \\
    DY E866 $\sigma^d_{\rm DY}/2\sigma^p_{\rm DY}$ (NuSea) \cite{H1:2018flt} & 15 & 0.8  & 0.8 & 0.8 & 0.9\\
    CDF $Z$ rapidity \cite{CDF:2010vek} & 15 & 0.7  & 0.8 & 0.7 & 0.9\\
    ATLAS $\sigma^{tot}_{tt}$ 7 TeV \cite{ATLAS:2016nqi} & 1 & 0.7  & 0.8 & 1.0 & {\bf 5.9}\\
    ATLAS $\sigma^{tot}_{tt}$ 8 TeV \cite{ATLAS:2017irc} & 1 & 0.7  & 0.7 & 1.0 & {\bf 5.3}\\
    ATLAS single $t$ 8 TeV ($1/\sigma d\sigma/dy_{t}$) \cite{ATLAS:2017rso} & 3 & 1.0  & 1.0 & 1.0 & {\bf 1.5}\\
    CMS $\sigma_{tt}^{\rm tot}$ 5 TeV \cite{CMS:2017zpm} & 1 &  0.7  & 0.9 & {\bf 1.2}  & {\bf 5.4}\\
    CMS $t\bar{t}$ 2D $2l$ 8 TeV ($1/\sigma d\sigma/dy_{t}dm_{t\bar{t}}$) \cite{CMS:2017iqf} & 15 & 0.7 & 0.8 &1.0  & {\bf 4.7}\\
    CMS single-inclusive jets 8 TeV \cite{CMS:2016lna} & 185 & 0.7  & 0.9 & {\bf 1.1} & {\bf 2.3}\\
    \hline
    Total (out-sample) & 823 & 0.9 & 0.9 & 1.0 & {\bf 2.8} \\
    \hline
    Total & 4616 & 0.9 & 0.9 & 1.0 & {\bf 2.5} \\
    \hline
  \end{tabular}
  \caption{Same as Table~\ref{tab:ratio_bv_dy} for the closure test of
    Sect.~\ref{subsec:jets}. As in Table~\ref{tab:ratio_bv_dy}, only
    in-sample datasets for which $R_b>1$ for at least one value of $\lambda$ are shown.} 
  \label{tab:ratio_bv_jets_cms_out}
\end{table}

 \begin{figure}[htb]
  \centering
  \includegraphics[width=0.8\textwidth]{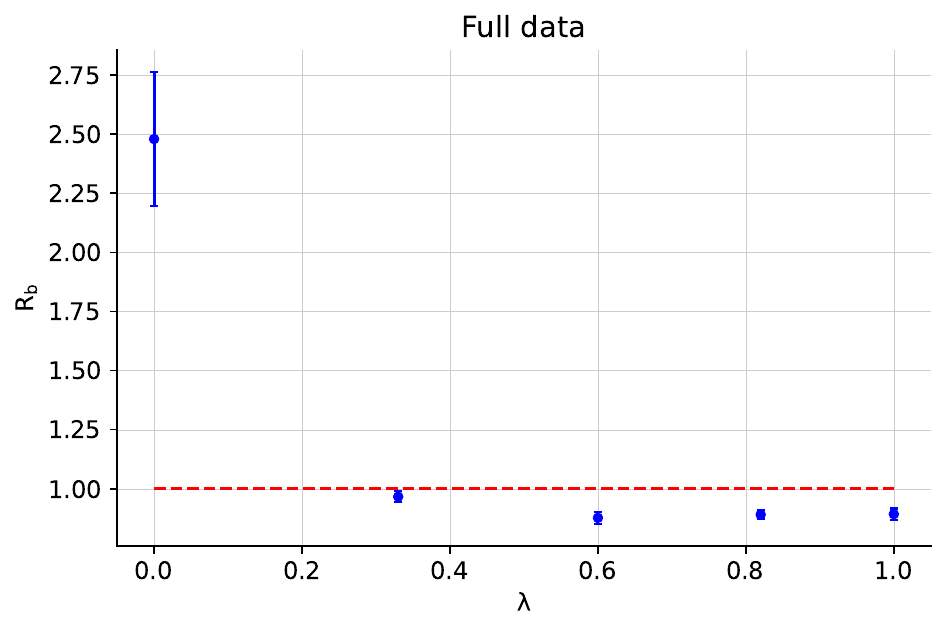}
    \caption{Same as Fig.~\ref{fig:rbv_scan_all_dis_data} for the
    closure test of Sect.~\ref{subsec:jets}.}
    \label{fig:glo_trend_cms_out}
 \end{figure}
\begin{figure}[htb]
  \centering
  \includegraphics[width=0.31\textwidth]{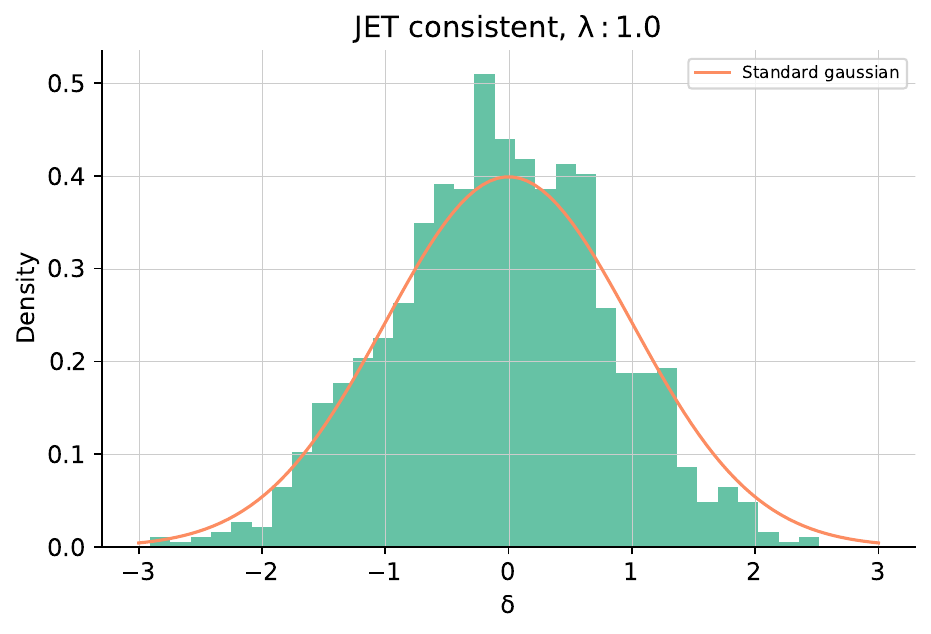}
  \includegraphics[width=0.31\textwidth]{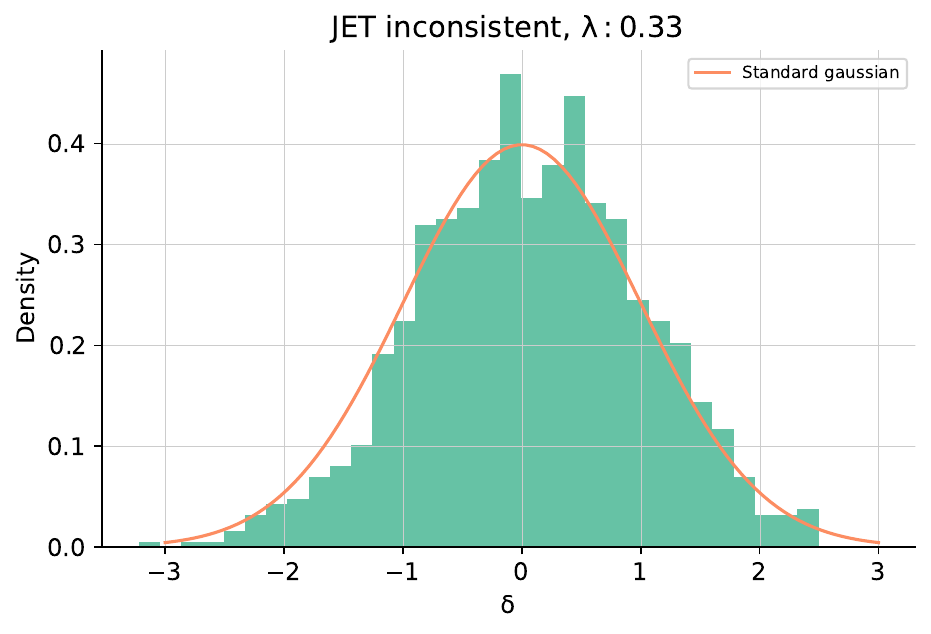}
  \includegraphics[width=0.31\textwidth]{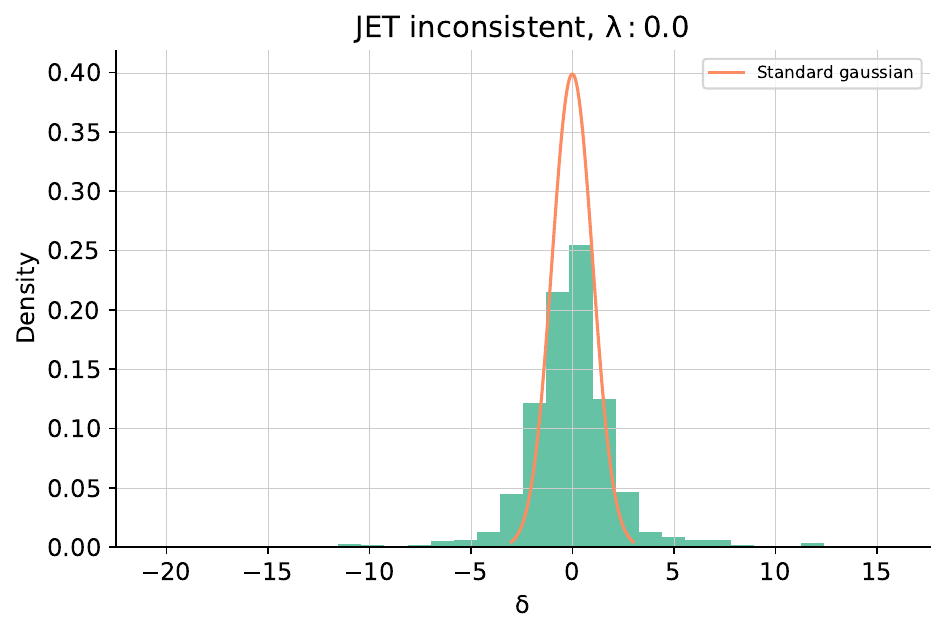}
    \caption{Same as Fig.~\ref{fig:delta_histograms_dis} for the
    closure test of Sect.~\ref{subsec:jets} (but with $\lambda=0.33$
    for the central plot).}
    \label{fig:delta_jets_cms_out}
\end{figure}

The values of the normalised bias $R_b$
are collected in Tab.~\ref{tab:ratio_bv_jets_cms_out}. The normalised
bias $R_{b}$ is plotted as a function of $\lambda$ in 
Fig.~\ref{fig:glo_trend_cms_out}, the values of the
quantile  $\xi_{1\sigma}$ are collected in
Table~\ref{tab:xi_jets_cms_out}, and the normalised $\delta_{i}^{(l)}$
distribution 
 is shown in Fig.~\ref{fig:delta_jets_cms_out} for $\lambda=1$,
$\lambda=0.4$, and $\lambda=0$.
In this case the dramatic change of behaviour of the normalised bias as
$\lambda$ decreases is even more apparent and in fact it extends to
even smaller $\lambda$: when $\lambda\gtrsim0.3$,
$R_b$ is completely flat and equal to its consistent value, while when
$\lambda=0$ it is quite high, in fact even larger than in the case of
bulk inconsistency shown in Fig.~\ref{fig:rbv_scan_all_dis_data}. The
same behaviour is displayed by  $\xi_{1\sigma}$ and  the
$\delta_{i}^{(l)}$ distribution.  
\begin{table}[htb]
  \centering
  \begin{tabular}{|c|c|}
    \hline
    $\lambda$ & $\xi_{1\sigma}$\\
    \hline
    $1.00$ & $0.75 \pm 0.02$\\
    $0.82$ &  $0.74 \pm 0.01$\\
    $0.60$ &  $0.74 \pm 0.02$\\
    $0.33$ &  $0.70 \pm 0.02$\\
    $0.00$ &  $0.45 \pm 0.04$\\
    \hline
  \end{tabular}
  \caption{Same as Tab.~\ref{tab:xi1sigma_dis_full_data} for the
    closure test of Sect.~\ref{subsec:jets}.}
  \label{tab:xi_jets_cms_out}
\end{table}

Coming now to individual datasets,
Table~\ref{tab:ratio_bv_jets_cms_out} shows that when $\lambda \gtrsim
0.3$ the normalised bias is always faithful, with only a marginal sign
of inconsistency (with uncertainties underestimated at the 10-20\%
level) when $\lambda=0.33$,  and then only for the inconsistent
in-sample dataset, and for a couple out-of-sample datasets that are
most strongly correlated to it: specifically the corresponding CMS
measurement, and one datapoint for the total top pair production
cross section.
This indicates that all PDFs are essentially unchanged
in this inconsistent case in comparison to the consistent result.
On the other hand, when $\lambda=0$ all processes that
are highly correlated to the gluon PDF, 
specifically jets and top pair production (see
App.~\ref{app:corr}),
and even (though to a
lesser extent) HERA DIS data, show large inconsistencies,
with uncertainties underestimated by a factor that can be as large as
five, indicating large inconsistencies in the gluon PDF.

 \begin{figure}
  \centering
  \includegraphics[width=0.45\textwidth]{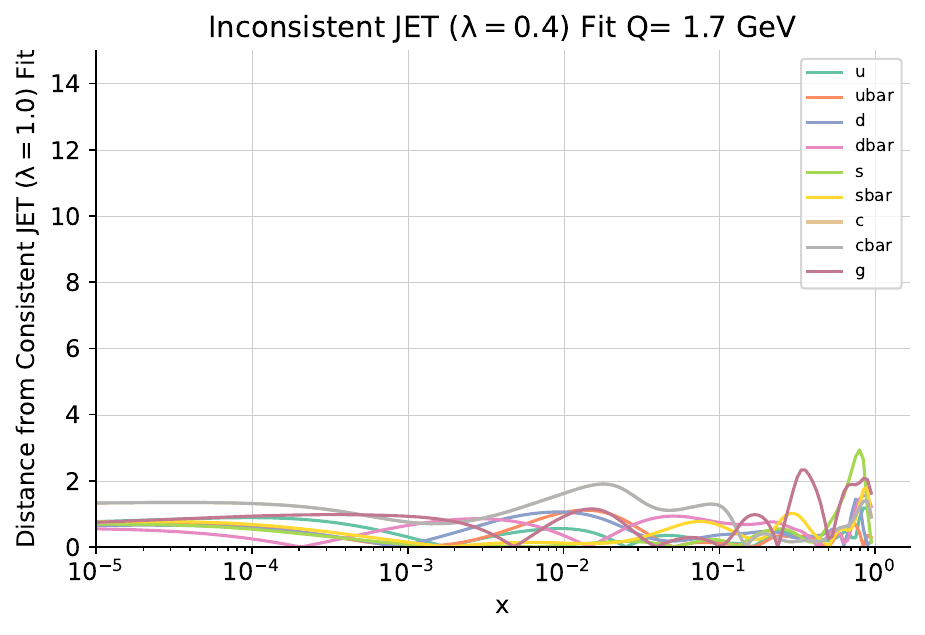}
  \includegraphics[width=0.45\textwidth]{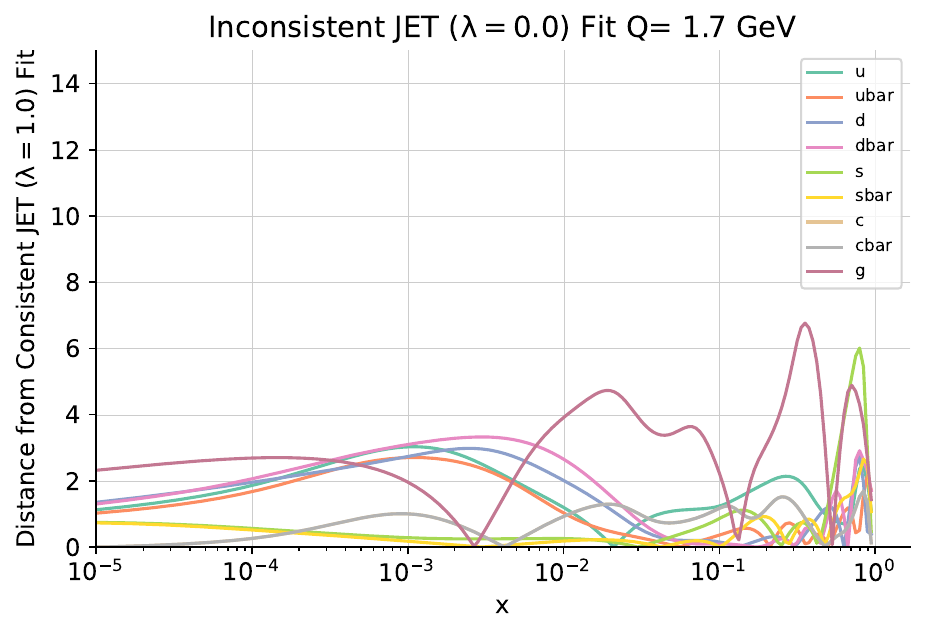}
  \caption{Same as Fig.~\ref{fig:pdf_dist_dis} for the
      closure test of  Sect.~\ref{subsec:jets}. 
    \label{fig:pdf_dist_jets}}
  \end{figure}
\begin{figure}[ht]
  \centering
  \includegraphics[width=0.45\textwidth]{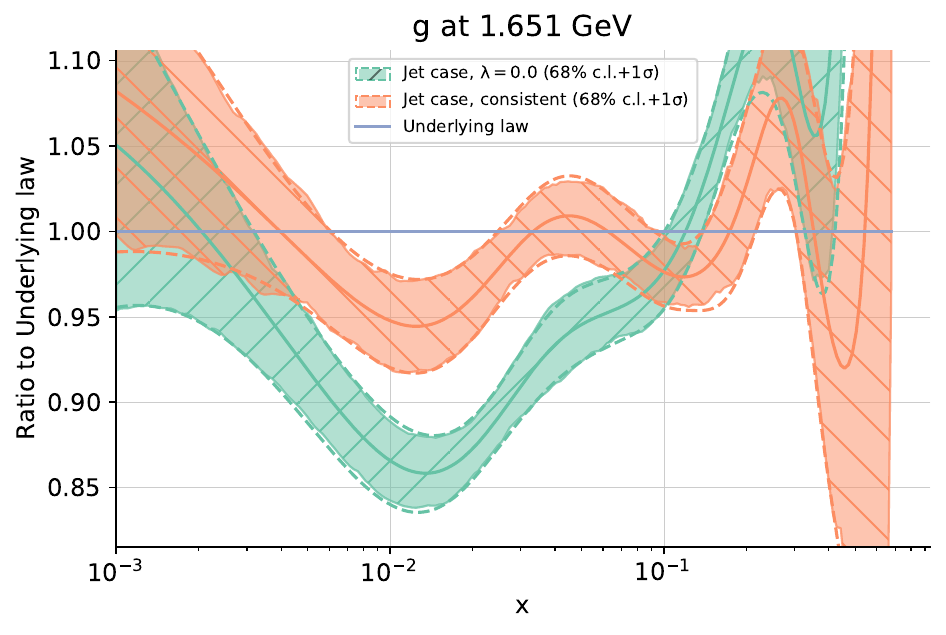}
  \includegraphics[width=0.45\textwidth]{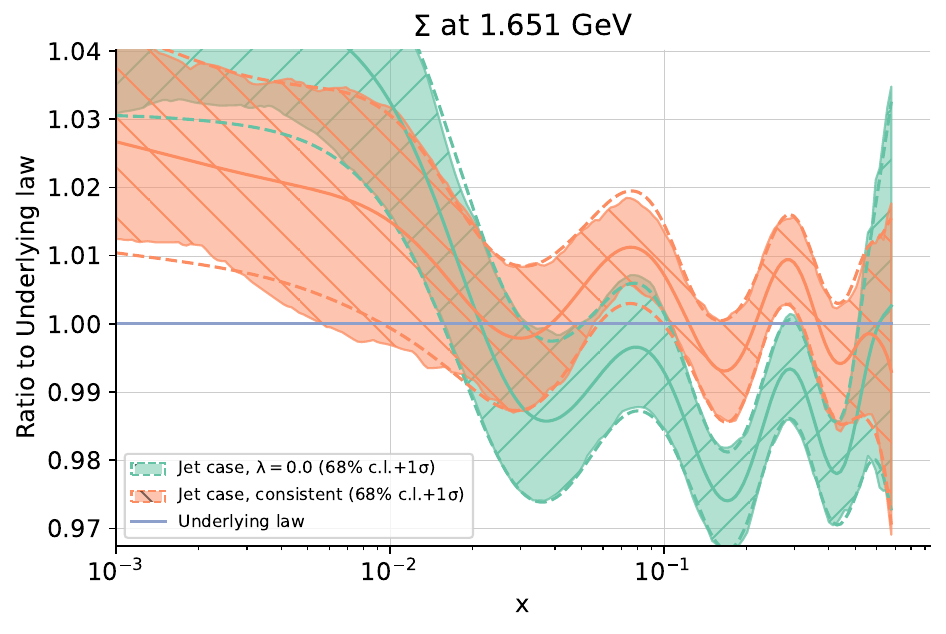}
    \caption{Comparison between the gluon PDF (left) and singlet PDF (right) obtained in the
inconsistent and consistent closure tests, shown as a ratio to the true underlying law.}
    \label{fig:gluon_pdf_jets}
  \end{figure}

This behaviour of the PDFs is indeed seen when determining the distance
between results found in the inconsistent and consistent cases (see
Fig.~\ref{fig:pdf_dist_jets}). When $\lambda=0.33$ (left) all
distances are compatible with a statistical fluctuation, while when
$\lambda=0$   distances as large as two- or three-sigma  are seen for
all PDFs. A direct comparison of the gluon PDF (see
Fig.~\ref{fig:gluon_pdf_jets}, left) shows that the large distance is
due to the fact that in the inconsistent case the gluon shows a large
deviation from its true value, but with no significant increase in its
uncertainty in comparison to the consistent case. A similar trend can be observed from 
the right plot of Fig.~\ref{fig:gluon_pdf_jets} in the singlet PDFs, $\Sigma = \sum_i (q_i+\bar{q}_i)$, 
which is  directly coupled to the gluon via DGLAP evolution equations, and therefore also shows a sizable deviation from the true value in the inconsistent case, 
although to the one-sigma level. 

%% file: sec-prescription.tex
\section{Flagging inconsistencies}
\label{sec:recipe}

The main result of the study, presented in Sect.~\ref{sec:results}, is
that in all the scenarios that we investigated the neural net corrects
for inconsistencies, leading to faithful uncertainties, unless the
inconsistency is extreme. 
Whereas it is unlikely that extreme
inconsistencies such as the one considered in
Sect.~\ref{subsec:dis} with $\lambda=0$ might be realistic, a
situation such as those considered in
Sects.~\ref{subsec:dy}-\ref{subsec:jets}, in which an individual
dataset is severely inconsistent, are well possible.
While the specific scenario we examined, in
which all correlated uncertainties are missed, is somewhat simplistic,
an extreme inconsistency could manifest itself, e.g. if a dominant
source of systematics was missed altogether. This then raises the
question of whether in such a case the inconsistency could be detected
in a real-life scenario.

Clearly, this cannot be done by looking  at  the normalised bias,
given that, as explained in Sect.~\ref{sec:definition}, this quantity
can only be computed if the underlying theory is known and more sets
of $L_1$ data, corresponding to different runs of the universe, are
available. Of course,  
in a real-life situation, the underlying law is not known and only one run of the universe
is available via the actual experimental data: so is it possible to
detect inconsistent datasets included in a global PDF analysis? 
In this section we address this question, and use the closure test as
a means to optimise a procedure that can be used in a realistic case,
taking as a starting point a procedure that was suggested in
Ref.~\cite{NNPDF:2021njg}.

\subsection{Testing for inconsistencies}
\label{subsec:test_setting}

In order to construct consistency indicators to be used in  an actual PDF fit 
we consider the closure test of Sect.~\ref{subsec:dis}, based on $N_{\rm fit}=25$ instances of
$L_1$ data and we compute $N_{\rm fit}$
independent values of $\chi^2_{i}$ for each dataset $i$. 
Next, we define the $n_{\sigma}$ estimator as the normalised  
deviation of the $\chi^2$ from its mean, {\it i.e.}
\begin{equation}
n_{\sigma}^{(i)} = \frac{\chi^2_i-N_{\rm data}^{(i)}}{\sqrt{2 /N_{\rm data}^{(i)}}}\, ,
\end{equation}
where $i$ is the dataset index (note that the standard deviation is
$\sqrt{2/N_{\rm data}}$ because the  $\chi^2$
Eq.~(\ref{eq:chi2_definition}) is normalised to the number of datapoints). In a closure test, we define
the mean value 
\begin{equation}\label{eq:muave}
    \mu_i = \left\langle n_{\sigma}^{(i)} \right\rangle_{N_{\rm fit}}
    \end{equation}
of $n_{\sigma}^{(i)}$ over the $N_{\rm fit}$ instances.

\begin{figure}[h]
    \centering
    \includegraphics[width=0.45\textwidth]{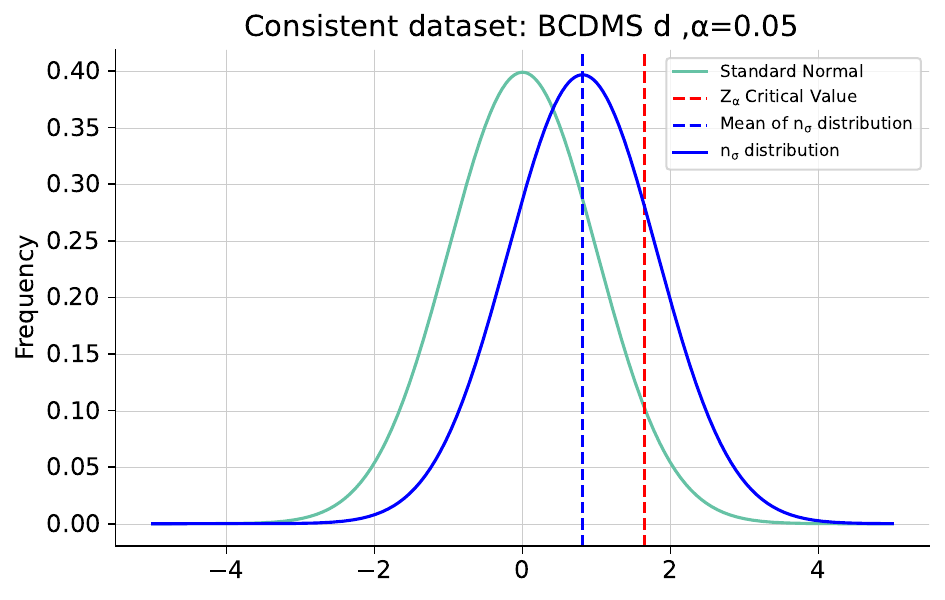}
    \includegraphics[width=0.49\textwidth]{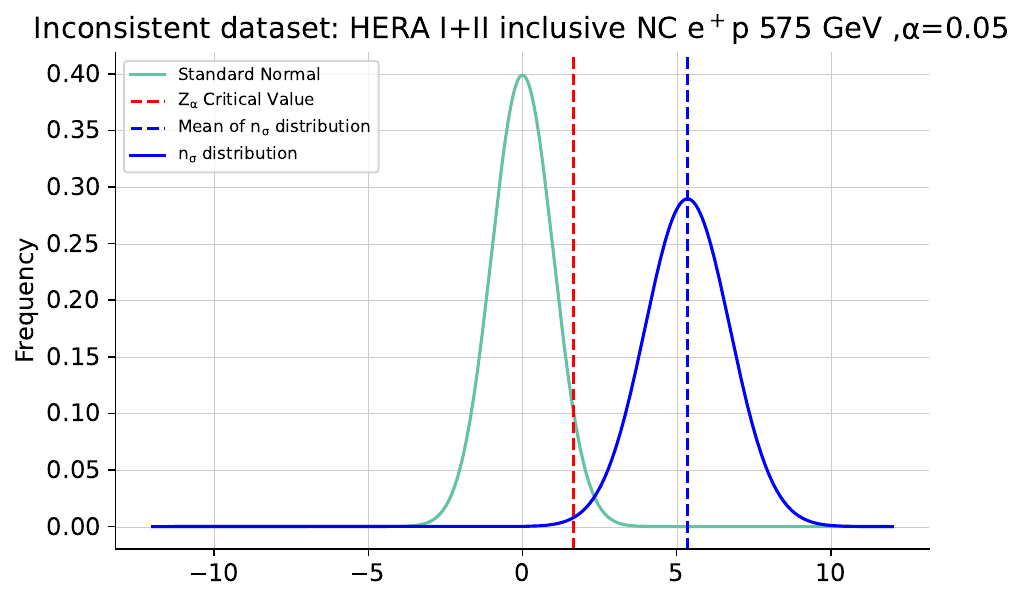}
    \caption{Distribution of $n_{\sigma}$ values in the closure test
      of Sect.~\ref{subsec:dis}, with maximal $\lambda=0$
      inconsistency. Results are shown both for a 
      consistent and an inconsistent dataset:
      BCDMS $F_2^d$~\cite{Benvenuti:1989rh} (consistent, left panel)  and HERA I+II
      $\sigma_{\rm NC}^{e^\pm p}$
      $E_p=575$~GeV~\cite{Abramowicz:2015mha} (inconsistent,
      right panel). An univariate standard normal is also shown for
      comparison with a  vertical line indicating its 95\% quantile.
    }
    \label{fig:nsigma_dist_vs_h0}
\end{figure}
We then flag a dataset as inconsistent if the condition $S_1$ is satisfied, namely
\begin{equation}\label{eq:incons}
S_1:\qquad  \mu_i > Z,
\end{equation}
where $Z$ is some suitable threshold value. 
Note that we consider $\mu_i>0$ because we are interested in
flagging cases in which uncertainties are underestimated. Given that the
distribution of $n_{\sigma}^{(i)}$ tends to a standard normal in the $N_{\rm data}^{(i)}\to\infty$ limit, 
we can think of $Z$ as being approximately a
standard normal quantile. 
In Fig.~\ref{fig:nsigma_dist_vs_h0} we
display the distribution of $n_{\sigma}^{(i)}$ values found in the
closure test of  Sect.~\ref{subsec:dis} both for a consistent (left)
and inconsistent (right) dataset in the case of maximal inconsistency, $\lambda=0$. 
For reference we also show a standard normal, with a
vertical line indicating its 95\% confidence quantile.

Clearly, a low value of $Z$ leads to flagging the
inconsistency in a larger fraction of cases (true positives,
henceforth), but at the cost of also flagging as inconsistent a
consistent dataset (false positives, henceforth), and conversely. We
accordingly wish to construct a selection procedure that maximises
true positives while minimising false positives. Inspection of
Fig.~\ref{fig:nsigma_dist_vs_h0} suggests that no choice of $Z$ leads
to a satisfactory compromise: for instance it is clear from the plot
that choosing a 95\% confidence level would lead to a very large
fraction of false positives, i.e. the consistent BCDMS experiment
would be flagged as inconsistent in a significant fraction of
cases. This suggests that a more efficient selection criterion is
needed, that does not merely rely on the value of $n_\sigma$.

\subsection{Dataset weighting}
\label{subsec:weightedfit}

In Ref.~\cite{NNPDF:2021njg} it was suggested to detect
inconsistencies based on a two-stage procedure. In a first stage, a
dataset is flagged as tentatively inconsistent if it satisfies the
criterion $S_1$, Eq.~(\ref{eq:incons}), with a suitable choice of $Z$ (as well as other criteria). In a
second stage a new PDF determination is performed, in which the
tentatively inconsistent dataset is assigned a large weight, i.e. the
loss function Eq.~(\ref{eq:chi2_definition}) is replaced by   the
weighted loss
\begin{equation}
  \chi^2_{\rm weighted}=\frac{1}{N_{\rm data}-N_{\rm
    data}^{(j)}}\,\sum_{j=1,j\neq i}^{N_{\rm exp}} \,N_{\rm
  data}^{(j)}\,\chi^2_j \, + \,w^{(i)}\chi^2_i,
  \end{equation}
  with
  \begin{equation}
    \label{eq:wgts}
    w^{(i)} = N_{\rm data}/N_{\rm data}^{(i)}.
  \end{equation}
This means that the putatively inconsistent $i$-th dataset now carries the same weight as all
the rest of the data.
  
If after minimisation the weighted loss remains above threshold, namely if the condition $S_2$ is satisfied,
\begin{equation}\label{eq:S2}
S_2:\qquad n_{\sigma}^{{\rm weighted},(i)} > Z,
\end{equation}
or if the loss  $\chi^2_j$ of any other dataset $j\neq i$ deteriorates more than
the given threshold, namely if the condition $S_3$ is satisfied,
\begin{equation}\label{eq:S3}
S_3:\qquad  n_{\sigma}^{{\rm weighted},(j)} - n_{\sigma}^{(j)} > Z \qquad \forall j\neq i,
 \end{equation}
then dataset $i$ is considered inconsistent. However, if neither
condition $S_2$ nor $S_3$ is satisfied, i.e. if  in the
weighted fit experiment $i$ is now well reproduced, without
significant deterioration of the fit quality for all other experiments, 
the flag is removed and dataset $i$ is  considered
consistent. Note that criterion $S_3$ is based on
the deviation between the weighted and unweighted values of the loss
for dataset $j$,
rather than its deviation from zero: namely,  what is tested is not the
absolute consistency of experiment $j$, but rather its compatibility
with experiment $i$. Note also that in Ref.~\cite{NNPDF:2021njg} a
precise threshold value in Eqs.~(\ref{eq:S2}-\ref{eq:S3}) was not
specified, and the criterion was applied in a somewhat loose way. Note
finally that in principle
one might choose a different $Z$ value in each
of these equations and also in Eq.~(\ref{eq:incons}), though for
simplicity we choose a single threshold value.

\begin{figure}[htb]
    \centering
    \includegraphics[width=0.49\textwidth]{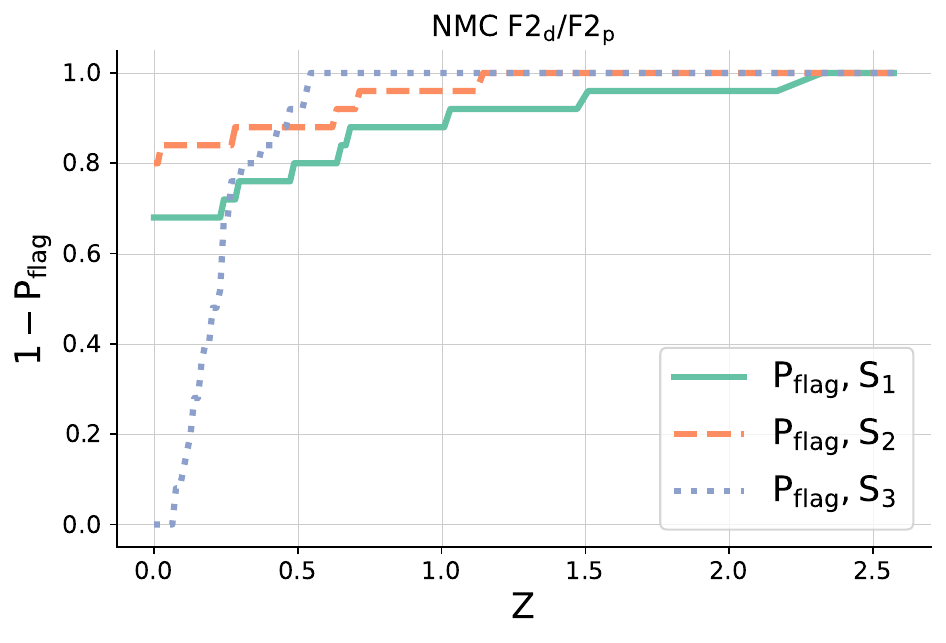}
    \includegraphics[width=0.49\textwidth]{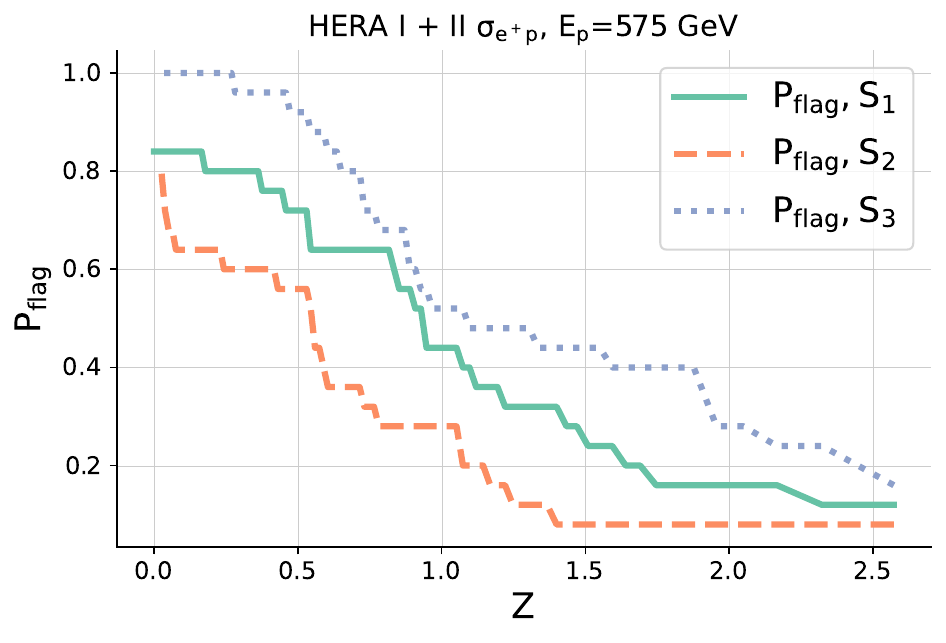}
    \caption{Probability of a true negative (left) and a true positive
      (right) as a function of the value of the $Z$ threshold. The true
      negative (consistent experiment not flagged as inconsistent)
      refers to  the NMC $F_2^d/F_2^p$ consistent
      dataset~\cite{Arneodo:1996kd}; the true positive (inconsistent
      experiment flagged as such) to the inconsistent HERA I+II $\sigma_{\rm NC}^{e^\pm p}$  $E_p=575$~GeV 
    dataset~\cite{Abramowicz:2015mha}. The three curves correspond to
    flagging a dataset as inconsistent when it satisfies either of the
    three conditions $S_1$
Eq.~(\ref{eq:incons}), $S_2$
Eq.~(\ref{eq:S2}) , or $S_3$
Eq.~(\ref{eq:S3}).}
    \label{fig:flags_s}
\end{figure}

In the context of a closure test we can assess quantitatively this
procedure.
To this purpose, we examine the closure test of  Sect.~\ref{subsec:dis} with
$\lambda=0$. Specifically, we consider 
the HERA I+II $\sigma_{\rm NC}^{e^\pm p}$
$E_p=575$~GeV~\cite{Abramowicz:2015mha} inconsistent dataset, and the consistent 
NMC \( F_2^d/F_2^p \) dataset~\cite{Arneodo:1996kd}. 
The specific choice of inconsistent dataset is motivated by the fact that the measurement taken with a proton 
beam energy of $E_p=575$ GeV is the least precise of the inconsistent
datasets, hence we expect the effect of the inconsistency  
to be more realistic. We then  look at the loss distribution for each of
these two datasets over the $N_{\rm fits}$ sets of $L_1$ data both in
the unweighted case, or when each of them is weighted in turn.
In Fig.~\ref{fig:flags_s} we plot as a function of the threshold $Z$
the probability of a true negative (consistent experiment not
flagged as inconsistent) and of a true positive  (inconsistent experiment 
flagged as inconsistent), when each of the conditions $S_1$
Eq.~(\ref{eq:incons}), $S_2$ Eq.~(\ref{eq:S2}) , or $S_3$
Eq.~(\ref{eq:S3}) is applied separately.
First, we see that indeed, as expected, and discussed at the end of
Sect.~\ref{subsec:test_setting}, there is a trade-off in that
increasing $Z$ increases the probability of true negatives but
decreases the probability of true positives, and
conversely. Furthermore, we note that condition $S_2$ is a more
restrictive version of condition $S_1$,  since giving a 
higher weight a given dataset can only improve its $\chi^2$. Indeed,
for fixed $Z$, $S_2$ in comparison to $S_1$ gives more true negatives,
but fewer true positives: effectively, condition $S_2$ behaves in a
way which is quite similar to that which is obtained using $S_1$, but
with a larger $Z$ value. 
Also as noted in Sect.~\ref{subsec:test_setting}, neither of these
selection criteria seems to lead to a reasonable compromise for any
value of $Z$: for instance, choosing $S_3$ with $Z\sim 0.5$ (or
equivalently $S_2$ with $Z\sim1$) leads to a
good fraction of true negatives, around 90\%, but an unacceptably
small fraction of true positives, smaller than 60\% .

On the other hand, selection based on
condition  $S_3$ seems rather more promising, as for most $Z$ values
(except very small ones $Z\lesssim 0.5$) it gives both a larger true
positive but also a larger true negative rate than conditions $S_1$
or $S_2$. This suggests that an optimal selection criterion can be
constructed by exploiting a suitable combination of criteria that
includes condition $S_3$.
\subsection{Optimal inconsistency detection}
\label{subsec:prescriptions}

As mentioned in Sect.~\ref{subsec:weightedfit}, in
Ref.~\cite{NNPDF:2021njg} it was suggested that inconsistencies could
be detected by sequentially applying criteria $S_1$ and then $S_2$ and
$S_3$. However, Fig.~\ref{fig:flags_s} suggests that a more effective
procedure could be to always check for $S_3$, even when a
dataset has not been flagged as inconsistent by  $S_1$.
Indeed, it is clear from the plot on the left of the figure that,
unless a very low value of $Z\lesssim0.3$ is 
chosen, there is little chance of $S_3$ leading to a false
positive even if one does not also require criterion $S_1$. On the
other hand,  whatever the value of $Z$, $S_3$ is always more
effective in catching true positives.

\begin{figure}[h]
  \centering
  \includegraphics[width=0.49\textwidth]{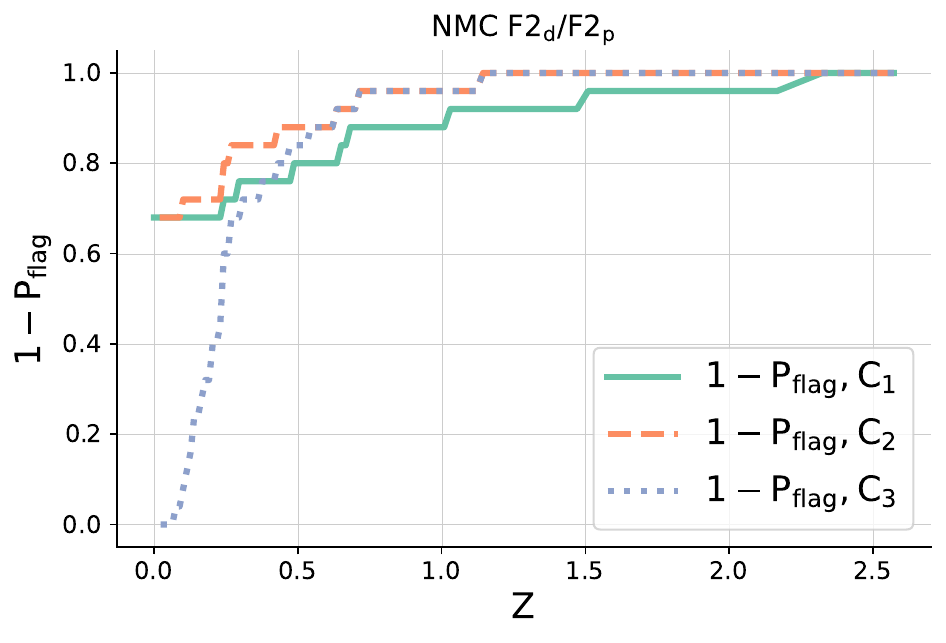}
  \includegraphics[width=0.49\textwidth]{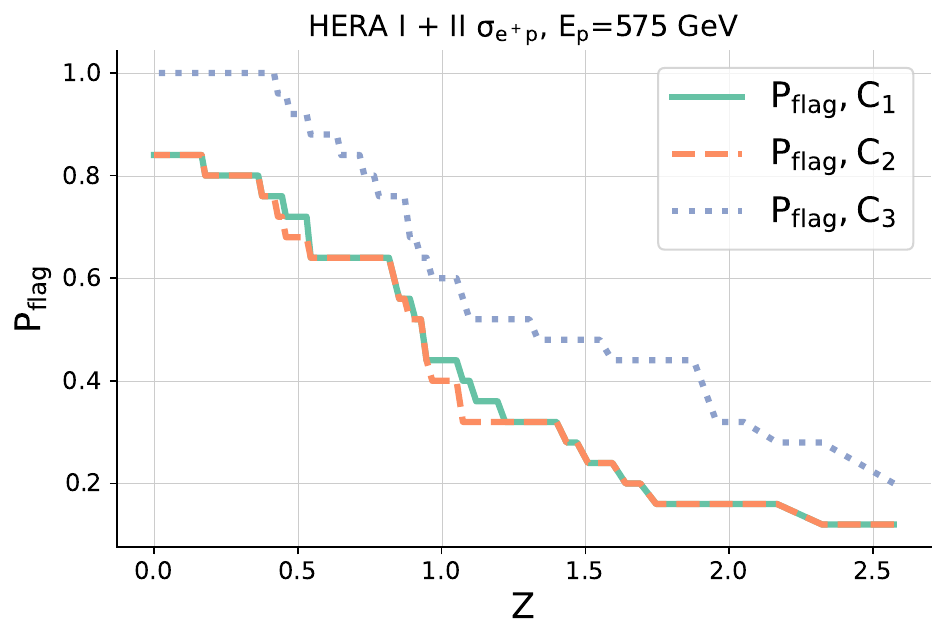}
  \caption{Same ad Fig.~\ref{fig:flags_s} but now flagging a dataset
    as inconsistent using the three criteria $C_1$-$C_3$ (see
    text). Criterion $C_1$ coincides with $S_1$ of
    Fig.~\ref{fig:flags_s}, while $C_2$ was adopted in  Ref.~\cite{NNPDF:2021njg}). 
} 
  \label{fig:flags_c}
\end{figure}

In order to illustrate this, we compare in Fig.~\ref{fig:flags_c} the
true and false positive rates based on using three different criteria,
namely: flag dataset $i$ as inconsistent if
\begin{itemize}
\item $C_1$:  condition $S_1$ is satisfied (same as shown in Fig.~\ref{fig:flags_s});
\item $C_2$: condition $S_1$ is satisfied, and in a weighted fit
  either $S_2$ or $S_3$ are satisfied (same as in 
  Ref.~\cite{NNPDF:2021njg});
\item $C_3$: in a weighted fit
  either $S_2$ or $S_3$ are satisfied.
\end{itemize}
  Note that the  difference between $C_2$ and $C_3$ is that in $C_2$
a weighted fit is only run for a given dataset
when condition $S_1$ is satisfied, while in
$C_3$ a weighted fit is always run for all datasets, and its result
are used to flag inconsistencies. Criterion $C_3$ is accordingly more
computationally intensive as it requires as many weighted fits as
there are datasets.

It is clear that  criterion $C_2$ used in Ref.~\cite{NNPDF:2021njg}
for $Z\gtrsim0.5$ is somewhat more efficient than purely using
criterion $C_1$ in terms of true negatives, though equally inefficient
in detecting true positives. On the other hand, criterion $C_3$ for
$Z\gtrsim0.5$ is equally efficient in detecting true negatives as
criteria $C_1$ and $C_2$, and only fails for very small $Z$ values,
where even a small deterioration in fit quality in other datasets upon
performing a weighted fit would lead to flagging the dataset as
inconsistent. But on the other hand there is a very considerable
improvement in efficiency in flagging true negatives, so that for
instance a value of $Z\approx 0.5$ would lead to 90\% efficiency both
for true positive and true negatives.

We conclude that selection based on criterion $C_3$ provides a
reliable way of identifying genuinely inconsistent datasets. Of
course, the precise value of $Z$ should be determined on a
case-by-case basis, it will in general depend on the nature of the
full global dataset, and should be determined by repeating this analysis for
all included datasets with a dedicated closure test.

%% file: sec-summary.tex
\section{Summary}
\label{sec:summary}

The possibility that tensions and incompatibilities  between 
experimental measurements may significantly affect the reliability of
PDFs and their uncertainty has been often stressed (see
e.g.~\cite{Ablat:2024muy}). In this work, we have investigated this
issue using the powerful
tool of closure testing, by studying the impact of data inconsistency
on PDF determination through the NNPDF methodology in a fully
controlled setting. While closure tests had been previously used as a
tool to validate the NNPDF methodology, this was done so far assuming perfect
theory and fully reliable data uncertainties. 
The analysis of situations in which an inconsistency is
built in the input dataset in a controlled way has enabled us perform
this validation under conditions that are closer to
real-world situations.
Indeed, the scenarios that we have investigated  cover a broad spectrum of
realistic situations,  
ranging from bulk inconsistency, in which the majority of the data that determine the PDFs are inconsistent, to 
an inconsistency localised in a single, though highly precise and relevant, dataset, or a single dataset 
that almost entirely determines one specific PDF.

Our results demonstrate that the NNPDF methodology manages to correct for
moderate to medium inconsistencies, leading to faithful PDFs and
uncertainties that generalise correctly to
data that are not part of the input dataset, even when they are
correlated strongly to the inconsistent data used for PDF
determination.
Focussing on the effect of inconsistencies of
experimental origin, we have learnt that the effect of the
inconsistency is
visible and distorts the PDFs only when the underestimate of 
experimental systematic uncertainties is strong, and then
only in regions in which  experimental measurements are
systematics-dominated. We have further developed a procedure in order
to detect cases in which uncorrected inconsistencies are present.
Taken together, these results provide a reassuring indication that 
as more data are
included in global PDF determination, thereby  making PDFs more
precise, it will be possible to achieve an accuracy that matches the level of precision. 

A natural development of our study  is to extend it to more realistic
situations, specifically by considering specific critical
experimental measurements and systematic uncertainties through a
direct involvement of the experimental community.
The other natural line of development is to use the same technique to
study the impact of inconsistencies of theoretical rather than
experimental origin, specifically the fact that predictions are always
obtained through a truncation of perturbation theory, thereby leading
to systematic shifts whose size is difficult to estimate reliably. 
These developments will be the object of future studies. In fact, the flexibility of the closure test methodology and the public
availability of the NNPDF code~\cite{NNPDF:2021uiq} will facilitate
these studies by any interested party.

%% file: app-comparison-est.tex
\section{The normalized bias and the bias-variance ratio}
\label{app:comp}

The main estimator that we used in this work in order to assess the 
faithfulness of uncertainties is the normalised bias $R_b$ defined  in
Eq.~\eqref{eq:bias_variance_ratio_definition}.
This differs from the bias-variance ratio used as an estimator in
previous
studies~\cite{Ball:2014uwa,NNPDF:2021njg,DelDebbio:2021whr}. The
bias-variance ratio is defined by first  introducing the bias and
variance, respectively given by (using the notation of Sect.~\ref{subsec:est}) 
\begin{align}
    B^{(l)}(C) & = \left(\expval_{\ep}{\cal G}(u_{*,k}) - f\right)^T
                 C^{-1} \left(\expval_{\ep}{\cal G}(u_{*,k}) -
                 f\right) , \label{eq:old_bias_definition}\\
      V^{(l)}(C) & = \expval_{\ep}\left[\left(\expval_{\ep}{\cal
                   G}(u_{*,k}) - {\cal G}(u_{*,k})\right)^T C^{-1}
                   \left(\expval_{\ep}{\cal G}(u_{*,k}) - {\cal
                   G}(u_{*,k})\right)\right] . \label{eq:old_variance_def}
\end{align}
One can then define a bias-variance ratio for the $l$-th instance of
$L_1$ data
\begin{equation}
  \label{eq:bvr}
   R^{(l)}_{bv} = \frac{ B^{(l)}(C) }{  V^{(l)}(C)},
    \end{equation}
and the bias-variance ratio indicator is then the average root-mean square
bias-variance ratio, namely
\begin{align}
\label{eq:bias_variance_ratio_old_definition}
    R_{bv} = \sqrt{\frac{\expval_\eta B^{(l)}(C)}{\expval_\eta V^{(l)}(C)}}.
\end{align}

Note that if one replaces the experimental
covariance matrix  $C$ with the average PDF covariance matrix
$\overline{C}_{\rm PDF}$ Eq.~(\ref{eq:avcovmat}) then the
bias-variance ratio coincides with the normalised bias. In fact, the
bias-variance ratio for the $l$-th instance is given by
\begin{align}
    \label{eq:bvr_diag}
        R_{bv}^{(l)} (\overline{C})  = \frac{1}{N_{\rm data}}\sum_{i=1}^{N_{\rm data}}
        \frac{(\hat\Delta^{(l)}_i)^2}{(\hat \sigma_{{\rm PDF} i})^2},
\end{align}
to be compared to Eq.~(\ref{eq:bias_diag}), where now  $\hat \sigma_{{\rm PDF} i}$  and $\hat\Delta^{(l)}_i$
are respectively the $i$-th eigenvalue and projection of the vector
$\left(\expval_{\ep}{\cal G}(u_{*,k}^{(l)})_j - f_j\right)$ along the $i$-th
normalised eigenvector $\hat v^{(i)}_j$
of $\overline{C}$. In other words, the
normalised bias Eq.~(\ref{eq:bias_diag}) and the bias variance ratio
Eq.~(\ref{eq:bvr_diag}) are both equal to
the ratio of the measured deviation from truth over its expected value,
but the former in the basis of eigenvectors of the PDF covariance
matrix, and the latter in the basis of eigenvectors of the
experimental covariance matrix.

The normalised bias provides a more stable measure of faithfulness
when there are datasets that  contain significantly different numbers
of datapoints. This can be understood considering a simple toy
scenario. Assume that there are only two uncorrelated datasets.
The bias is then given by
\begin{equation}
  B^{(l)}(C)= B_1^{(l)}(C)+ B_2^{(l)}(C),
\end{equation}
where $B^{(l)}_i$ is the bias computed for the $i$-th dataset. Assume
now that dataset one 
contains 3 datapoints and dataset two contains 300 datapoints. The
purely statistical  fluctuation of the bias between different $L_1$
instances in a closure test is ten times larger for dataset one than
it is for dataset two. In a particular instance in which the bias for
dataset one is ten times larger than that of dataset two it will
completely dominate the determination of the bias despite having a
much smaller number of datapoints.
This does not happen when using the normalised bias because in this
case the fluctuations about the underlying truth are directly measured
in the PDF eigenvector basis, and thus correctly normalised by the
Principal Component Analysis in terms of the projection of the bias
for individual datapoints along the  direction of the underlying
relevant PDF eigenvectors.

%% file: app-bootstrap_def.tex
\section{The bootstrap algorithm}
\label{app:bootstrap_def}

All closure test indicators, such as the normalized bias $R_b$,
have been obtained using 25 samples of 100
replicas. In order to estimate the uncertainty related to the finite
number of samples and to the finite
size of each replica sample we have used the bootstrap
method~\cite{Efron:1979bxm,Efron:1986hys}. We summarise here its implementation.
Given $n$ samples of $m$ replicas $F_i$
the algorithm involves the following steps:
\begin{enumerate}
\item Bootstrap sample generation: randomly select with replacement 
$n$ samples out of the $n$ available samples. Within each selected
  sample  randomly select again with replacement, 
  $m$ replicas out of the $m$ available replicas.
  This process creates a bootstrap sample comprising $n$ samples of
  $m$ replicas $F^*_i$.

\item Bootstrapped calculation: compute the value  $R_b^*$ of the desired
  quantity, e.g. the normalised bias $R_b$ using the bootstrapped
  sample of bootstrapped replicas. 

\item Repetition: repeat steps 1 and 2 for a total of $k$  iterations,
  generating $k$ instances of $R_b^{*(k)}$ of  $R_b^*$. We choose $k=100$.

\item Inference: estimate the bootstrapped value  and uncertainty of
  the indicator as the means and variance of the sample of bootstrapped instances:
\begin{align}
 \expval^*[R_b^*] &= \frac{1}{k}\sum_{i=1}^{k}{R_{b}^{*(k)}}, \\
 \text{Var}^*(R_{b}^*) & = \frac{1}{k-1}\sum_{i=1}^{k}\bigg({R_{b}^{*(k)}} -\expval^*[R_b^*]\bigg)^2 .
\end{align}
\end{enumerate}
%


%% file: app-pdfobscorr.tex
\section{Correlation between PDFs and observables}
\label{app:corr}

In order to understand how the inconsistency in different data
potentially affects individual PDFs it is useful to consider the
correlation between PDFs and observables,  defined in
Ref.~\cite{Carrazza:2016htc} (see also Ref.~\cite{Ball:2021dab}):
\begin{equation}
\label{eq:correlations}
    \rho(j,x,{\cal O})\equiv\frac{N_{\rm rep}}{N_{\rm rep}-1}
    \left(\frac{\langle f_j(x,Q)\,{\cal O}\rangle_{\rm rep}-\langle f_j(x,Q)\rangle_{\rm rep}\langle {\cal O}\rangle_{\rm rep}}{\Delta_{\rm PDF}f(x,Q)\,\Delta_{\rm PDF}{\cal O}}\right),
\end{equation}
where  $f_j(x,Q)$ is a PDF of the $j$-th flavor and ${\cal O}$ is a
specific PDF-dependent observable.

\begin{figure}
 \centering
 \includegraphics[width=0.48\textwidth]{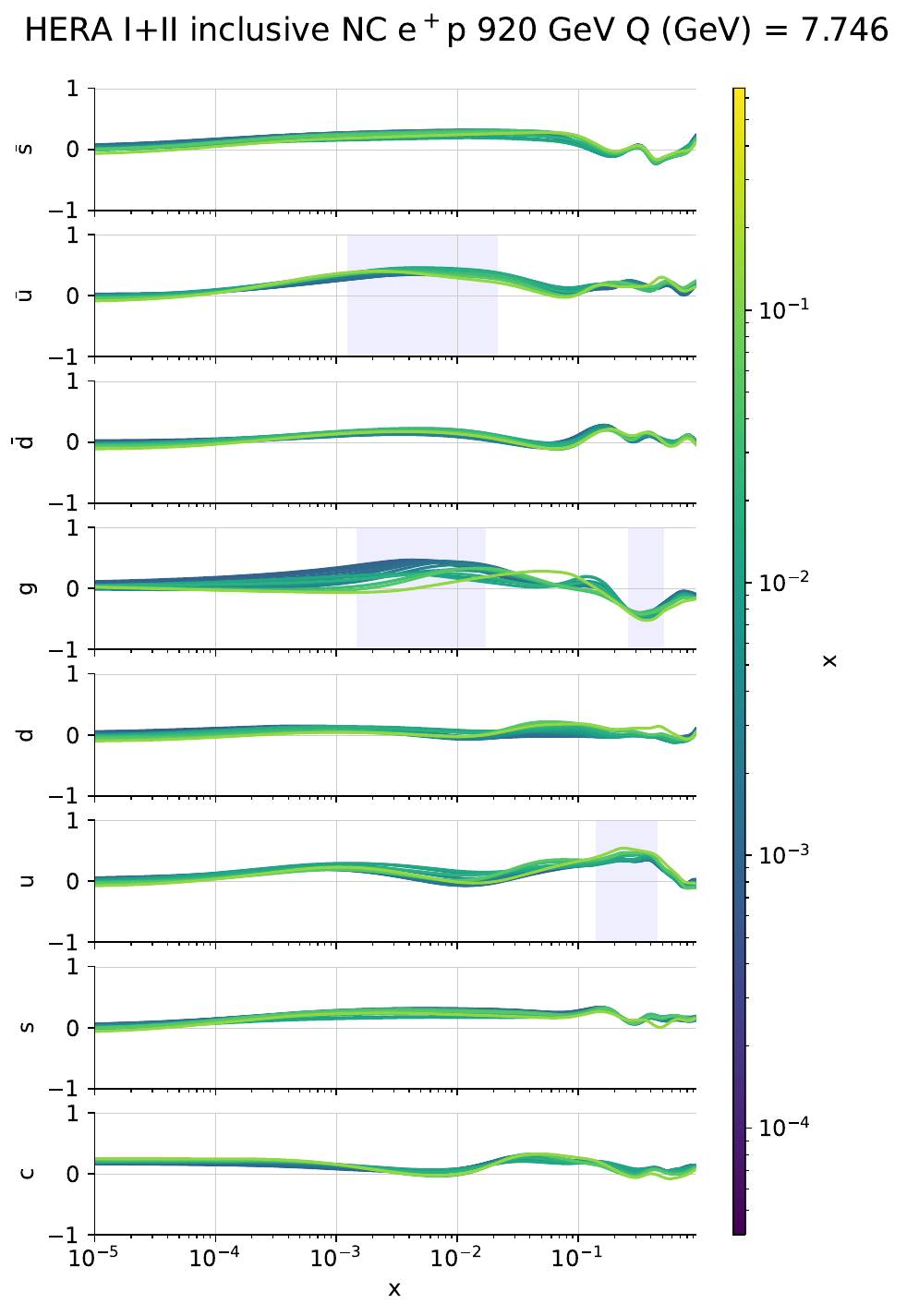}
 \includegraphics[width=0.48\textwidth]{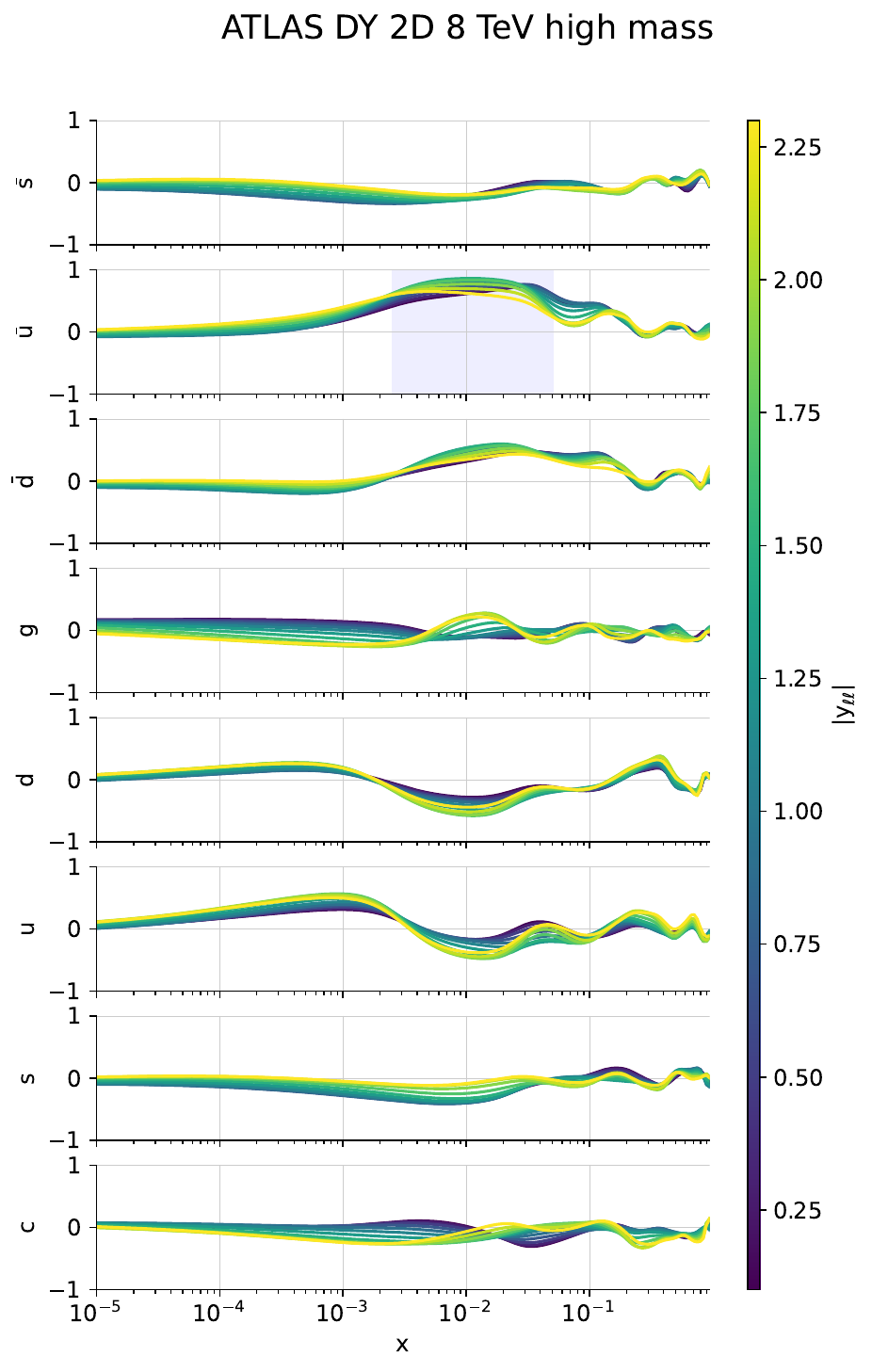}
   \caption{The correlation coefficient $\rho$
     Eq.~\eqref{eq:correlations} between the closure test  PDFs of the
     $\lambda=1$  set of  Sect.~\ref{subsec:dis}, 
     and the neutral-current deep-inelastic reduced cross-section $,{\cal
  O}=\sigma_{\rm NC}^{e^+ p}(x,Q^2)$. The correlation coefficient is
     plotted vs. the $x$ value of the PDF for PDFs at the
     parametrisation scale 
     $Q_0 = 1.651$~GeV. Each curve corresponds to a datapoint in
     the HERA I+II dataset of Tab.~\ref{tab:ratio_bv_dis}, with
     $Q^2=60$~GeV$^2$ ($E_p=920$~GeV data, left panel) or
     $10\leq Q^2\leq30$~GeV$^2$ ($E_p=460$~GeV data, right panel) and the
     value of $x$ shown by the color coding.
     The region in which $\rho > 0.7 \rho_{\rm max}$ is highlighted.}
   \label{fig:obs_corr_dis}
\end{figure}

In  Fig.~\ref{fig:obs_corr_dis} we display this correlation choosing
as an observable 
the neutral-current deep-inelastic reduced cross-section, ${\cal
  O}=\sigma_{\rm NC}^{e^+ p}(x,Q^2)$. The correlation is computed between
PDFs and datapoint in the HERA I+II $E_p=920$~GeV dataset (see
Tab.~\ref{tab:ratio_bv_dis}); each curve provides the correlation
with an individual datapoint,  with $Q^2=60$ GeV$^2$  (left) and
$Q^2=75$ GeV$^2$ (right) and the $x$ value for each datapoint shown
as a color scale, plotted versus the PDF $x$ value with  $Q=1.651$~GeV$^2$. The chosen dataset is among
those in which an inconsistency was introduced, and the bins shown are
those in which the larges values of the normalized bias was found: see
Sect.~\ref{subsec:dis} for a discussion. The correlation has been
computed using the $\lambda=1$ PDF set of Sect.~\ref{subsec:dis}.

\begin{figure}
  \centering
  \includegraphics[width=0.48\textwidth]{images/final/plot_smpdf_1.pdf}
  \includegraphics[width=0.48\textwidth]{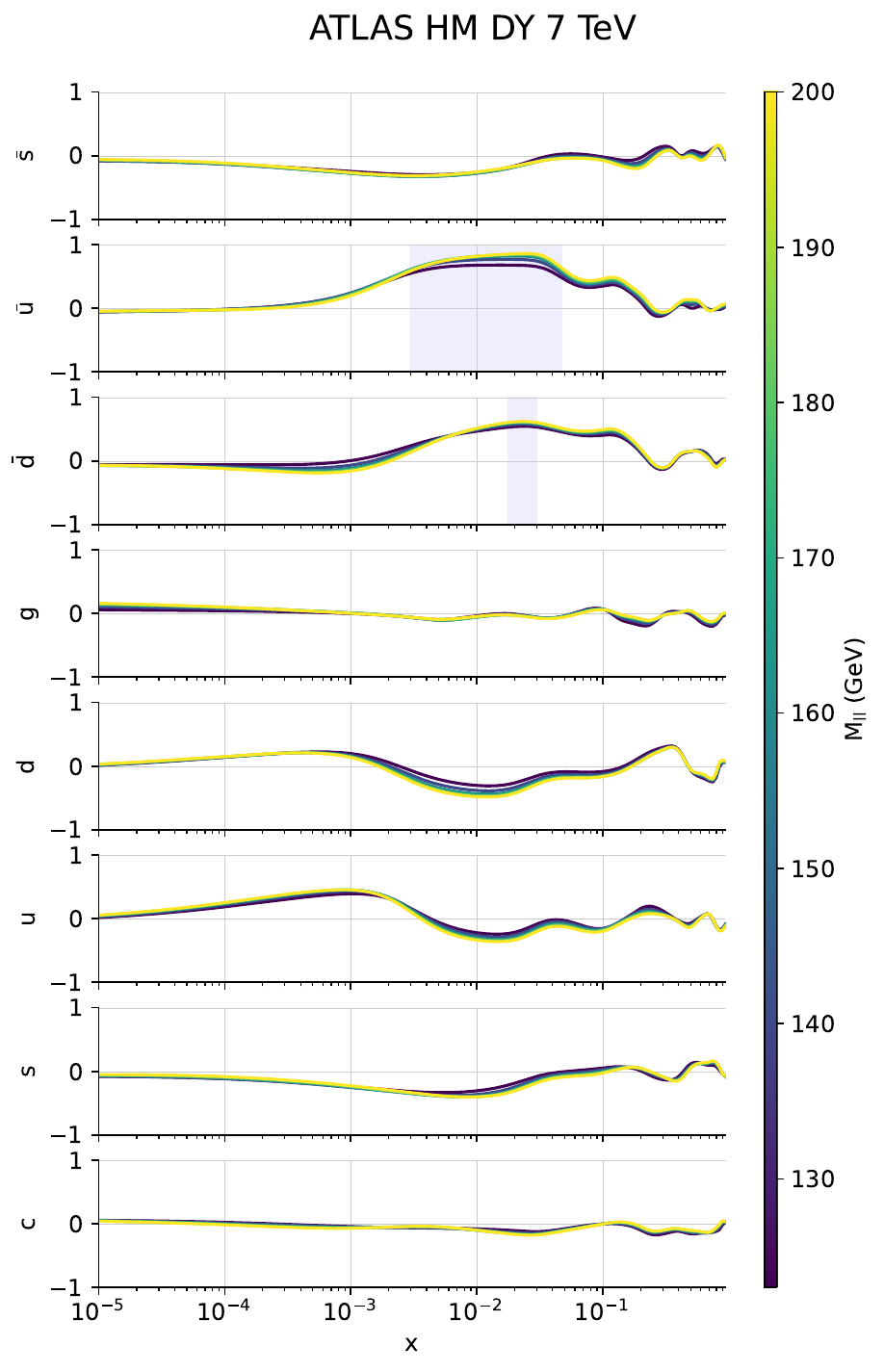}
  \caption{Same as Fig.~\ref{fig:obs_corr_dis}, for the closure test
    PDFs of Sect.~\ref{subsec:dy} and ATLAS high-mass
Drell-Yan measurements at $8$ and $7$~TeV respectively. For the
$8$~TeV dataset (left) each curve refers to a
dilepton rapidity bin, averaged over all values of the gauge boson
virtuality, and for the $7$~TeV dataset (right) to an invariant mass bin.}
    \label{fig:obs_corr_dy}
 \end{figure}

In Fig.~\ref{fig:obs_corr_dy} we display the correlation computed using
the $\lambda=1$ PDF set of Sect.~\ref{subsec:dy}, now shown for the
single dataset in which the inconsistency is introduced, and for the
out-of-sample 
dataset that is mostly affected by it, namely
the ATLAS high-mass
Drell-Yan measurements at $8$ and $7$~TeV respectively, also shown in
Fig.~\ref{fig:ratio_bv_dy_oos}. In this case, each curve for the
$8$~TeV dataset refers to a
dilepton rapidity bin, averaged over all values of the gauge boson
virtuality, and for the $7$~TeV to an invariant mass bin. 

\begin{figure}
  \centering
  \includegraphics[width=0.48\textwidth]{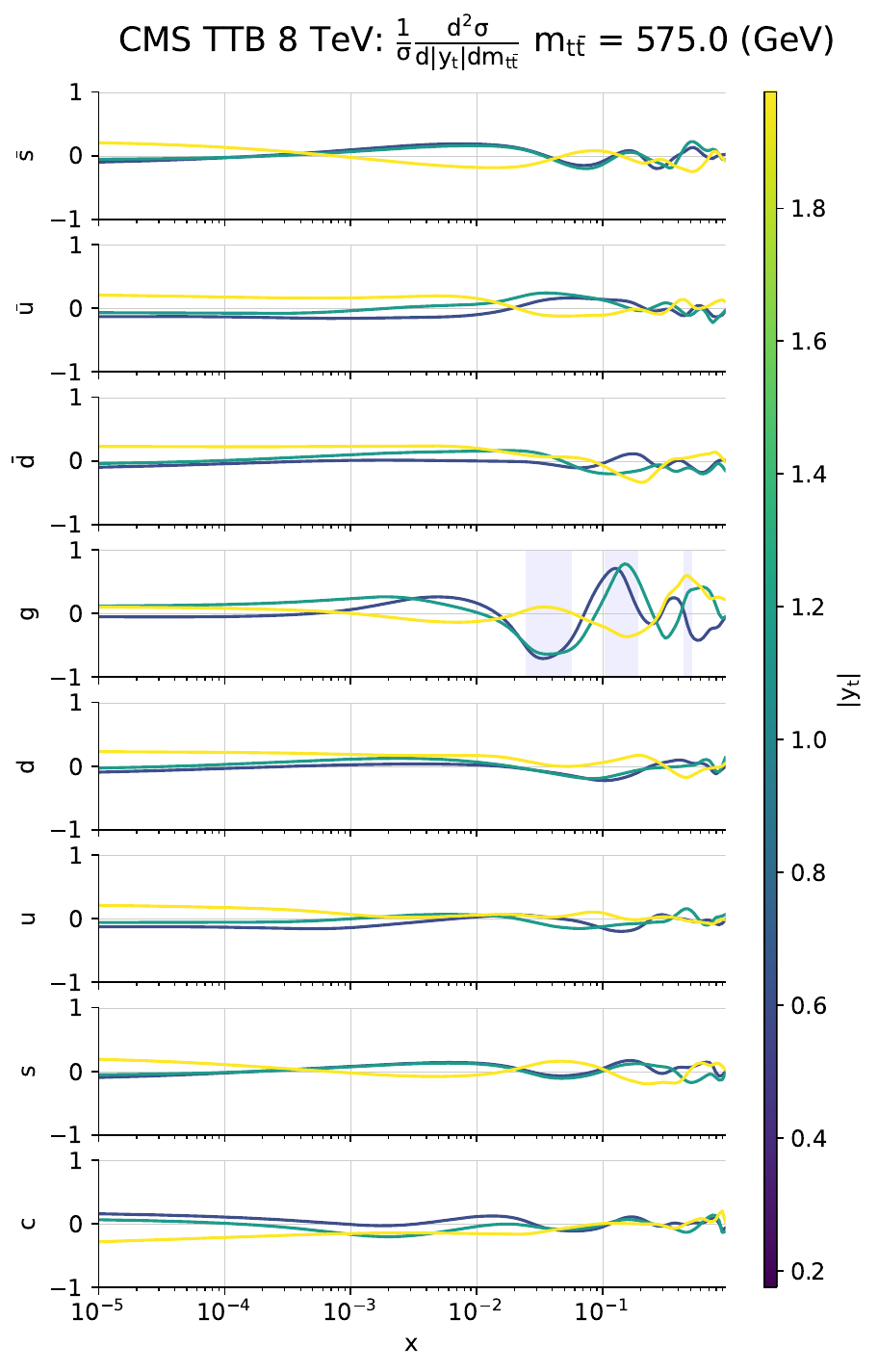}
  \includegraphics[width=0.48\textwidth]{images/final/plot_smpdf_2.pdf}
    \caption{Same as Fig.~\ref{fig:obs_corr_dis}, for the closure test
      PDFs of Sect.~\ref{subsec:jets} and
ATLAS
single-inclusive  jets (left) and CMS $t\bar{t}$ double differential
cross section at $8$~TeV (right), with individual curves corresponding
to fixed $k_T$ in the former case, and to fixed top rapidity averaged
over top pair invariant mass in the latter case.}
    \label{fig:obs_corr_jet}
 \end{figure}

We finally show in Fig. \ref{fig:obs_corr_jet} the correlations for the
closure test 
PDFs of Sect.~\ref{subsec:jets}, also in this case for the
dataset in which the inconsistency is introduced, and for the
out-of-sample 
dataset that is mostly affected by it, namely respectively ATLAS
single-inclusive  jets and the CMS $t\bar{t}$ double differential
cross section at $8$~TeV. 
